\documentclass[prd,10pt]{revtex4}
\usepackage[dvips]{graphicx}
\usepackage{dcolumn}
\usepackage{bm}
\usepackage{amsmath,amsfonts}

\newcommand{\nc}{\newcommand}
\nc{\be}[1]{\begin{equation} \mbox{$\label{#1}$}}
\nc{\ee}{\end{equation}}
\nc{\bea}[1]{\begin{eqnarray} \mbox{$\label{#1}$}}
\nc{\eea}{\end{eqnarray}}
\nc{\lr}{\Leftrightarrow}
\nc{\set}[1]{\left\{#1\right\}}
\nc{\bset}[1]{\left(#1\right)} \nc{\sbset}[1]{\left[#1\right]}
\nc{\state}[1]{\left<#1\right>}
\nc{\eq}[1]{\mbox{Eq.~(\ref{#1})}}
\nc{\eqs}[2]{\mbox{Eqs.~(\ref{#1}, \ref{#2})}}
\nc{\fig}[1]{\mbox{Fig.~\ref{#1}}}
\nc{\tbl}[1]{\mbox{TABLE~\ref{#1}}}
\nc{\figs}[2]{\mbox{Figs.~\ref{#1}, \ref{#2}}}
\nc{\nb}{\nonumber}
\nc{\order}[1]{\mathcal{O}(#1)}
\nc{\half}{\frac{1}{2}}
\nc{\tisig}{\tilde{\sigma}}
\nc{\hatsig}{\hat{\sigma}}
\nc{\tirhoE}{\tilde{\rho}_E}
\nc{\dsig}{\delta\sigma}
\nc{\dtheta}{\delta\theta}
\nc{\p}{\prime}
\nc{\ka}{\frac{\mathbf{k}}{a}}
\nc{\spc}{\hspace*{10pt}}
\nc{\tirhoQ}{\tilde{\rho}_Q}
\nc{\tiu}{\tilde{u}}

\begin{document}

\title{Affleck-Dine dynamics, $Q$-ball formation and thermalisation}

\author{Mitsuo I. Tsumagari\footnote{ppxmt@nottingham.ac.uk}} 
\affiliation{\vspace{.5cm}\\ School of Physics and Astronomy, University of Nottingham, University Park, Nottingham NG7 2RD, UK}
\begin{abstract}
We present both analytically and numerically the consistent analysis from the Affleck-Dine (AD) dynamics to the subsequent semiclassical evolution in both gravity-mediated and gauge-mediated models. We obtain analytically the elliptic motions in the AD dynamics as the analogy of the well-known Kepler-problem, and by solving the equations of motion in a lattice, we find that the semiclassical evolution goes through three distinct stages as a nonequilibrium process of reheating the Universe: \emph{pre-thermalisation}, \emph{bubble collisions} and \emph{thermalisation}. We report that the second stage of our case lasts rather long compared to the second stage of the reheating case, and the thermalisation process is unique due to the presence of ``thermal $Q$-balls''.
\end{abstract}
\vskip 1pc 
\maketitle
%
\section{Introduction}
The present baryon asymmetry in the Universe is one of the most mysterious problems in cosmology and particle physics, see a review \cite{Dine:2003ax}. Within the Standard Model (SM), the electroweak baryogenesis was suggested to explain the inequality between baryon and anti-baryon number, and the recent developments are shifted into constructing the theory of reheating the Universe \cite{Kofman:1997yn}. The electroweak baryogenesis satisfies the well-known Sakharov's three conditions required for successful baryogenesis \cite{Sakharov:1967dj}, namely baryon number production, charge parity (CP) violation and the process taking place out-of-equilibrium; however, the predicted CP violation is too small to explain the present observed baryon number. By satisfying the above three conditions, the Affleck-Dine (AD) baryogenesis \cite{Affleck:1984fy}, which was proposed in the theoretical framework beyond the SM, namely, the Minimal Super-symmetric Standard Model (MSSM), is a more successful scenario to tackle this puzzle, since it may solve problems of gravitino and moduli overproduction and give rise simultaneously to the ordinary matter and dark mater in the Universe. The MSSM has many gauge-invariant flat directions along which R parity is preserved. The flat directions are lifted by super-symmetry (SUSY) breaking effects arising from nonrenormalisable terms, which give a U(1) violation through A-terms. In the original scenario of the AD baryogenesis, one can parametrise one of the flat directions in terms of a complex scalar field known as an AD field (or AD condensate which consists of a combination of squarks and/or sleptons fields). The AD field evolves to a large field expectation value during an inflationary epoch in the early Universe. After inflation, the orbit of the AD field can be kicked along the phase direction due to the A-terms which generate the U(1) charge (baryon/lepton number), and then the A-terms become negligible, where the AD field rotate towards the global minimum of the scalar potential. Hence, the generated global U(1) charge is fixed and the orbit of the AD field rotates around the origin of the complex field-space, \textit{c.~f.} the anomaly mediated models \cite{Randall:1998uk}. After the AD condensate decays into usual baryons and leptons, AD baryogenesis becomes complete.

The trajectory of the AD field is identical to the planetary orbits in the well-known Kepler-problem as we will show later, replacing the Newtonian potential by an isotropic harmonic oscillator potential \cite{Tort:1989}. This coincidental classical-mechanics reduction was noted for the orbits of a probe brane in the Branonium system \cite{Burgess:2003qv, Rosa:2007dr}. As general relativity predicted that planetary orbits precess by adding the relativistic correction to the Newtonian potential, we will see the similar events occur for the orbits of AD fields, which are disturbed by quantum and nonrenormalisable effects. 

By including quantum corrections \cite{Enqvist:1997si, Nilles:1983ge} and/or thermal effects \cite{Allahverdi:2000zd} in the mass term of the standard AD scalar potentials, the AD condensate is classically unstable against spatial perturbations due to the presence of negative pressure \cite{McDonald:1993ky}, and fragments to bubble-like objects, eventually evolving into $Q$-balls \cite{Coleman:1985ki}. Lee pointed out \cite{Lee:1994qb} that $Q$-balls may form due to bubble nucleation (first order phase transition) \cite{Coleman:1977py}, even in the case that the condensate is classically stable against the linear spatial perturbations.

A $Q$-ball is a nontopological soliton \cite{Friedberg:1976me} whose stability comes from the existence of a continuous global or local charge $Q$ (see a review \cite{Lee:1991ax} and references therein). Tsumagari \textit{et.~al.} \cite{Tsumagari:2008bv, Copeland:2009as} showed previously the complete stability analysis of $Q$-balls at zero-temperature in both polynomial potentials and MSSM flat potentials. Laine \textit{et.~al.} \cite{Laine:1998rg} investigated the stability of $Q$-balls in a thermal bath. The stability of the thermal SUSY $Q$-balls is different from the one of the standard ``cold'' $Q$-balls, since they suffer from evaporation \cite{Laine:1998rg}, diffusion \cite{Banerjee:2000mb}, dissociation \cite{Enqvist:1998en}, and decays into light/massless fermions \cite{Cohen:1986ct}. Therefore, most SUSY $Q$-balls are generally not stable but long-lived, and it may thermalise the Universe by decaying into baryons on their surface \cite{Enqvist:2002rj}, which could solve the gravitino and moduli over-production problems without fine-tuning. The SUSY $Q$-balls in gravity-mediated (GRV-M) models are quasi-stable decaying into the lightest SUSY particles (LSP dark matter), and the fraction of the baryons from the $Q$-balls may give the present baryon number, which can explain the similarity of the energy density between the observed baryons and dark matter \cite{Kusenko:1997si, Enqvist:1998en}. The SUSY $Q$-balls in gauge-mediated (GAU-M) models, however, can be extremely long-lived so that those $Q$-balls are a candidate of cold dark matter \cite{Kusenko:1997si} and may give the present observed baryo-to-photon ratio \cite{Laine:1998rg}.

The dynamics and formation of $Q$-balls have been investigated numerically. With different relative phases and initial velocities, the authors \cite{Axenides:1999hs} found a charge transfer from one $Q$-ball to the other and interesting ring formation after the collision. It has been found \cite{Radu:2008pp} that similar ring-like solutions are responsible for the excited states from the ground state ($Q$-ball) by introducing extra degrees of freedom: spatial spins \cite{Volkov:2002aj} and twists \cite{Axenides:2001pi}. The formation of $Q$-balls after inflation have been extensively investigated in both GRV-M models \cite{Kasuya:2000wx} and GAU-M models \cite{Kasuya:1999wu, Kasuya:2001hg}, in which SUSY is broken by either gravity or gauge interactions. As we will show, the $Q$-ball formation involves a nonequilibrium dynamics which is related to reheating problem in cosmology.

The reheating process after the inflation period involves nonlinear, out-of-equilibrium, and nonperturbative phenomena so that it is extremely hard to construct a theory for the whole process, see the 2 particle irreducible effective action as a remarkable approach \cite{Berges:2004yj}. In the first stage of reheating (\emph{preheating}), it is currently well known that the fluctuations at low momenta are amplified, which leads to explosive particle production. After preheating, the subsequent stages towards equilibrium are described by the wave kinetic theory of turbulence; Micha \textit{et.~al.} \cite{Micha:2004bv} recently estimated the reheating time and temperaute. These turbulent regimes appear in a large variety of nonequilibrium process, and indeed, the evolution of $Q$-ball formation experiences the active turbulence at which stage, many bubbles collide as observed in the next stage of tachyonic preheating \cite{Felder:2000hj}. During this bubble-collision stage within the reheating scenario, it is believed that gravitational waves may be emitted from the stochastic motion of heavy objects \cite{Felder:2006cc, GarciaBellido:2007af}. The problem of gravitational wave emissions has been discussed only in the fragmentation stage of $Q$-ball formation so far \cite{Kusenko:2008zm}, but not in the collision stage as opposed to the preheating cases.

In this paper, we show analytically and numerically that in GRV-M and GAU-M models the approximate trajectory of the AD fields is, respectively, precessing spiral or shrinking trefoil due to the quantum, nonrenormalisable, and the Hubble expansion effects. Moreover, we explicitly present the exponential growth of the linear spatial perturbations in both models. By introducing $3+1$ (and $2+1$)-dimensional lattice simulations, we identify that the evolution in $Q$-ball formation involves nonequilibrium dynamics, including turbulent stages. Following the pioneering work on the turbulent thermalisation by Micha \textit{et.~al.} \cite{Micha:2004bv}, we obtain scaling laws for the evolution of variances during the $Q$-ball formation.

The paper is divided as follows. We explore both analytically and numerically the dynamics of the AD field in Sec. \ref{sectorbit}. In Sec. \ref{sectinst}, we study the late evolution of the AD fields and the process of $Q$-ball formation, introducing detailed numerical lattice results. Finally, we conclude and discuss our results in Sec. \ref{concl}. Two appendices are included. We obtain the equations of motion and their perturbed equations for multiple scalar fields in Appendix \ref{MULTI}. In Appendix \ref{App2}, we find elliptic forms for the orbits of AD fields.
\section{The Affleck-Dine dynamics}\label{sectorbit}
In this section, we investigate the equation for the orbit of an AD condensate. This orbit coincides with the well-known orbit equation in the centre force problem in classical dynamics, i.e. planetary motions so that we call the AD condensate ``AD planet'' sometimes. For the bound orbits, the effective potential should satisfy the condition where the curvature at the minimum of the effective potentials should be positive. In the presence of Hubble expansion, the effective potential depends on time; thus, the full solution of the orbit equations can be obtained numerically except the case that the AD field is trapped by a quadratic potential. In Appendix \ref{App2}, we obtain the exact orbit in this exceptional case when Hubble expansion is assumed to be small and adiabatic. The orbit of the AD planet, or more precisely an eccentricity of the elliptic motion in the complex field-space, is determined by the initial charge and energy density. In order to obtain analytic expressions of the orbit in more general potential cases in which we are more interested, we restrict ourself to work in Minkowski spacetime and on the orbit which should be nearly circular. In Appendix \ref{App2}, we also obtain the perturbed orbit equation and necessary conditions for closed orbits where the orbits come back to their original positions after some rotations around the minimum of the effective potential. In Bertrand's theorem \cite{Bertrand}, there are only two potential forms allowed to be closed orbits: isotropic harmonic force and the inverse-squared force. Each of the central forces gives dynamical symmetries, namely Fradkin tensor \cite{Fradkin} and Runge-Lenz vector \cite{Laplace-Runge-Lenz}, respectively. These dynamical charges are obtained both classically  by the algebra of Poisson bracket \cite{Stehle:1967} and quantum-mechanically by the corresponding Lie algebra in the abelian case \cite{Higgs:1978yy} as well as nonabelian case \cite{Leemon:1978yz}. By approximating phenomenologically motivated models that appear in the MSSM and using the results in Appendix \ref{App2}, we present, in this section, analytic motions of the nearly circular orbits and the pressure of the AD planets. Further, we check these analytic results with full numerical solutions.

\vspace*{10pt}

\begin{figure}[!ht]
  \begin{center}
	\includegraphics[angle=-90, scale=0.31]{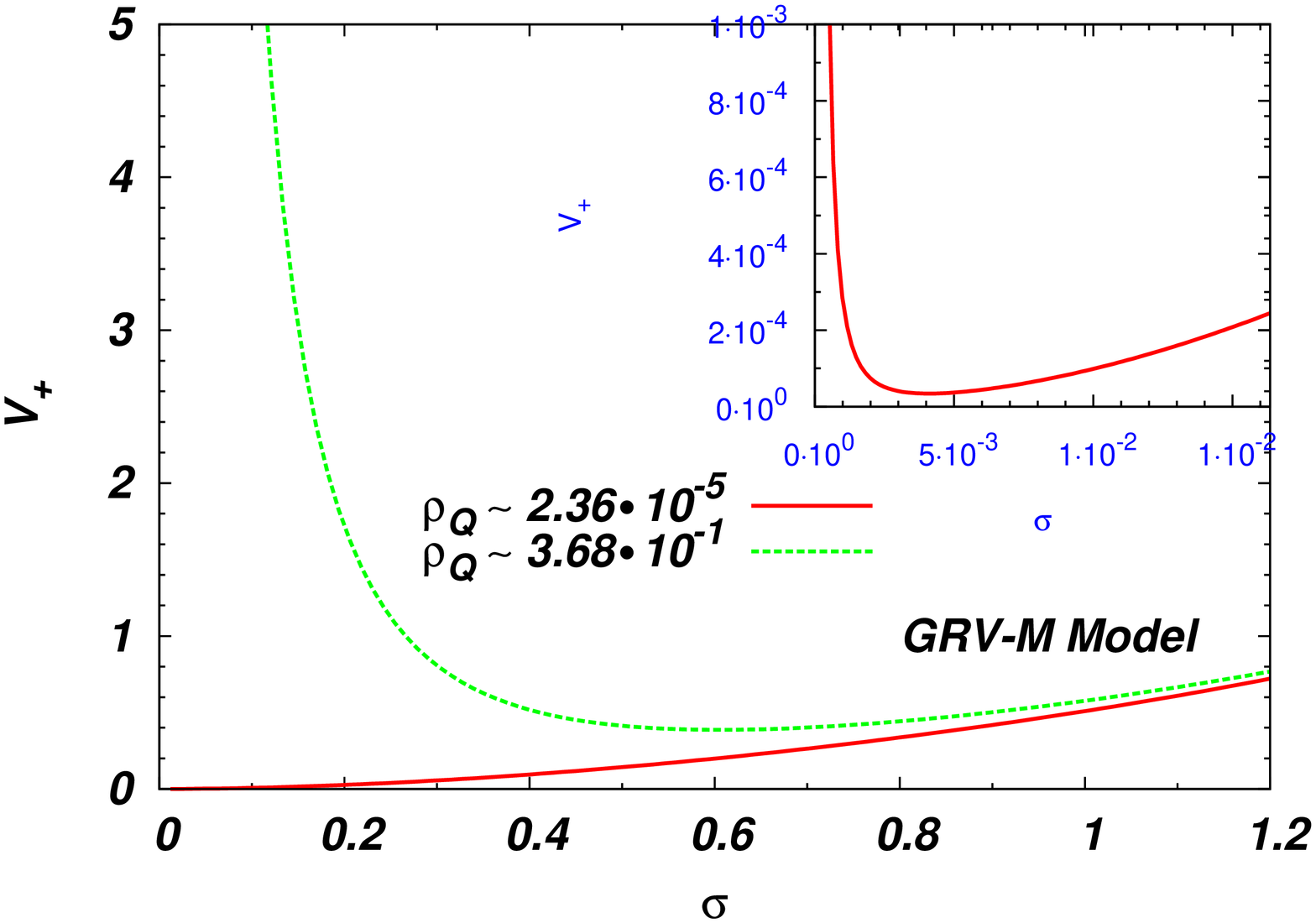}
	\includegraphics[angle=-90, scale=0.31]{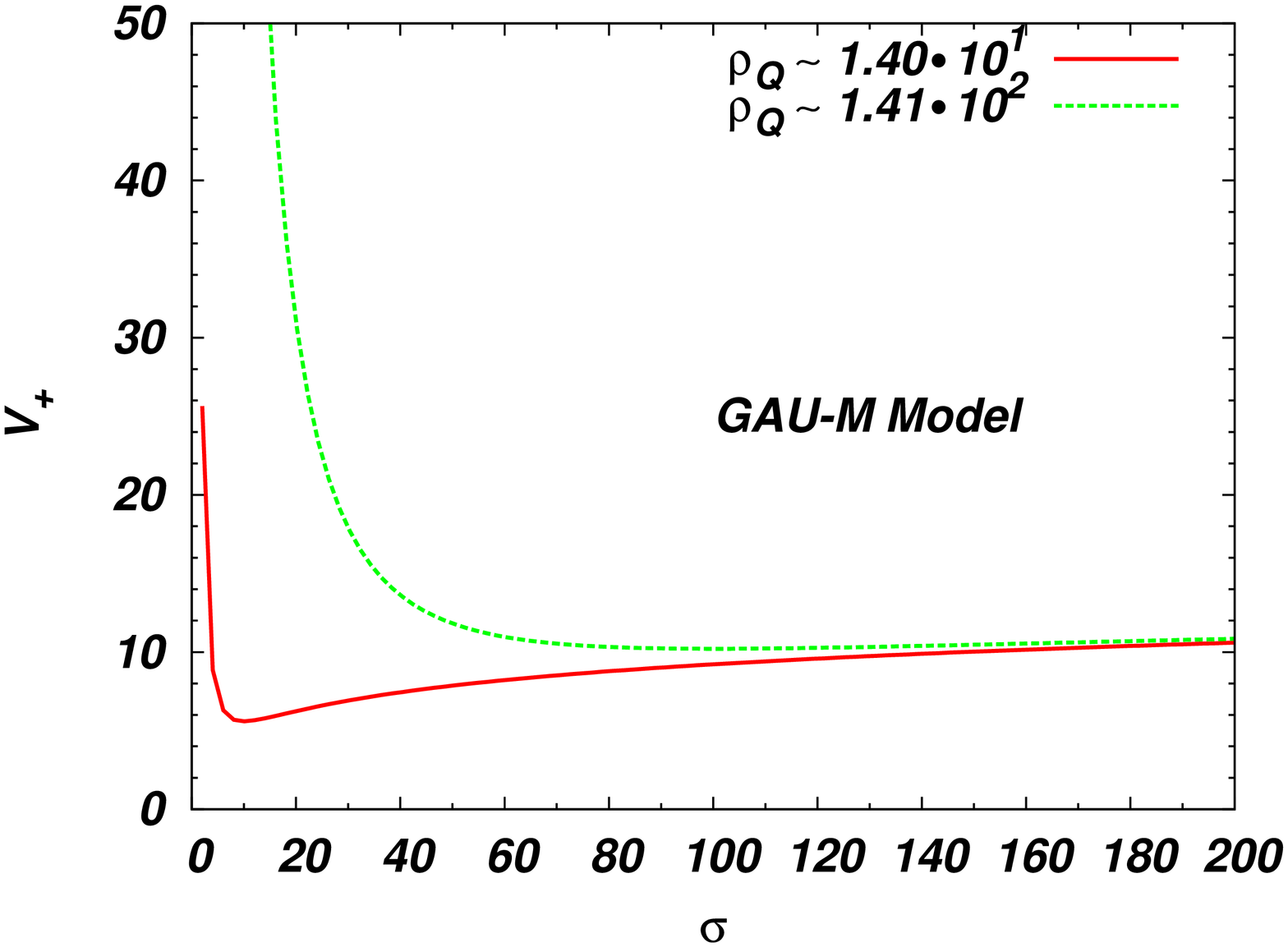}
  \end{center}
\caption{ \textbf{(Color online)} We show the effective potentials, $V_+\equiv V(\sigma)+\frac{\rho^2_Q}{2\sigma^2}$, against $\sigma$ in two types of potentials which we call gravity mediated model (GRV-M Model) on left and gauge mediated model (GAU-M Model) on right. The potential in GRV-M Model has the following form, $V(\sigma)=\half \sigma^2 \bset{1-|K|\ln\sigma^2}+b^2_*\sigma^6$, where, we set $|K|=0.1$ and $b^2_*=\frac{|K|}{4e}\sim 9.20 \times 10^{-3}$. The potential in GAU-M Model is $V(\sigma)=\ln\bset{1+\sigma^2}+b^2 \sigma^6$, where we set $b^2\sim10^{-30}$. We choose the following values of $\rho_Q$: red-solid line for $\rho_Q\sim 2.36\times 10^{-5}$ and green-dashed line for $\rho_Q=1/e\sim 3.68\times 10^{-1}$ in GRV-M Model and red-solid line for $\rho_Q\sim 1.40\times 10^1$ and green-dashed line for $\rho_Q \sim 1.41\times10^2$ in GAU-M Model.}
  \label{fig:adpot}
\end{figure}

Let us consider a motion of AD fields in an expanding universe with scale factor $a(t)$ and Hubble parameter $H=\dot{a}/a$, where a over-dot denotes the time derivative. We investigate the AD field after they start to rotate around the origin of the effective potentials and the value of the U(1) charge $\rho_Q$ is fixed due to negligible contributions from A-terms. By decomposing the complex (AD) field $\phi$ as $\phi(t)=\sigma(t) e^{i\theta(t)}$, where $\sigma$ and $\theta$ are real scalar fields, the equations of motion for $\sigma(t)$ and $\theta(t)$ (see \eqs{sigeom}{thetaeom} in Appendix \ref{MULTI}) are
\bea{radhomo}
\ddot\sigma +3H\dot\sigma+\frac{dV_+}{d\sigma}&=&0,\\
\label{phshomo}\ddot\theta+3H\dot\theta+\frac{2}{\sigma}\dot\sigma\dot\theta&=& 0\spc \lr\spc \frac{d\rho_Q}{dt}=0,
\eea
where the conserved comoving charge density is defined by $\rho_Q\equiv a^3 \sigma^2 \dot\theta$, and the effective potentials are $V_\pm=V(\sigma) \pm \frac{\rho^2_Q}{2a^{6}\sigma^2}$. Note that we will use $V_-$ shortly. From \eq{epcom}, the energy density $\rho_E$ and pressure $p$ are given by
\be{rhop}
\rho_E=\half {\dot{\sigma}}^2+V_+,\hspace*{30pt} p=\half {\dot{\sigma}}^2 -V_-.
\ee
With various values of the charge density $\rho_Q$, \fig{fig:adpot} shows typical effective potentials $V_+$ in Minkowski spacetime where we set $a=H=1$. The models shown in \fig{fig:adpot} will be used later. 

Given an initial charge and energy density (or equivalently initial momenta and positions), the AD field oscillates around the value $\sigma_{cr}$, which is defined by 
\be{sigcr}
\left.\frac{dV_+}{d\sigma}\right|_{\sigma_{cr}}=0,
\ee
where the orbit becomes circular when it starts from there, i.e. $\sigma(0)=\sigma_{cr},\; \dot{\sigma}(0)=0$. This orbit is bounded when the curvature is positive
\be{condbound}
W^2\equiv \left.\frac{d^2V_+}{d\sigma^2}\right|_{\sigma_{cr}}>0.
\ee
For example, given a power-law potential such that $V= \lambda_1\sigma^l$ where $\lambda_1$ is a coupling constant and $l$ is real and a power of the homogeneous field $\sigma$, the condition given by \eq{condbound} implies that bound orbits exist for $l<-2,\; 0<l$ if $\lambda_1>0$ and for $-2<l<0$ if $\lambda_1<0$, where we used \eq{sigcr}. Another example is the case that a scalar potential is logarithmic, i.e. $V=\lambda_2 \ln\sigma$ where the coupling constant $\lambda_2$ is positive. In this case, \eq{condbound} is automatically satisfied. We investigate these two cases in more detail in appendix \ref{App2}.

Let us rescale the field $\sigma(t)$ as $\sigma(t)=\bset{\frac{a_0}{a(t)}}^{3/2}\tisig(t)$ where $a_0$ is the value of $a(t)$ at an initial time. It follows that the equations of motion in \eqs{radhomo}{phshomo} are
\be{modrad}
\ddot{\tisig}-\bset{\frac{3}{4}H^2+\frac{3}{2}\frac{\ddot{a}}{a}}\tisig-\frac{\tirhoQ^2}{\tisig^3}+\bset{\frac{a}{a_0}}^3\frac{dV(\sigma)}{d\tisig}=0, \hspace*{20pt} \frac{d\tirhoQ}{dt}=0,
\ee
where we defined $\tirhoQ\equiv \tisig^2\dot{\theta}=a^{-3}_0\rho_Q$, and the terms involving $H^2$ and $\ddot{a}/a$ are negligible under the assumption of an adiabatic Hubble expansion, i.e. $H^2 \ll 1,\; \ddot{a}\ll a$. 

By introducing a new variable, $\tiu(t)\equiv 1/\tisig(t)$, and using the second expression in \eq{modrad}, the first expression in \eq{modrad} becomes the well-known orbit equation in the centre force problem such that
\be{orbiteq}
\frac{d^2\tiu}{d\theta^2}+\tiu=-\frac{1}{\tirhoQ^2}\bset{\frac{a}{a_0}}^3\frac{dV}{d\tiu}\equiv J(\tiu,t).
\ee
Notice that $J(\tiu,t)$ depends on time caused by the Hubble expansion, whereas  the time-dependence in $J$ vanishes when the potential $V$ is given by a quadratic mass term, $\half M^2 \sigma^2$, where $M$ is a mass of the AD field, $\phi$. 

\subsection{Model A and Model B for MSSM flat potentials}\label{MODEL-ABC}

Let us introduce two models which appear in the MSSM in which SUSY is broken due to either gravity or gauge interactions. The former case, so-called, the gravity-mediated (GRV-M) model, has a scalar potential
\be{gravity-pot}
V=\half m^2\sigma^2\bset{1+K\ln\frac{\sigma^2}{M^2_*}} + \frac{\lambda^2}{m^{n-4}_{pl}} \sigma^n,
\ee
where $m$ is of order of the SUSY breaking scale, which could be the gravitino mass scale $m_{3/2}$ evaluated at the renormalisation scale $M_*$ \cite{Nilles:1983ge}. Also, $\lambda$ is a coupling constant for the nonrenormalisable term, which is suppressed by a high energy scale, e.g. the Planck scale $m_{pl}\sim 10^{18}$ GeV. Here, $K$ is a factor from the gaugino-loop correction, whose value is typically $K\simeq -[0.01-0.1]$ when the flat direction does not have a large top quark component \cite{Enqvist:2000gq, Enqvist:1997si}; thus, we concentrate on the case of $K<0$ from now on. The power $n$ of the nonrenormalisable term depends on flat directions. As examles of the directions involving squarks, the $u^cd^cd^c$ direction has $n=6$, whilst the $u^cu^cd^ce^e$ direction is $n=10$. For $|K|\ll \order{1}$, the first two terms in \eq{gravity-pot} can be approximated by $\frac{m^2M^{2|K|}_*}{2}\sigma^{2-2|K|}$, we then find that
\be{modelA}
V(\sigma)\simeq \frac{M^2}{2}\sigma^l+\frac{\lambda^2}{m^{n-4}_{pl}}\sigma^n \hspace*{15pt} \textrm{for}\; \; n>l
\ee
which we call 'Model A', where we set $M^2 \equiv m^2M^{2|K|}_*$ and $M$ has a mass-dimension, $\frac{4-l}{2}\simeq 1$, since $l\equiv 2-2|K|$ for $|K|\ll \order{1}$. For small values of $\sigma$, we confirm that the power $l$ is not approximately $2-2|K|$, so we will find a value of $l$ numerically in that case later.
\vspace*{5pt}

In another scenario in which SUSY is broken by gauge integration, so-called, gauge-mediated (GAU-M) model, the scalar potential has the curvature with the electroweak mass at a low energy scale, whilst it grows logarithmically at the high energy scale (which means that the potential is nearly flat as similar as the case of $l=0$ in \eq{modelA}). Therefore,
\be{gauge-pot}
V\simeq m^4_{\phi} \ln\bset{1 + \bset{\frac{\sigma}{M_s}}^2}+\frac{\lambda^2}{m^{n-4}_{pl}}\sigma^n,
\ee
where $M_s$ is the messenger scale ($\sim 10^4$ GeV) above which the potential grows logarithmically and $m_{\phi}$ is the same scale as $M_s$. We, thus, set $M_s =m_{\phi}$ for later convenience. Then, the scalar potential at high energy scale is approximately given by \cite{Kusenko:1997si}
\be{modelC}
V\simeq m^4_{\phi} \ln\bset{ \frac{\sigma}{m_{\phi}}}^2+\frac{\lambda^2}{m^{n-4}_{pl}} \sigma^n.
\ee
What follows is that we assume the orbit of the AD condensate is determined by the high energy scale where $\sigma_{cr}\gg m_{\phi}$, calling this case, \eq{modelC}, 'Model B'.

Using the results in appendix \ref{App2}, we obtain the following quantities, $W,\; \Phi$ and $\state{w}$ by assuming that the dominant contribution in Model A and B is, respectively, either power-law term or logarithmic term, each of which corresponds to the first term in \eqs{modelA}{modelC}, respectively. Here, we defined $\Phi$ as a phase difference when the radial field $\sigma$ goes from the minimum value through the maximum one and back to the same minimum point, see \eq{Phi}; in addition, $\state{w}$ is given by a value of the equation of state averaged over a rotation of the orbit, see \eq{eosg}. The sub-dominant terms (nonrenormalisable terms) perturb the orbits by introducing infinitesimally small quantities $\epsilon_{A,\; B}$ where the subscripts correspond to the names of models introduced above. Thus, the main contributions are either \eqs{powerPhi}{powerpress} or \eqs{logPhi}{presslog}.

\vspace*{10pt}

\subsubsection{Model A -- $V(\sigma)=\frac{M^2}{2}\sigma^l+\frac{\lambda^2}{m^{n-4}_{pl}}\sigma^n$}

By recalling \eq{condbound}, we obtain the following relations for $n>l$:
\be{modelA-W}
W^2= \frac{l(l+2)M^2\sigma^{l-2}_{cr}}{2}\bset{1+\epsilon_A},
\ee
where we defined a positive parameter, $\epsilon_A\equiv \frac{n(n+2)}{l(l+2)}\frac{2\lambda^2}{M^2m^{n-4}_{pl}}\sigma^{n-l}_{cr}\ll 1$, which is assumed to be infinitesimally small. We also obtain $\beta^2\simeq (l+2)\left(1+\frac{n-l}{n+2}\epsilon_A\right)>0$, where $\beta$ is defined in \eq{betasq}. Substituting $\beta$ into \eqs{Phi}{eosg}, we obtain $\Phi$ and $\state{w}$:
\bea{modelA-P}
\Phi&\simeq& \frac{\pi}{\sqrt{l+2}}\bset{1+\frac{l-n}{2(n+2)}\epsilon_A}, \\ \label{modelA-w}\state{w}&=&\frac{(l-2)\bset{1+\epsilon_A\frac{l(l+2)(n-2)}{n(n+2)(l-2)}}}{(l+2)\bset{1+\epsilon_A\frac{l}{n}}}\simeq \frac{l-2}{l+2}\bset{1+\epsilon_A\frac{4l(n-l)}{n(n+2)(l-2)}}.
\eea
From \eq{modelA-P}, the orbits for $l=2-2|K|\simeq 2$ are nearly closed, but it is perturbed by the nonrenormalisable term involved with $\epsilon_A$. It results in that the periapsis appears to precess where the precession rate can be obtained from \eq{modelA-W}. The reader should notice that the denominator of the term involving $\epsilon_A$ in the second expression of \eq{modelA-w} has $l-2\simeq -2|K|\ll \order{1}$, which implies that it would be possible to have the non-negligible contribution from the term, even though $\epsilon_A\ll \order{1}$. From now on, we restrict ourself that this is not the case; therefore, the dominant contributions appear as the leading orders in \eqs{modelA-W}{modelA-P} and \eq{modelA-w}, which correspond to \eqs{powerPhi}{powerpress} and \eq{compress}. From \eq{modelA-w} with $\epsilon_A\simeq 0$, our results recover the result published in \cite{Enqvist:1997si}, $\state{w}\simeq -\frac{|K|}{2}$.

\vspace*{10pt}

\subsubsection{Model B -- $V(\sigma)=m^4_{\phi} \ln\bset{\sigma/m_{\phi}}^2+\frac{\lambda^2}{m^{n-4}_{pl}}\sigma^n$}
By introducing another infinitesimally small positive parameter, $\epsilon_B\equiv \frac{n(n+2)\lambda^2\sigma^n_{cr}}{4m^4_{\phi}m^{n-4}_{pl}}\ll 1$, we find
\bea{ModelC-W}
W^2&\simeq& \frac{4m^4_{\phi}}{\sigma^2_{cr}}\bset{1+\epsilon_B}, \spc \Phi\simeq \frac{\pi}{\sqrt{2}}\bset{1-\frac{n}{2(n+2)}\epsilon_B},\\
\label{ModelC-w} \state{w}&=& \frac{ 1-2\ln\bset{\frac{\sigma_{cr}}{m_{\phi}}} +\frac{2(n-2)}{n(n+2)}\epsilon_B }{1+2\ln\bset{\frac{\sigma_{cr}}{m_{\phi}}} +\frac{2}{n}\epsilon_B} \gtrsim -1.
\eea
Since we are working in the high-energy regime, $\sigma_{cr}\gg m_{\phi}$, the pressure of the AD condensate is likely to be negative, see \eq{ModelC-w}. From the second expression for $\Phi$ of \eq{ModelC-W}, the orbits are not closed and it should look like trefoil, see \eq{logPhi}.

\vspace*{10pt}

In an expanding universe, the above orbits for Model A and B suffer from the Hubble damping so that the orbits are naively expected to be precessing spiral or shrinking trefoil in the field space, respectively.

\subsection{Numerical results}\label{numAD}

In this subsection, we present numerical results to check the analytic results which we found in the previous subsection. To do so, we use the full potentials, \eqs{gravity-pot}{gauge-pot}, instead of \eqs{modelA}{modelC}, and then solve \eq{radhomo} numerically in Minkowski spacetime as well as in an expanding universe. We adopt the 4th order Runge-Kutta method with various sets of initial conditions, such as $\rho_Q$ and $\varepsilon^2$. Since our analytical work holds as long as $\varepsilon^2\ll \order{1}$, we are concern with the two cases: a nearly circular orbit with $\varepsilon^2=0.1$ and a more elliptic orbit with $\varepsilon^2=0.3$. First of all, we parametrise \eqs{gravity-pot}{gauge-pot} by introducing dimensionless variables: $\mathring{\sigma}=\sigma/M_*,\; b^2_*=\frac{\lambda^2M^{n-2}_*}{m^{n-4}_{pl}m^2}=|K|e^{-1}/4,\; \mathring{t}=mt,\; \mathring{\mathbf{x}}=m\mathbf{x}$ in GRV-M Model and $\mathring{\sigma}=\sigma/M_s,\; b^2=\frac{\lambda^2M^{n-4}_s}{m^{n-4}_{pl}},\; \mathring{t}=M_st,\; \mathring{\mathbf{x}}=M_s\mathbf{x}$ in GAU-M Model. Since we know that $m\sim 10^2$ GeV, $M_*\sim 10^{10}$ GeV, $m_{pl}\sim 10^{18}$ GeV; hence, we can set $b^2_*\sim 9.20\times 10^{-3}\sim \order{10^{-2}}$ in GRV-M Model, where we choose $|K|=0.1$. Notice that these choices are same as the ones used in \cite{Copeland:2009as}. On the other hand, we know that $m_{\phi}\sim M_s \sim 10^4$ GeV; hence, we can set $b^2\sim 10^{-30}$ in GAU-M Model, where we choose $\lambda\sim 10^{-2}$ as used in the GRV-M case. Notice that we can obtain the rescaled charge density $\mathring{\rho}_Q$ and energy density $\mathring{\rho}_E$, such that $\rho_Q=mM^2_*\mathring{\rho}_Q,\; \rho_E=m^2M^2_*\mathring{\rho}_E$ in GRV-M Model and $\rho_Q=M^3_s\mathring{\rho}_Q,\; \rho_E=M^4_s\mathring{\rho}_E$ in GAU-M Model.

Therefore, our rescaled potentials in GRV-M and GAU-M models for a $n=6$ flat-direction are, respectively,
\bea{numGRV}
V&=&\half \sigma^2\bset{1-2|K|\ln \sigma} +b^2_* \sigma^6,\\
\label{numGAU} V&=&\ln\bset{1+\sigma^2}+b^2\sigma^6,
\eea
where we omit over-rings for simplicity. The reader should notice that these variables that appear within the rest of this sub-section are dimensionless. We can also obtain the ratio defined by an energy density divided by a mass multiplied by a charge density, where the mass corresponds to $m$ or $M_s$ in either GRV-M or GAU-M Model, respectively.

In order to obtain appropriate initial values of $\sigma(0),\; \dot{\sigma}(0)$ and $\dot{\theta}(0)$ satisfying the conditions $\epsilon_A,\; \epsilon_B\ll \order{1}$ and giving not too small charge densities, we shall show that we need to choose only the initial values of $\dot{\theta}(0)$ in both GRV-M and GAU-M models. First, by ignoring the nonrenormalisable term in \eq{numGRV} for GRV-M Model, we can obtain $\sigma_{cr}=\exp\bset{-\frac{1}{2|K|}\bset{\dot{\theta}^2(0)+|K|-1}}:=\sigma(0)$ from \eq{sigcr}, where we set $\sigma_{cr}:=\sigma(0)$, which implies that we are setting that the initial phase is $3\pi/2$. Since $\dot{\sigma}$ has the maximum value at $\sigma=\sigma_{cr}$, we can set $\dot{\sigma}(0):=\varepsilon^2\sigma(0)\sqrt{\dot{\theta}^2(0)-|K|/2}$ from \eq{deltaeq}, which imply that $\epsilon_A\sim 12 b^2_* \sigma^4(0)$ from the definition. We notice that $\sigma(0)\gg \order{1}$ for $\dot{\theta}(0)\ll \order{1}$; hence, it breaks the condition, $\epsilon_A \ll \order{1}$. We can also see that $\sigma(0)\ll \order{1}$ for $\dot{\theta}(0)\gg \order{1}$, so the charge density is suppressed exponentially. Therefore, we are concern with the following two cases: $\dot{\theta}(0)=\sqrt{2}$ and $1.0$, which give, respectively, $\epsilon_A\sim 1.20\times 10^{-11},\; \rho_Q \sim 2.36\times 10^{-5}  $ and $\epsilon_A \sim 1.58\times 10^{-2},\; \rho_Q \sim 3.68\times 10^{-1}$. Similarly, in GAU-M Model, we choose that $\sigma_{cr}=\sqrt{\frac{2}{\dot{\theta}^2(0)}-1}:=\sigma(0)$, $\dot{\sigma}(0):=\varepsilon^2\sqrt{1-\frac{3}{4}\dot{\theta}^2(0)}$ and $\epsilon_B=12b^2\sigma^6(0)$ from the definition of $\epsilon_B$. Here, we also set the initial phase is $3\pi/2$ due to $\sigma_{cr}:=\sigma(0)$. With this fact and the approximation, $\sigma_{cr}\gg \order{1}$, we need to have $\dot{\theta}(0)\ll \order{1}$. In addition, we should have $\sigma(0)< \order{10^5}$ due to the condition, $\epsilon_B < \order{1}$. Therefore, we choose $\dot{\theta}(0)=\sqrt{2}\times 10^{-1}$ and $\sqrt{2}\times 10^{-2}$ which gives, respectively, $\epsilon_B\sim 1.16\times 10^{-23},\; \rho_Q \sim 1.40\times 10^1$ and $\epsilon_B \sim 1.20\times 10^{-17},\; \rho_Q \sim 1.41\times 10^2$. 

Upon the above initial conditions, we initiate the numerical simulations with 8 different sets of the initial values in GRV-M Model and GAU-M Model summarised in TABLE \ref{parameterSET} where we call each of the parameter-sets 'SET-1, SET-2,..., and SET-8'. In \fig{fig:adpot}, we also show,  with the various charges which we introduced above, the effective potentials $V_+$ for the GRV-M potential given by \eq{numGRV} in the left panel and for the GAU-M potential given by \eq{numGAU} in the right panel. Our time-step, $dt$, in the numerical simulations is $dt=1.0\times 10^{-4}$ in the GRV-M case and $dt=1.0\times 10^{-3}$ in the GAU-M case.

\begin{center}
\begin{table} [!ht]
\begin{tabular} { |c|c|c|c|c|c|c|c| }

\hline
SET & Model & $\dot{\theta}(0)$ & $\sigma(0)$ & $\rho_Q$ &   $\epsilon_A$ or $\epsilon_B$ & $\varepsilon^2$ & $\rho_E/\rho_Q$ \\
\hline \hline
1 & & & & & & 0.1  & 1.46\\ 
2 & & \raisebox{1.5ex} {$\sqrt{2}$} & \raisebox{1.5ex} {$\sim 4.09\times 10^{-3}$} & \raisebox{1.5ex} {$\sim 2.36\times 10^{-5}$}&  \raisebox{1.5ex} {$\sim 1.20\times 10^{-11}$} & 0.3 & 1.51 \\
3 & \raisebox{1.5ex} {GRV-M} & & &  &  & 0.1 & 1.06 \\
4 & & \raisebox{1.5ex} {1.0} & \raisebox{1.5ex} {$\sim 6.07\times 10^{-1}$}  & \raisebox{1.5ex} {$\sim 3.68\times 10^{-1}$}&  \raisebox{1.5ex} {$\sim 1.58\times 10^{-2}$}  &  0.3 & 1.09\\ 
\hline \hline
5 & & &  & & & 0.1 & $4.00\times 10^{-1}$ \\ 
6 & & \raisebox{1.5ex} {$\sqrt{2}\times 10^{-1}$} & \raisebox{1.5ex} {$\sim9.95$} & \raisebox{1.5ex} {$\sim 1.40 \times 10^1$} &  \raisebox{1.5ex} {$\sim 1.16\times 10^{-23}$} & 0.3 & $4.03\times 10^{-1}$\\
7 & \raisebox{1.5ex} {GAU-M}  & & & &  & 0.1  & $7.22\times 10^{-2}$ \\
8 & & \raisebox{1.5ex} {$\sqrt{2}\times 10^{-2}$} &  \raisebox{1.5ex} {$\sim1.00\times 10^{2}$} & \raisebox{1.5ex} {$\sim 1.41\times 10^2$}  &  \raisebox{1.5ex} {$\sim 1.20\times 10^{-17}$}  &  0.3 & $7.25\times 10^{-2}$\\ 
\hline
\end{tabular}
\caption{We show 8 different parameter sets in both the GRV-M and GAU-M cases, where we call each of the parameter-sets 'SET-1, SET-2,..., and SET-8'. The initial parameters of $\sigma(0)$ and $\dot{\sigma}(0)$ can be obtained by the values of $\dot{\theta}(0)$. We also set $\theta(0)=\frac{3\pi}{2}$ in all cases, and show the values of $\epsilon_A$ for GRV-M Model and the values of $\epsilon_B$ for GAU-M Model. By substituting these values and choosing the values of the third eccentricity $\varepsilon^2=0.1$ and $0.3$, we obtain the dimensionless energy-to-(mass multiplied by charge) ratios, $\rho_E/\rho_Q$.}
\label{parameterSET}
\end{table}
\end{center}

\subsubsection{The orbit of an Affleck-Dine ``planet'' in Minkowski spacetime}

First, we present numerical results in Minkowski spacetime in order to check our analytical results. We then give ans\"{a}tze which are motivated by our analytic solutions in an expanding universe in the next sub-subsection.

\vspace*{10pt}

\paragraph*{\underline{\bf The motion of $\sigma^2(t)$}}

\begin{figure}[!ht]
  \begin{center}
	\includegraphics[angle=-90, scale=0.31]{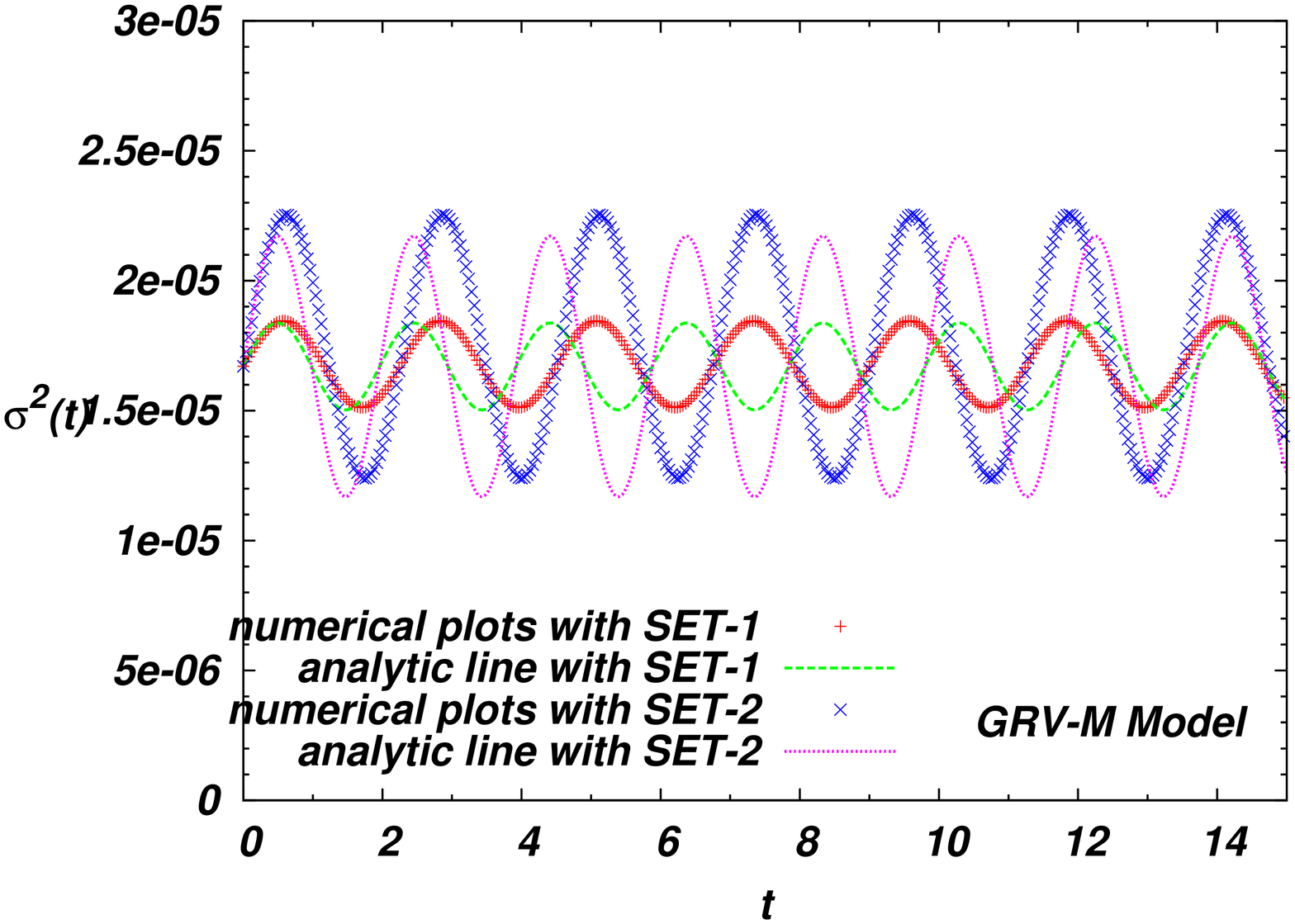}
	\includegraphics[angle=-90, scale=0.31]{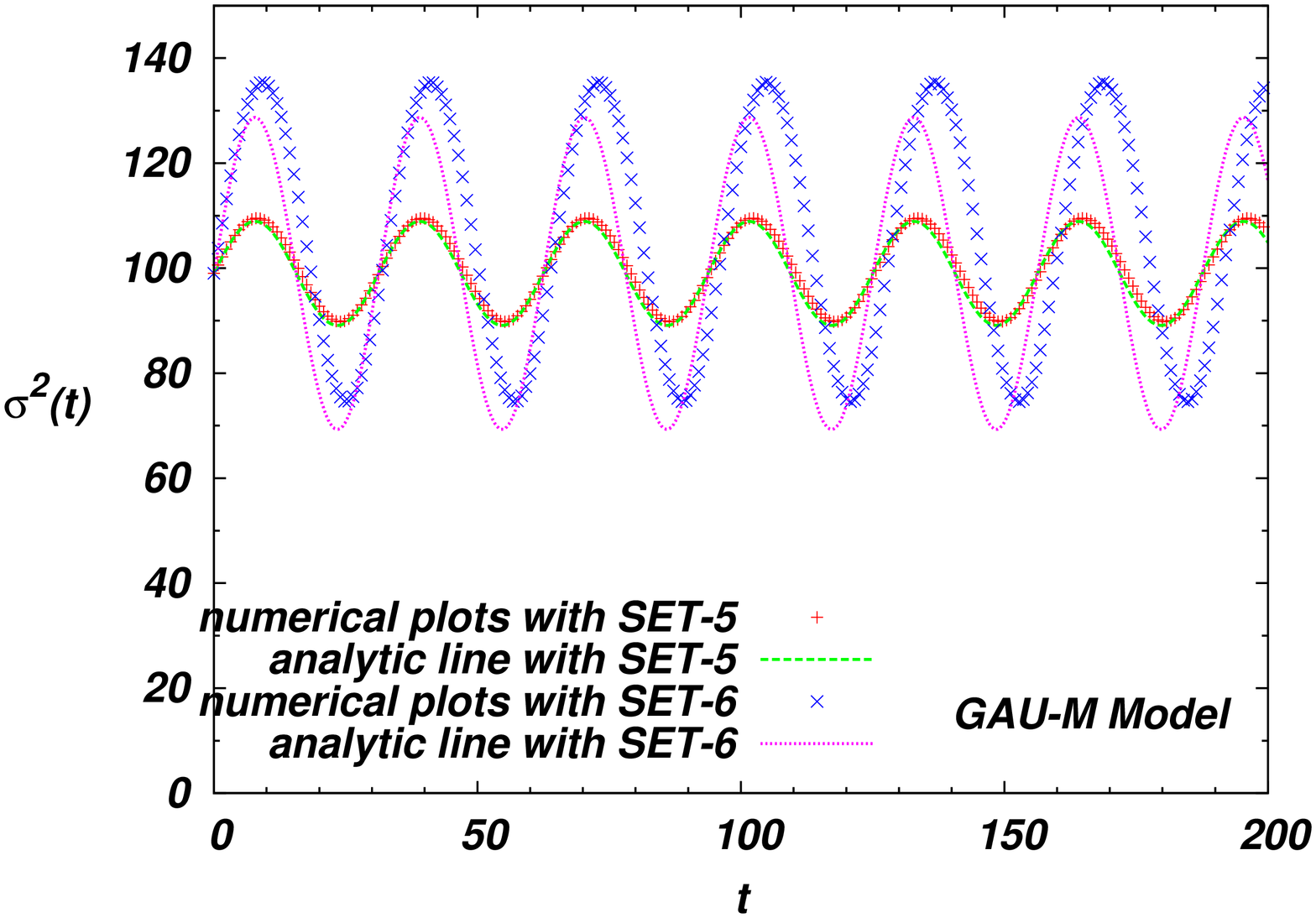}\\
	\includegraphics[angle=-90, scale=0.31]{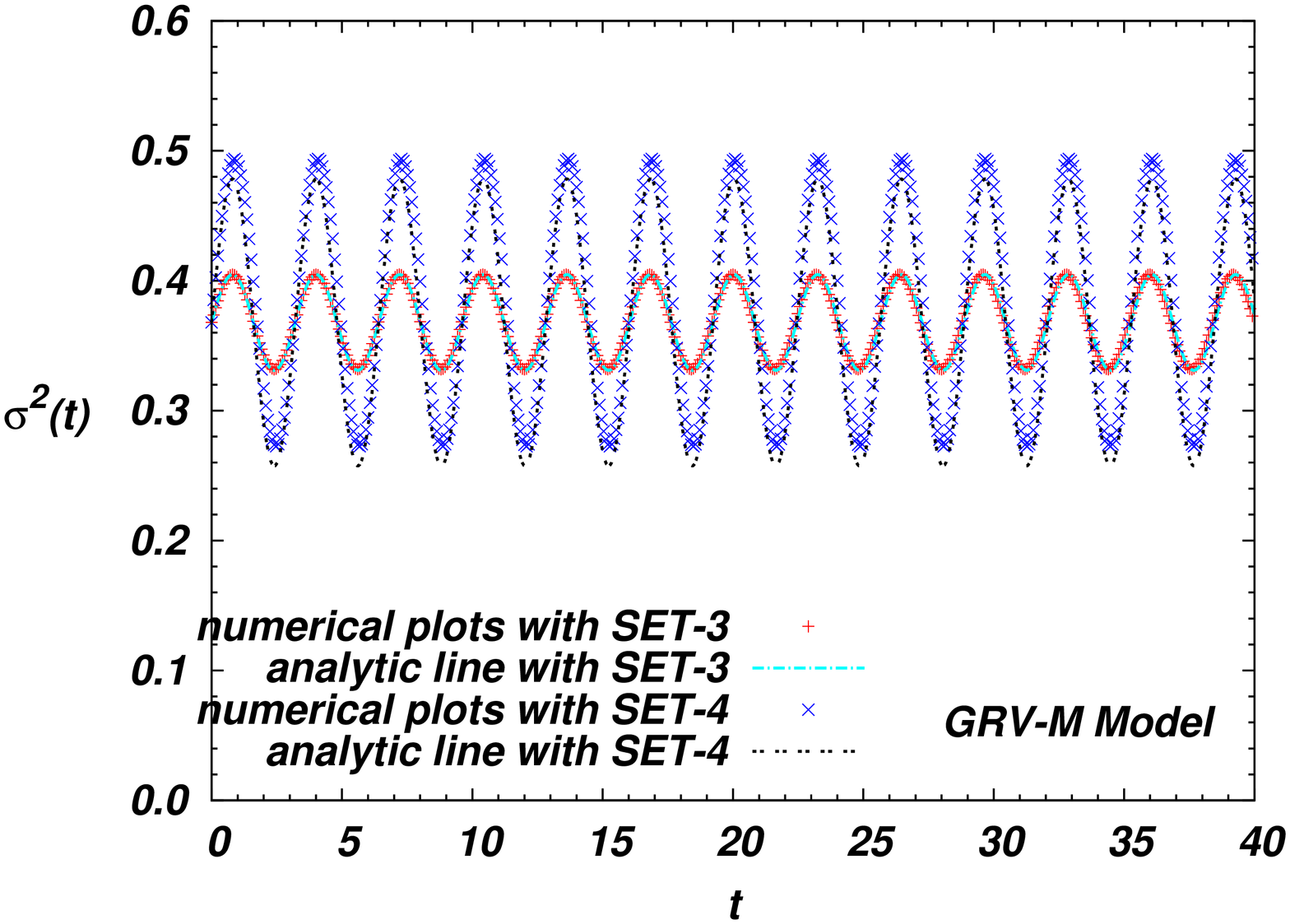}
	\includegraphics[angle=-90, scale=0.31]{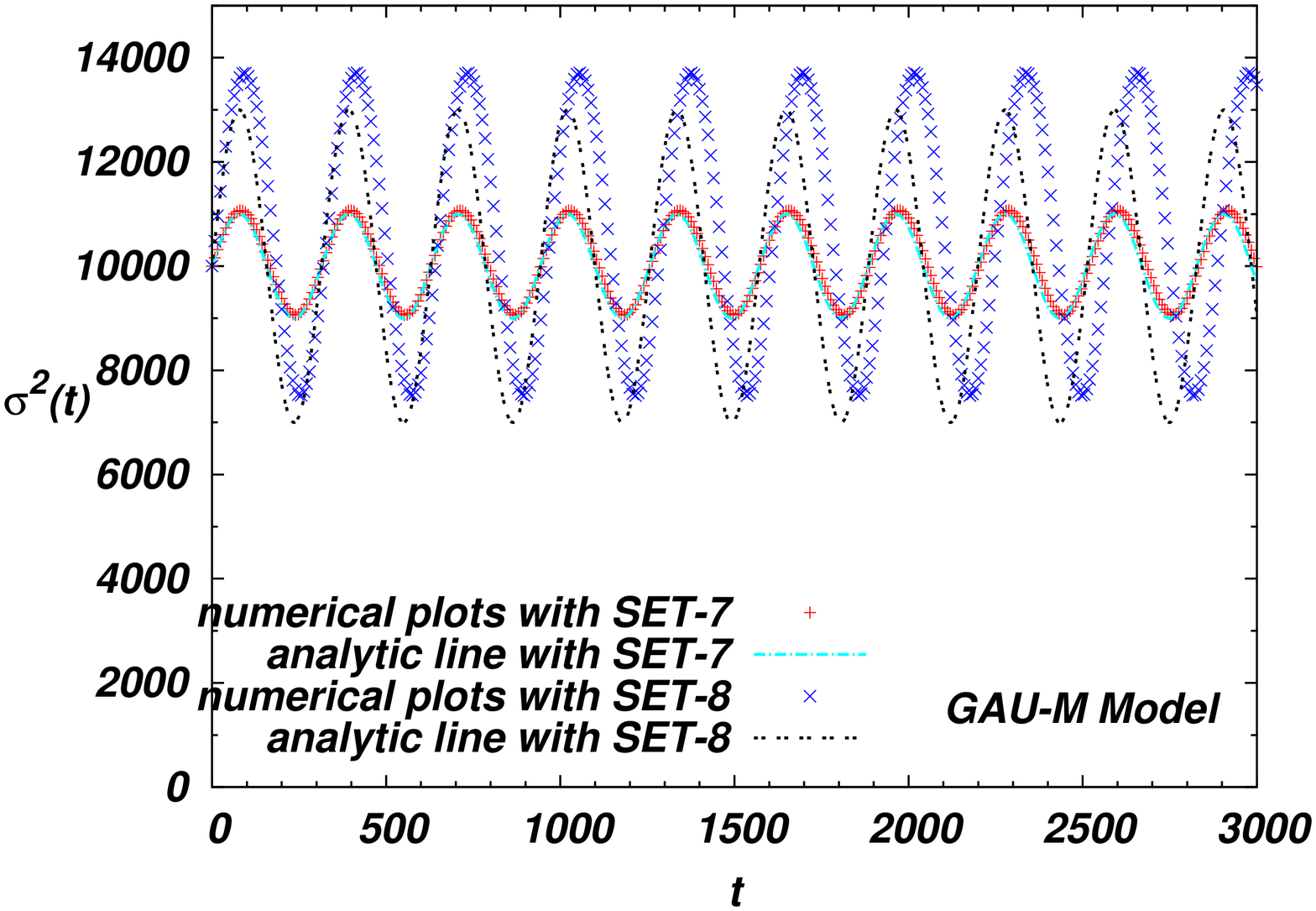}
  \end{center}
  \caption{ \textbf{(Color online)} Using the parameter sets summarised in TABLE \ref{parameterSET}, we plot the numerical evolution for $\sigma^2(t)$ in both GRV-M Model (left) and GAU-M Model(right). In all of the panels except the case for the left-top panel, the numerical plots (red-plus dots and blue-cross dots) agree well with the corresponding analytic lines, which are obtained from Sec. \ref{MODEL-ABC}. The disagreements between the numerical and analytic plots in the left-top panel come from the artifact that the analytical estimated value, $l=1.8$, in \eq{modelA}.}
  \label{fig:sigsq}
\end{figure}

In \fig{fig:sigsq}, we show the numerical solutions using the GRV-M potential with \eq{numGRV} (left) and using the GAU-M potential with \eq{numGAU} (right), and compare them with the corresponding analytic solutions which are given by \eq{sigsol}. Using the initial values whose parameter sets can be seen in TABLE \ref{parameterSET}, we plot the numeric and analytic solutions in \fig{fig:sigsq}. In the top-left panel, the numerical plots (red-plus dots for SET-1 and blue-cross dots for SET-2) have the same amplitudes as the analytical ones (gree-dashed line for SET-1 and purple-dotted-dashed line for SET-2), we, however, can see the significant differences for the frequencies of each oscillation. We notice that these discrepancies come from the artifact of our choice as $l=2-2|K|$ in \eq{modelA}, since the choice is not appropriate for $\sigma\ll \order{1}$, recalling $\sigma(0)\sim 4.09\times 10^{-3}$ in SET-1 and SET-2. Shortly, we will obtain numerically this power $l$, and the obtained semi-analytic solutions match with the numerical ones. With SET-3 and SET-4, we can see that $\sigma(0)$ is not so small as opposed to the previous cases, i.e. $\sigma(0)\sim 6.07\times 10^{-1}$; thus, in the left-bottom panel of \fig{fig:sigsq} we can see a nice agreement between numerical plots (red-plus dots for SET-3 and blue-cross dots for SET-4) and analytic plots (skyblue-dotted-dashed line for SET-3 and black-dotted line for SET-4).

Similarly,  we show the numerical and analytic plots for the GAU-M potential in the right-panels of \fig{fig:sigsq} using the parameter-sets: for SET-5 and SET-6 in the right-top panel and for SET-7 and SET-8 in the right-bottom panel. By changing the values of the third eccentricity $\varepsilon^2$ (see TABLE \ref{parameterSET}), the numerical plots deviate slightly from our analytic lines in the right-top and right-bottom panels of \fig{fig:sigsq} as we can expect; in particular, we can see that our analytic values of both the frequencies and amplitudes of $\sigma^2(t)$ are larger than the numerical ones, and this difference can be significantly reduced when the orbits of the AD planets is nearly circular with $\varepsilon^2=0.1$.

\begin{figure}[!ht]
  \begin{center}
	\includegraphics[angle=-90, scale=0.4]{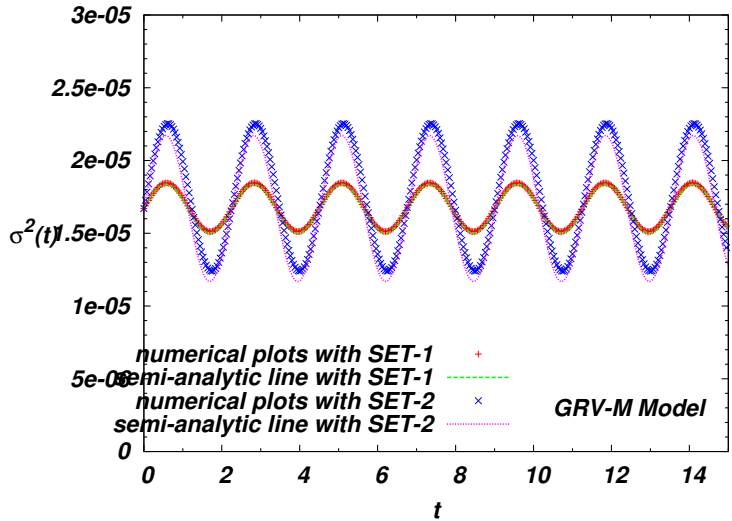}
  \end{center}
  \caption{ \textbf{(Color online)} Substituting the numerical value, $l=1.86002$, into \eq{modelA}, we plot the semi-analytic evolution for $\sigma^2(t)$. Our semi-analytic solutions agree with the numerical solutions.}
  \label{fig:sigsqSEMI}
\end{figure}

As we have seen in the left-top panel of \fig{fig:sigsq}, our analytic value, $l=1.8$, in \eq{modelA} are not good enough to reproduce the numerical solutions since $\sigma(t)\ll \order{1}$. Therefore, we set a trial function, $f(\sigma)=\half \sigma^{\alpha}+b^2_*\sigma^6$, where a numerical value $\alpha$ is found by using the 'fit' command in the numerical software called 'gnuplot'. We find that $\alpha=1.86002:=l$ is the best value of $\alpha$, where we fitted this trial function $f(\sigma)$ onto the numerical full potential in \eq{numGRV} for the range of $\sigma \in [1.0\times 10^{-2}-1.0\times 10^{-3}]$, recalling $\sigma(0)\sim 4.09\times 10^{-3}$ in SET-1 and SET-2. Using this value of $\alpha$ as the value of $l$ instead of $l=1.8$, we plot the semi-analytic evolution for $\sigma^2(t)$ in \fig{fig:sigsqSEMI} (green-dashed line for SET-1 and purple-dotted-dashed line for SET-2) against the corresponding numerical plots (red-plus dots for SET-1 and blue-cross dots for SET-2). Now, our semi-analytic solutions match with the numerical solutions.

\vspace*{5pt}

\paragraph*{\underline{\bf The average values of $w(t)$}}

Using \eqs{modelA-w}{ModelC-w}, we show both numerical values $\state{w_{num}}$ and (semi-)analytical values $\state{w_{ana}}$ of the averaged equation of state in \tbl{tbl:eos}. For all cases, the AD condensate has a negative pressure and one can say that the numerical values are of the same order as analytic values.

\begin{center}
\begin{table} [!ht]
\begin{tabular} { |c||c|c|c|c||c|c|c|c| }

\hline 
$\state{w}$ & \multicolumn{4}{c||}{GRV-M Model v.s. Model A} & \multicolumn{4}{c|}{GAU-M Model v.s. Model B}\\
\hline \hline
 & SET-1 & SET-2 & SET-3 & SET-4 & SET-5 &  SET-6 & SET-7 & SET-8 \\
\hline 
$\state{w_{num}}$ & \multicolumn{2}{c|}{$-2.42\times 10^{-2}$} & $-4.47\times 10^{-2}$ & $-4.45\times 10^{-2}$ &$-6.43\times 10^{-1}$ & $-6.45\times 10^{-1}$ & \multicolumn{2}{c|}{$-8.00\times 10^{-1}$} \\ 
\hline
$\state{w_{ana}}$ &  \multicolumn{2}{c|}{$-3.63\times 10^{-2}$} & \multicolumn{2}{c||}{$-5.00\times 10^{-2}$} & \multicolumn{2}{c|}{$-6.43\times 10^{-1}$} & \multicolumn{2}{c|}{$-8.04\times 10^{-1}$}\\ 
\hline
\end{tabular}
\caption{Using \eqs{modelA-w}{ModelC-w}, we show the both numerical values $\state{w_{num}}$ and analytical values $\state{w_{ana}}$ for the averaged equations of state. The values of $\state{w_{ana}}$ in SET-1 and SET-2 are semi-analytically obtained by substituting $l=1.86002$ into \eq{modelA}.  For all cases, the AD condensate has a negative pressure, and these analytical estimates are reasonable against the numerical values.}
\label{tbl:eos}
\end{table}
\end{center}

\vspace*{5pt}

\paragraph*{\underline{\bf The values of $\Phi$}}

In TABLE \ref{parameterPHI}, we show the numerical and (semi-)analytic values of $\Phi$ in both GRV-M Model and GAU-M Model, which are analytically obtained in Sec. \ref{MODEL-ABC}. Our analytical values agree well with the numerical values. These values suggest that the orbits in GRV-M Model and GAU-m model are nearly either elliptical or trefoil, respectively.

\begin{center}
\begin{table} [!ht]
\begin{tabular} { |c||c|c|c|c||c|c|c|c| }

\hline 
$\Phi$ & \multicolumn{4}{c||}{GRV-M Model v.s. Model A} & \multicolumn{4}{c|}{GAU-M Model v.s. Model B}\\
\hline \hline
 & SET-1 & SET-2 & SET-3 & SET-4 & SET-5 &  SET-6 & SET-7 & SET-8 \\
\hline 
$\Phi_{num}$ & 1.591 & 1.590 & 1.605 & 1.604 & 2.210 & 2.206 & 2.221 & 2.217 \\ 
\hline
$\Phi_{ana}$ &  \multicolumn{2}{c|}{1.612 (analytic) or 1.599 (semi-analytic)} & \multicolumn{2}{c||}{1.605} & \multicolumn{2}{c|}{2.221} & \multicolumn{2}{c|}{2.221}\\ 
\hline
\end{tabular}
\caption{We show the numerical and (semi-)analytic values of $\Phi$ in both GRV-M Model and GAU-M Model, which are analytically obtained in Section \ref{MODEL-ABC}.}
\label{parameterPHI}
\end{table}
\end{center}

\subsubsection{The orbit of an Affleck-Dine ``planet'' in an expanding universe}

We carry out our numerical simulation in an expanding universe when the inflaton field, which is trapped by a quadratic potential, starts to coherently oscillate around the vacuum during the reheating era. Then the evolution of Hubble expansion, $H(t)$, and scale factor, $a(t)$, follow as an ordinary nonrelativistic (zero-pressure) matter, see \eq{compress}. For $l=2$, we find $H=\frac{2}{3(t+t_0)}$ and $a(t)=a_0\bset{\frac{t+t_0}{t_0}}^{2/3}$, where $a_0$ is given by the value of $a(t)$ at $t=0$ and we set $a_0=0.1$. We also set the initial time as $t_0=4\times 10^2$ for GRV-M Model and $t_0=4\times 10^4$ for GAU-M Model. Notice that with this choice of $t_0$ our simulation starts from the same physical time because we rescaled the time by either $m\sim 10^2$ GeV or $M_s\sim 10^4$ GeV, respectively. We again solve the equation of motion, \eq{radhomo}, numerically using the 4th order Runge-Kutta method and compare them with the following ans\"{a}tze. In order to see the significant effects from the Hubble expansion, we use SET-3 in GRV-M Model and SET-7 in GAU-M Model as the initial parameters.

In an expanding spacetime, one can guess that our analytical results in Minkowski spacetime should be changed. In particular, the amplitude of $\sigma(t)$ may decrease due to the Hubble damping as we saw in the quadratic case in Appendix \ref{sec2}, and similarly the frequency $W$ in \eq{condbound} should be changed. Hence, the orbit of the AD planet is precessing spiral or shrinking trefoil in either GRV-M or GAU-M Model as one can see \cite{Jokinen:2002xw}. Let us give an ansatz for $\sigma^2(t)$,
\be{sigantz}
\sigma^2(t)=\bset{\frac{t_0}{t+t_0}}^{\alpha_1}\tisig^2 \bset{1+\varepsilon^2\cos{\bset{\widetilde{W} \cdot \bset{\frac{t_0}{t+t_0}}^{\alpha_2}\cdot t+\frac{3\pi}{2}}}}.
\ee
Here, we use the Minkowskian values of $\tisig$ and $\widetilde{W}$, and will obtain the possible values of $\alpha_{1,\ 2}$ in both models. From \eqs{sigcr}{condbound}, by ignoring the nonrenormalisable term and recalling $a(t)=a_0\bset{\frac{t+t_0}{t_0}}^{2/3}$, we can find the following proportion relations: $\sigma_{cr}(t)\propto (t+t_0)^{-4/(l+2)} \simeq (t+t_0)^{-2/(2-|K|)}$ and $W(t)\propto (t+t_0)^{-\frac{2(l-2)}{l+2}}\simeq (t+t_0)^{\frac{2|K|}{2-|K|}}$ in Model A, where we used $l=2-2|K|$. In Model B, we obtain $\sigma_{cr}\propto (t+t_0)^{-2}$ and $W(t)\propto (t+t_0)^2$. Therefore, we set $\alpha_1=\frac{4}{2-|K|},\;\alpha_2=-\frac{2-|K|}{2|K|}$ in Model A, and $\alpha_1=4,\;\alpha_2=-2$ in Model B. We believe that our ans\"{a}tze are valid as long as the nonrenormalisable term does not play a role and the frequency of the coherent rotation, $\order{W(t)}$, is rapid compared to Hubble expansion rate, $\order{H}$. The latter restriction implies that the rotation time scale is much shorter than the time scale of the Hubble expansion, i.e. $W^{-1}(t)\gg H^{-1}$ \cite{Turner:1983he}.

\vspace*{10pt}

\paragraph*{\underline{\bf The motion of $\sigma^2(t)$}}

\begin{figure}[!ht]
  \begin{center}
	\includegraphics[angle=-90, scale=0.31]{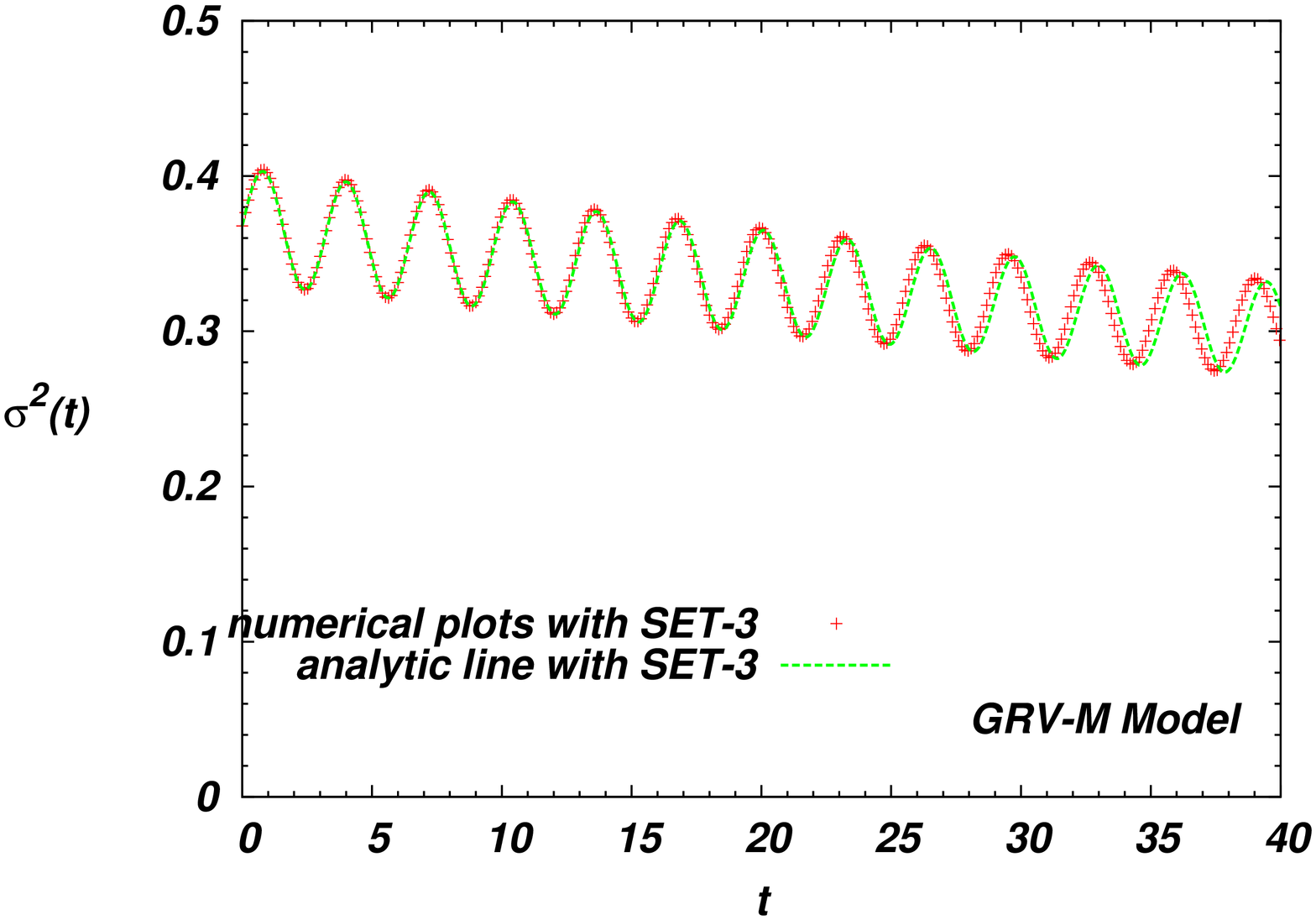}
	\includegraphics[angle=-90, scale=0.31]{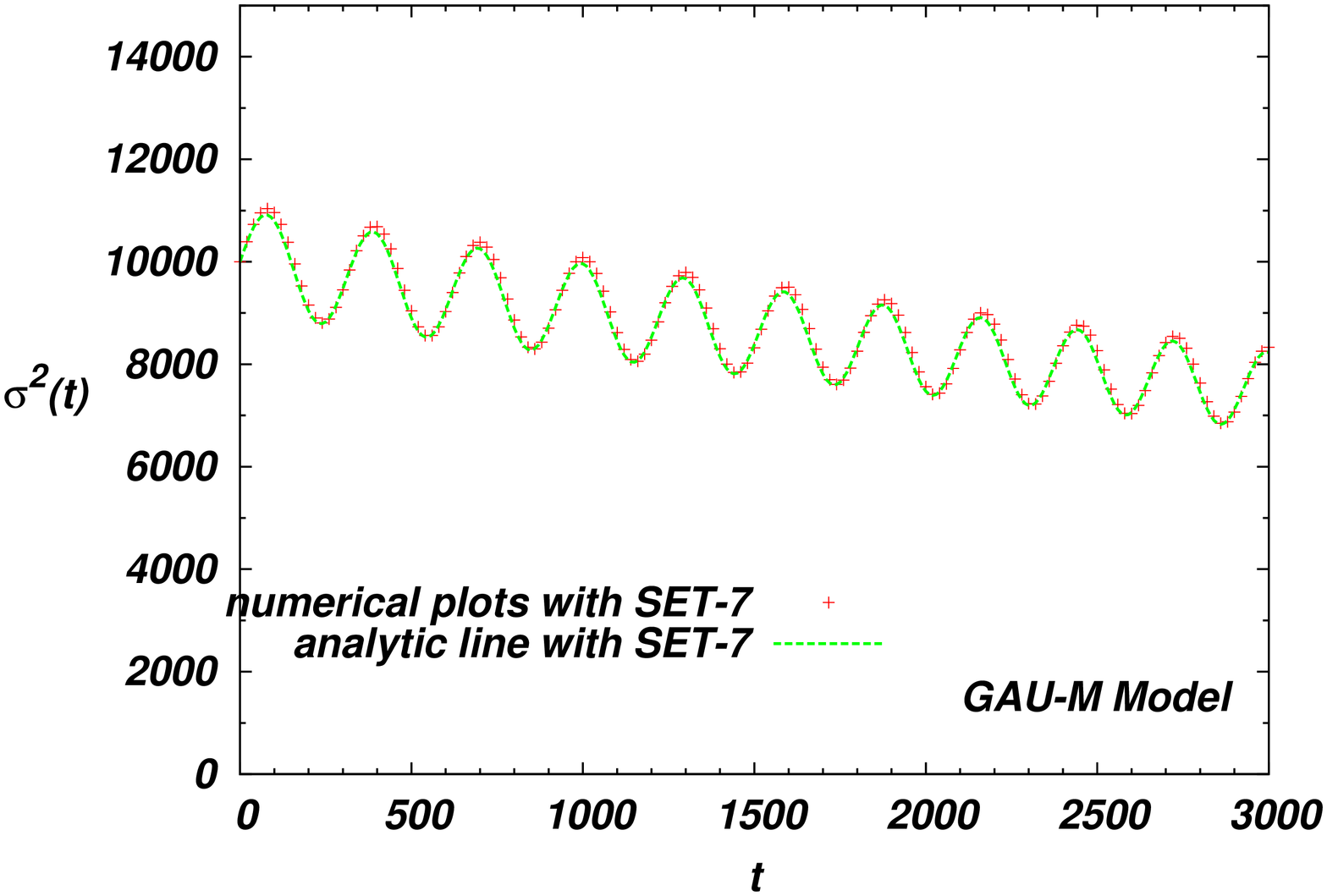}
  \end{center}
  \caption{ \textbf{(Color online)} We plot the evolution of $\sigma^2(t)$ with the numerical data (red-plus dots) for GRV-M Model (left) and for GAU-M Model (right) and with the analytic data (green-dotted lines) by using our ans\"{a}tze introduced in \eq{sigantz}.}
  \label{fig:sigexp}
\end{figure}

In \fig{fig:sigexp}, we plot the evolution of $\sigma^2(t)$ with the numerical data (red-plus dots) for GRV-M Model (left) and for GAU-M Model (right) and with the analytic data (green-dotted lines) using our ans\"{a}tze \eq{sigantz}. The readers should compare the Minkowskian cases of SET-3 (left bottom panel) and SET-7 (right bottom panel) in \fig{fig:sigsq} with the corresponding expanding background cases. For both cases, the amplitudes of $\sigma^2(t)$ decrease in time as we expected, and our analytic plots excellently agree with the corresponding numerical results. In the left panel of \fig{fig:sigexp}, the difference between the analytic line and the numeric plots arises for the late time. We believe that this comes from the artifact of the approximation on $l=2-2|K|$ in GRV-M Model, \eq{numGRV}, since the values of $\sigma^2(t)$ decrease to the region where the above approximation does not hold, i.e. for $\sigma\ll \order{1}$ as we saw in the left-top panel of \fig{fig:sigsq}.

\vspace*{10pt}

\paragraph*{\underline{\bf The motion of the equation of state: $w(t)=p(t)/\rho_E$}}

\begin{figure}[!ht]
  \begin{center}
	\includegraphics[angle=-90, scale=0.31]{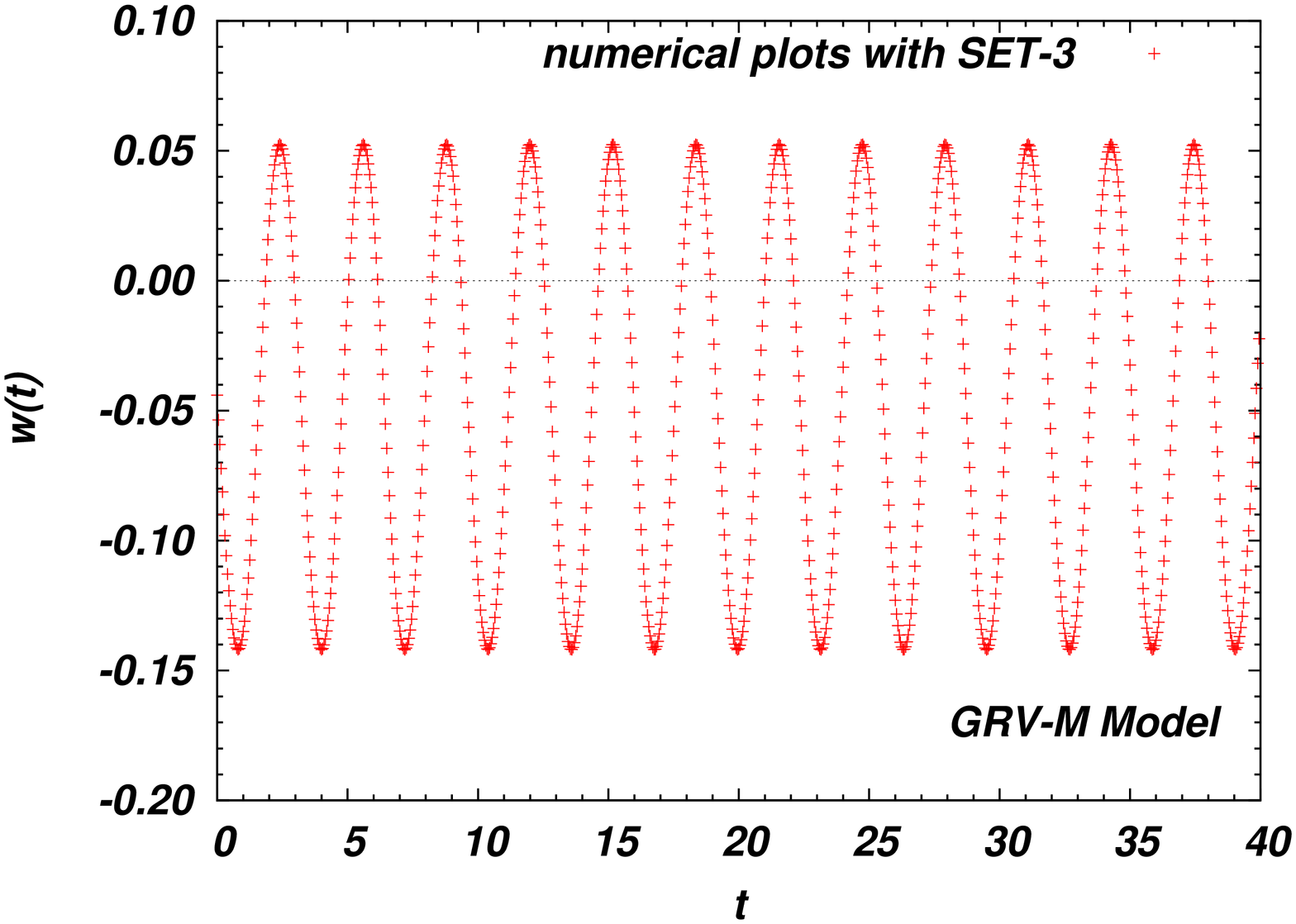}
	\includegraphics[angle=-90, scale=0.31]{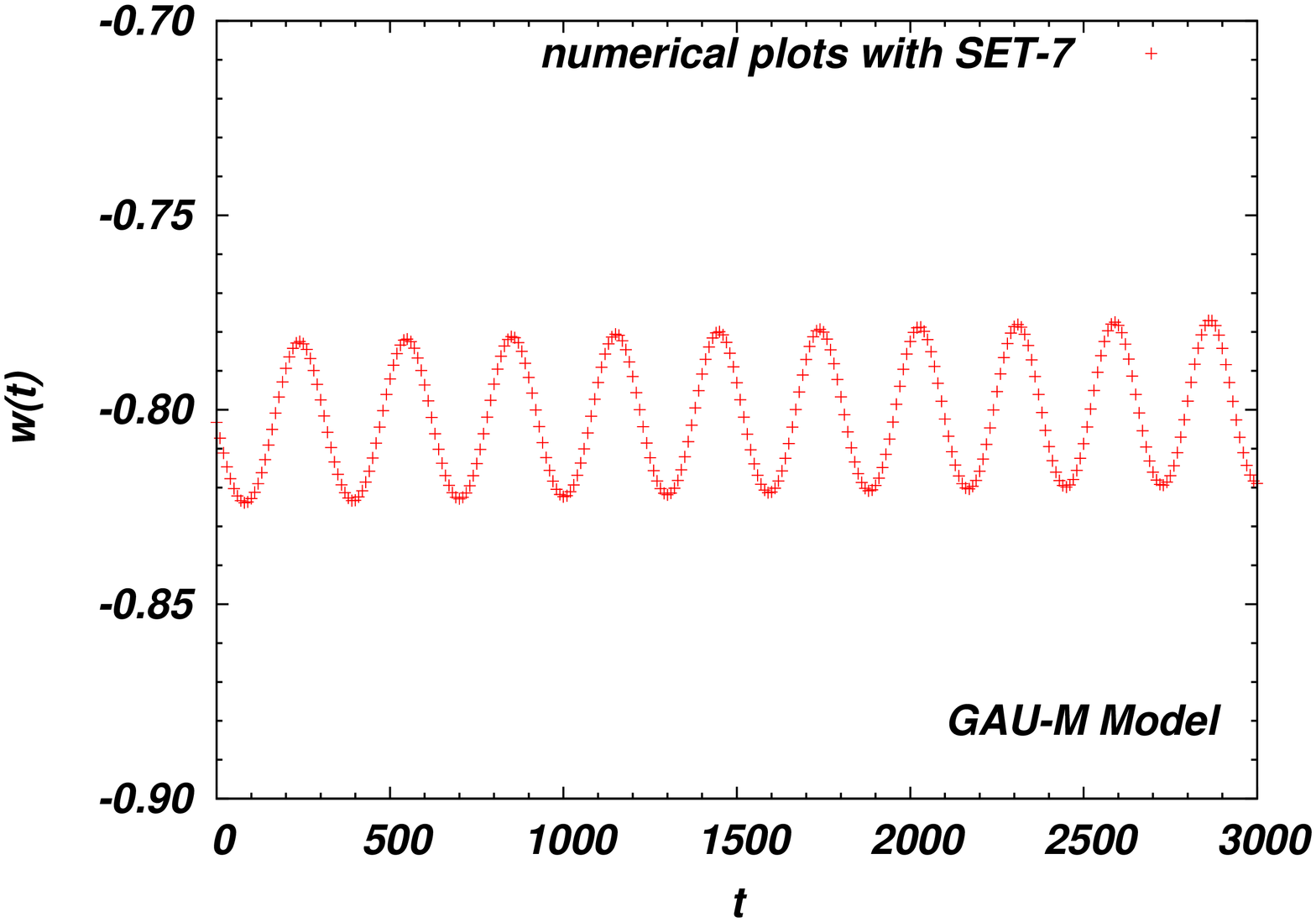}
  \end{center}
  \caption{ \textbf{(Color online)} Using the initial conditions called SET-3 (right-panel) and SET-7 (left-panel) in TABLE \ref{parameterSET}, we plot the numerical values of the equation of state which are given by $w(t)\equiv p(t)/\rho_E$, where $p(t)$ and $\rho_E$ are the pressure and energy density of the AD condensate.}
  \label{fig:eqsexp}
\end{figure}

In \fig{fig:eqsexp}, we plot the numerical values of the equation of state, which is given by $w(t)\equiv p(t)/\rho_E$, where $p(t)$ and $\rho_E$ in \eq{rhop} are the pressure and energy density of the AD condensate. The averaged pressure over the rotations seems to be negative in GRV-M Model, see the left panel; whereas, the pressure in GAU-M Model is always negative, see the right panel. The frequencies of the rotation for $w(t)$ in both cases are, respectively, similar as the corresponding frequencies of $\sigma^2(t)$, see \fig{fig:sigexp}; however, the phases are different from the phases of $\sigma^2(t)$ by $\pi$.

\vspace*{10pt}

In summary, we have analytically obtained the nearly circular orbits for both GRV-M Model and GAU-M Model in \eqs{numGRV}{numGAU} approximated by \eqs{modelA}{modelC}. We then checked that the (semi-)analytic results in \eqs{modelA-W}{modelA-P} and \eqs{ModelC-W}{ModelC-w} and our ans\"{a}tze in \eq{sigantz} agree well with the corresponding numerical results obtained by numerically solving \eqs{radhomo}{rad}. In the rest of this paper, we investigate the late evolution for the AD condensates once the spatial perturbations generated by quantum fluctuations or thermal noise from the early oscillation \cite{Allahverdi:2000zd} become non-negligible due to the negative pressure presented in \tbl{tbl:eos} and \fig{fig:eqsexp}.

\section{$Q$-ball formation and the thermalisation in Minkowski spacetime}\label{sectinst}
In this section, we analyse the late evolution of the AD condensates in both GRV-M and GAU-M models, in which we find that the spatial perturbations are amplified exponentially due to the presence of the negative pressure, and the presence of negative pressure supports the existence of nontopological solitons, i.e. $Q$-balls. As a process of reheating the Universe, the dynamics of the $Q$-ball formation is nonequilibrium, nonperturbative, and nonlinear process, and it includes three distinct stages: \emph{pre-thermalisation} (linear perturbation), \emph{driven turbulence} (bubble collisions), and \emph{thermalisation} towards thermal equilibrium. As opposed to the reheating process, we report that the driven turbulence stage lasts longer and the subsequent thermalisation process is different, which is caused by the presence of nontopological soliton solutions. During the turbulent stages, we find scaling laws for the variances of fields and for the spectra of the charge density. In addition, we adopt numerical lattice simulations to solve classical equations of motion, where our numerical code is developed from LATfield \cite{latfield-web-page}, and we present detailed nonlinear and nonequilibrium dynamics (some videos are available \cite{mit-web-page}).

\subsection{Linear evolution -- Pre-thermalisation}

The late evolution, after the AD condensate forms, depends on the properties of models. In the standard AD baryogenesis scenario \cite{Affleck:1984fy}, the condensate govern by the quadratic potential, \eq{quad}, decays into thermal plasma which may give our present baryons/leptons in the Universe. By including quantum and/or thermal corrections in the mass term as in \eqs{gravity-pot}{gauge-pot}, the subsequent evolution may be different from the standard AD scenario since the AD condensate has a negative pressure. The negative pressure, which causes the attraction force among particles in the condensate, amplifies exponentially the linear spatial fluctuations. We see this exponential growth for the linear perturbations in nearly circular orbit cases with the growth rate $\dot{S}_m$, and obtain the most amplified wave-number $k_m$, which give a rough estimate on the nonlinear time $t_{NL}$ and the radii of bubbles created just after the system enters into a nonlinear regime. As long as the perturbations are much smaller than the background field values, we call this initial linear perturbation stage '\emph{pre-thermalisation}'.

\subsubsection{Arbitrary and circular orbits}

Let us consider the linear spatial instability for an AD condensate in Minkowski spacetime. First, we perturb the AD field $\phi$ with the linear fluctuations, $\dsig$ and $\dtheta$. Equations of motion for $\dsig$ and $\dtheta$ are given by \eqs{pertsig}{pertth},
\bea{A8}
 \ddot{\dsig}-\bset{\nabla^2+\dot{\theta}^2-V^{\p\p}}\dsig-2\sigma\dot{\theta}\dot{\dtheta}=0,\\
\label{A9} \ddot{\dtheta}+\frac{2\dot{\sigma}}{\sigma}\dot{\dtheta}-\nabla^2\dtheta+\frac{2\dot{\theta}}{\sigma^2}\bset{\sigma\dot{\dsig}-\dot{\sigma}\dsig}=0.
\eea
Let us rescale $\dsig$ and $\dtheta$ in the following form
\be{scdsig}
\dsig\sim \dsig_0 e^{S(t)+i\mathbf{k}\cdot\mathbf{x}},\spc \dtheta \sim \dtheta_0 e^{S(t)+i\mathbf{k}\cdot\mathbf{x}}.
\ee
Notice that both of the exponents $S(t)$ should be the same in terms of a function of the wave number $\mathbf{k}$, because we are concerning with linear perturbations. Substituting \eq{scdsig} into \eqs{A8}{A9}, we obtain
\be{mat}
\left(
    \begin{array}{cc}
      \ddot{S}+\dot{S}^2+\mathbf{k}^2 -\dot\theta^2+V^{\p\p} & -2\dot\theta\dot{S} \\
      2 \dot\theta \bset{\dot{S}-\frac{\dot{\sigma}}{\sigma}} & \dot{S}^2+\frac{2\dot{\sigma} \dot{S}}{\sigma} +\mathbf{k}^2 \\
    \end{array}
  \right)
    \left(
    \begin{array}{cc}
        \dsig \\ \sigma\dtheta \\
        \end{array}
    \right) \simeq 0,
\ee
where $V^{\p\p}\equiv \frac{d^2
V}{d \sigma^2}$ and we ignore the terms $\ddot{S}$, assuming that the linear evolution is adiabatic, i.e. $\dot{S}^2 \gg \ddot{S}$ (WKB approximation). Notice that this assumption is violated only at the beginning of this linear evolution as we will see in the numerical subsection, Sec. \ref{qb_num}. The nontrivial solution for $\dot{S}$ can be obtained by taking the determinant of the matrix in \eq{mat}, namely
\begin{eqnarray}
\nb F(\dot{S}(k),k^2)&\equiv& \dot{S}^4 + \frac{2\dot{\sigma}}{\sigma}\dot{S}^3+\bset{2\mathbf{k}^2 +3\dot{\theta}^2+V^{\p\p}}\dot{S}^2\\ \label{FS} \label{FSK} &&+\frac{2\dot{\sigma}}{\sigma}\bset{\mathbf{k}^2-3\dot{\theta}^2+V^{\p\p}}\dot{S} + \mathbf{k}^2 \bset{\mathbf{k}^2- \dot{\theta}^2+V^{\p\p} }=0.
\end{eqnarray}
Notice that the terms involving $\dot{\sigma}$ vanish if the orbit of the AD field is exactly circular. By looking for the most amplified mode $k^2_m$, which is defined by $\left.\frac{\partial F}{\partial k^2}\right|_{k^2_m}=0$ from \eq{FS}, it implies that
\be{kmost}
k^2_m=\frac{\dot{\theta}^2-V^{\p\p}}{2}-\dot{S}\bset{\dot{S}+\frac{\dot{\sigma}}{\sigma}}>0,
\ee
where the inequality comes from the reality condition for $k_m$.
By concerning with this mode in \eq{kmost} and by solving $F(\dot{S}(k),k^2_m)=0$ in \eq{FS}, the solution for $\dot{S}_m\equiv \dot{S}(k=k_m)$ is
\be{Sevo}
\dot{S}_m=\frac{\frac{\dot{\sigma}}{\sigma}\bset{5\dot{\theta}^2-V^{\p\p}} \pm 2 \dot{\theta}\sqrt{\bset{\dot{\theta}^2-V^{\p\p}}^2+2\bset{\frac{\dot{\sigma}}{\sigma}}^2\bset{3\dot{\theta}^2-V^{\p\p}} }}{2\bset{4\dot{\theta}^2-\bset{\frac{\dot{\sigma}}{\sigma}}^2}},
\ee
in which we are interested in the growing mode, i.e. $Re(\dot{S}_m)>0$. Substituting \eq{Sevo} into \eq{kmost}, we may obtain the most amplified mode. Although it is rather hard to analytically solve \eq{FSK}, we know that the only one instability band exists for exactly circular orbits where $\dot{\sigma}=0$;
\be{instability}
0<\mathbf{k}^2< \dot{\theta}^2-V^{\p\p}(\sigma),
\ee
where $\dot{\theta}$ and $\sigma=\sigma_{cr}$ are time-independent due to the circular orbits.

\vspace*{10pt}

In addition, we can estimate a possible nonlinear time $t_{NL}$ when the spatial averaged variance, Var$(\sigma)$, becomes comparably large to the corresponding homogeneous mode $\sigma$. Here, we defined Var$(\sigma)\equiv \overline{\bset{\hat{\sigma}(\mathbf{x}, t)-\overline{\sigma}}^2}$, and a hat and a bar denote an original field and a spatial average of the field, respectively. Notice that the nonlinear time in \cite{Enqvist:1998en, Pawl:2004vi} is defined by the time when the linear fluctuation $\delta\sigma$ for the most amplified mode becomes comparable large to the homogeneous-mode; however, our definition is of more advantages as we see in the numerical subsection, Sec. \ref{qb_num}. The nonlinear time with our definition can be given by
\bea{nonlintime}
Var(\sigma)&\sim& \delta \sigma^2_0 \exp\bset{2 N \overline{\state{\dot{S}}}\tau} \sim \delta \sigma^2_0 \exp\bset{\int^{t_{NL}}_{t_*} 2 \state{\dot{S}_m}}\sim \sigma^2_0,\\
\label{nlt}\lr t_{NL} & \sim & t_* + \frac{1}{\state{\dot{S}_m}}\ln\bset{\frac{\sigma_0}{\delta \sigma_0}}.
\eea
Here, we approximated that $\overline{\state{\dot{S}}}\sim \state{\dot{S}_m}$ and that the orbits over $N$ rotations with the period $\tau$, \eq{periodgen}, can be expressed by the integral form as shown in \eq{nonlintime}. As we assumed, the spatial averaged variance of this field is not fully developed over all modes except $k=k_m$ until $t\sim t_*$, where $t_*$ is a typical time scale when the variance starts to grow with the growth rate $\overline{\state{\dot{S}_m}}$.

\vspace*{10pt}

Our main interest in this pre-thermalisation stage is the evolution of the number of particles in terms of modes, so that we consider $\rho_Q$ as the particle number here. For a free field theory, both of the positive and charged particle occupation number develop equally. The present case, however, gives different consequences due to the presences of nonlinear interactions and the initial inequality of a charge density (baryon asymmetry). Without loss of generality, we can focus on the case where the positive charge is initially present. Since the charge density is given by $\rho_Q=\hat{\sigma}^2\dot{\hat{\theta}}$, we can approximately obtain the evolution in the linear regime using \eqs{radhomo}{phshomo},
\be{evochrg}
\dot{\rho}_Q\simeq \sigma^2(t)\nabla^2\delta\theta.
\ee
Hence, the charge density evolves due to the linear fluctuation of the phase field. Let $n^{\pm}_k(t)$ be the amplitude of Fourier-transformed positive and negative charge density, $n^{\pm}(\mathbf{x}, t)$, which are defined through the following decomposition, $\rho_Q=n^+(\mathbf{x}, t)-n^-(\mathbf{x}, t)$. Notice that the Fourier transformed functions, $n^\pm_k$, are related to, but are potentially different from the corresponding quantum mechanical expressions, $\tilde{n}^+_k\equiv a^\dag_k a_k,\; \tilde{n}^-_k\equiv b^\dag_k b_k$ and $Q=\int d^3x \rho_Q=\int \frac{d^3k}{(2\pi)^{3/2}} \bset{\tilde{n}^+_k - \tilde{n}^-_k}$. Here, $\tilde{n}^\pm_k$ are occupation numbers for positive and negative charged particles in a free field theory, and $a_k,\; a^\dag_k,\; b_k$ and $b^\dag_k$ are the annihilation/creation operators for both of the particles, respectively. Since we are interested in the growing mode for the positive charge density $n^+_k(t)$ in \eq{evochrg} which is initially zero except the zero-momentum mode, it implies that using \eq{scdsig}
\bea{n+k0}
\nb n^+_k(t) &\simeq& k^2|\delta\theta_0|\int^{t}_{t_0} d\tilde{t} \sigma^2(\tilde{t})e^{\state{\dot{S}(k)}\tilde{t}},\\
\label{n+k} &\sim&  k^2|\delta\theta_0| \sigma^2_{cr} \frac{e^{\state{\dot{S}}(t-t_0)}}{\state{\dot{S}}} \propto e^{\state{\dot{S}}(t-t_0)},
\eea
where $t_0$ is a numerical value and we assumed $\sigma^2(t)\sim \sigma^2_{cr}$, going from the first line to the second one. Therefore, the evolution of the positive charged particle number for a mode $k$ is proportional to $e^{\state{\dot{S}(k)}(t-t_0)}$.

\vspace*{10pt}

Our results, \eqs{kmost}{Sevo}, are generalisations of the known results \cite{Kasuya:2001hg, Enqvist:2002si}, in which the orbit of the Affleck-Dine condensate is exactly circular. We also obtained the nonlinear time $t_{NL}$ in \eq{nlt} and the exponential growth of the particle number in \eq{n+k}.

\subsubsection{Nearly circular orbits in Model A and B}\label{NCO}

Using the results obtained in the previous subsection, we can compute the most amplified mode $\state{\mathbf{k}^2_m}$ and the growing mode $\state{\dot{S}_m}$ averaged over one rotation of the nearly circular orbits for the models introduced in Section \ref{MODEL-ABC}, i.e. Model A and Model B. We shall confirm that these values are the same as the cases when the orbits are exactly circular, which implies that the instability band, \eq{instability}, could exist even for the present nearly circular orbit cases.

\vspace*{10pt}
\paragraph*{\underline{\bf Model A:}}
Substituting the expressions, $\dot{\sigma}/\sigma,\; \dot\theta^2$ and $V^{\p\p}$ (\textit{c.f.} \eqs{sigsol}{dottheta} and \eq{modelA-W}), into \eq{Sevo}, we obtain the averaged growing factor for Model A where $M^2>0$:
\bea{ext11}
\state{\dot{S}_m}&\simeq& \pm \frac{(2-l)M}{4}\sqrt{\frac{l\sigma^{l-2}_{cr}}{2}}\bset{1 + \frac{(l+2)(2n-l-2)}{2(n+2)(l-2)} \epsilon_A   },\\
\label{ext13} \state{\mathbf{k}^2_m}&\simeq& \frac{M^2l(2-l)(l+6)\sigma^{l-2}_{cr}}{32}\bset{1+\frac{(l+2)(4n-12-l^2+2nl)}{(n+2)(l-2)(l+6)}\epsilon_A},
\eea
where we substituted \eq{ext11} into \eq{kmost} to obtain $\state{\mathbf{k}^2_m}$ and these results are consistent with the case for the exactly circular orbit. In order to satisfy $\state{\mathbf{k}^2_m}>0$, we should have $l<-6,\; 0<l<2$, and \eq{ext11} implies that the condensate is unstable against spatial fluctuations when the pressure is negative with $0<l<2$, see \eq{powerpress}.

We can check the results \cite{Enqvist:2002si} that $\state{\dot{S}_m}\simeq \frac{m|K|}{2}\bset{1+\frac{|K|}{2}}$ and $\state{\mathbf{k}^2_m}\simeq m^2|K|\bset{1-\frac{|K|}{4}}$ by setting $l=2-2|K|$ in \eqs{ext11}{ext13} and by ignoring the nonrenormalisable term as did in \cite{Enqvist:2002si}, i.e. $\state{\dot{S}_m} \simeq \frac{|K|M}{2}\bset{1-\frac{|K|}{2}}\sigma^{-|K|}_{cr}$ and $\state{\mathbf{k}^2_m}\simeq |K|M^2\bset{1-\frac{5|K|}{4}}\sigma^{-2|K|}_{cr}$. These are of the same order as their results, recalling that $\sigma^{-2|K|}_{cr}\sim \order{1}$ due to $|K|\ll \order{1}$.

\vspace*{10pt}
\paragraph*{\underline{\bf Model B:}}

Similarly, we can also obtain the averaged growing factor for Model B from \eq{ModelC-W}
\be{ext12}
\state{\dot{S}_m}\simeq \frac{m^2_{\phi}}{\sqrt{2}\sigma_{cr}}\bset{1-\frac{n-1}{n+2}\epsilon_B},\hspace*{5pt}
\state{\mathbf{k}^2_m}\simeq \frac{3m^4_{\phi}}{2\sigma^2_{cr}}\bset{1-\frac{2(n-3)}{3(n+2)}\epsilon_B}
\ee
whose leading orders reproduce the results \cite{Kasuya:2001hg} in which case that the AD orbit is assumed to be exactly circular and ignored the nonrenormalisable term.

\vspace*{10pt}
Before we finish this subsection, let us remark the classical and absolutely stability of AD condensates. Lee found \cite{Lee:1994qb} that the dispersion relations for the waves of linear fluctuations from \eq{FSK} when the orbits of the AD field are bounded. For the longwave-length limits, there exist one massive and one massless modes. The massless mode can be interpreted as the sound wave whose sound speed should be real for the classical stability reason, and the squared value of the sound speed is related to the value of $\state{w}$ in \eq{eosg}. Therefore, this stability condition for the sound waves corresponds to the sign of the pressure in the AD condensates. In other words, the AD condensate has a negative pressure when the sound speed is imaginary; equivalently, it is classically unstable against spatial fluctuations. The zero-pressure AD condensate whose energy density is minimised with respect to any degrees of freedom is equivalent to the $Q$-matter phase as Coleman discussed in \cite{Coleman:1985ki}, where the absolutely stable $Q$-matter can be excited by classically stable sound waves. 

\subsection{Non-linear evolution and nonequilibrium dynamics}

\subsubsection{Driven (Stationary) and free turbulence}

Even when the perturbations are fully developed to support the nonlinear solutions, the system is still far from equilibrium. Indeed, the system enters into more stochastic stages, 'turbulence regimes', where the strength of the turbulent behaviour depends on the ``Reynolds'' number \cite{Grana:2001ms}. As a theory of reheating Universe, a general nonequilibrium system goes through two different turbulence stages, going from driven turbulence to free turbulence stages. A major energy transfer from the zero mode takes place during driven turbulence. Garcia-Bellido \textit{et. al.} \cite{GarciaBellido:2007af} observed that bubbles form and collide during this stage in tachyonic preheating, and they suggested that the bubble collisions can be an active source of gravitational waves \cite{Khlebnikov:1997di}. In the usual reheating scenarios, this stage terminates when the energy left out in the zero-mode becomes smaller than the energy stored in other modes (created particles). Since the energy exchange between zero-mode and other modes becomes negligible, the particle distribution is self-similar in time (free turbulence) and evolves towards thermal equilibrium. In the free turbulence stage, the quantum effects change the late evolution significantly, and the created particles are distributed followed by Bose-Einstein statistics rather than by a classical manner. As long as an active and stable energy source exists in momentum space, we expect that the driven turbulence stage lasts for a long time. In the case of $Q$-ball formation, we expect that this active energy source corresponds to the excited states of $Q$-balls; hence, we can expect that the driven turbulence stage lasts longer compared to the linear perturbation regime as opposed to the usual reheating Universe scenarios. Note that during this thermalisation stage the transition from the classical to quantum regime becomes important \cite{Micha:2004bv}; in the rest of this paper we concentrate on the case where the system is govern by classical evolution all the time.
In turbulent stages, the scaling law can be found \cite{Micha:2004bv}:
\be{variation}
\textrm{Var}(\sigma) \propto t^{p},
\ee
where the power $p$ depends on the parameters of models, e.g. the relativistic values of $p$ are $p=\frac{1}{2m-1}$ in the driven turbulence regime and $p=-\frac{2}{2m-1}$ in the free turbulence regime. Here, $m$ is the number with which particles interact. For the free turbulence regime, the particle number distribution follows a scaling law from the time $t_{free}$ when the free turbulence turns on, namely
\be{nmbdist}
n_k(t)=t^{-\frac{4}{2m-1}}n_{k_*}(t=t_{free}),
\ee
where $k_*\equiv kt^{-\frac{1}{2m-1}}$.

\subsubsection{Thermal equilibrium state in the presence of nontopological solitons}

In this sub-subsection, we show that the condition of the negative pressure is the same as the existence condition of $Q$-balls,  known in \cite{Coleman:1985ki}. This does not always mean that the spatially unstable condensate evolves towards to $Q$-balls; with given initial conditions, the condensate may evolve into other thermo-dynamically favour states in which the free energy is minimised. 

The ansatz of $Q$-balls claims that $\dot{\hat{\theta}}$, which corresponds to the ``chemical potential'' $\omega$, is constant, and that the radial field $\hat\sigma$ should be time-independent and depend on the radius $r$ of the $Q$-ball, i.e. $\hat{\phi}=\hat{\sigma}(r)e^{i\omega t}$. Hence, the existence condition of $Q$-balls at zero-temperature is
\be{exist}
\textrm{min}\bset{\frac{2V}{\hat\sigma^2}}\leq \omega^2 < \left.\frac{d^2V}{d\hat\sigma^2}\right|_{\hat\sigma=0}.
\ee
This condition implies that the potential should grow less quickly than a quadratic term; thus, it is equivalent to the fact that the AD condensate has a negative pressure for $l<2$ in \eq{modelA}, see \eq{powerpress}. Notice that this condition only tells that $Q$-balls may appear after a unstable AD condensate fragments. The evolution to the thermal equilibrium state is rather hard to compute analytically, and it is related to stability problems of the $Q$-balls \cite{Copeland:2009as, Laine:1998rg}. Therefore, we conduct numerical lattice simulations that give the entire processes of nonlinear as well as nonequilibrium evolution.

\subsection{Numerical results}\label{qb_num}

In this subsection, we present detailed numerical results involved with lattice simulations for both GRV-M and GAU-M models with the parameter sets, SET-3 and SET-7 shown in \tbl{parameterSET}; we then check our analytical results obtained in the previous sections. In order to solve the second-order partial differential equations, $\frac{d^2\hat{\phi}}{dt^2} -\nabla^2 \hat{\phi} + \frac{dV}{d\hat{\phi}^\dag}=0$, with the potentials introduced in \eqs{numGRV}{numGAU}, we use the following appropriate parameters: $dx=0.2,\; dt=0.02$ in GRV-M Model and $dx=5.0,\; dt=0.2$ in GAU-M Model, which minimise the numerical errors. Here, $dx$ is the fundamental lattice space and $dt$ is the time step. Note that the variables in this subsection are normalized by appropriate energy scales as in Sec. \ref{numAD}. We then conduct $3+1$ (and $2+1$)-dimensional lattice simulations with $512^3$ (and $512^2$) lattice units, imposing a periodic boundary condition. Our initial conditions are, $\hat\phi_0=\phi_0+\delta\phi_0$ and $\dot{\hat{\phi}}_0=\dot{\phi}_0+\delta\dot{\phi}_0$, where the initial fluctuations, $\delta\phi_0$ and $\delta\dot{\phi}_0$, are of a Gaussian noise, which are responsible for ``quantum'' fluctuations. Their fluctuations, $\delta\phi_0$ and $\delta\dot{\phi}_0$, are of order $10^{-5}$ in GRV-M case and of order $10^{-3}$ in GAU-M case. In order to visualise these detailed evolution, we use a 3D software, 'VAPOR' \cite{vapor-web-page}, and some videos of our numerical results are available in \cite{mit-web-page}.

\subsubsection{Pre-thermalisation}

\paragraph*{\underline{\bf The initial evolution --Non-adiabaticity:}}

\begin{figure}[!ht]
  \begin{center}
	\includegraphics[angle=-90, scale=0.31]{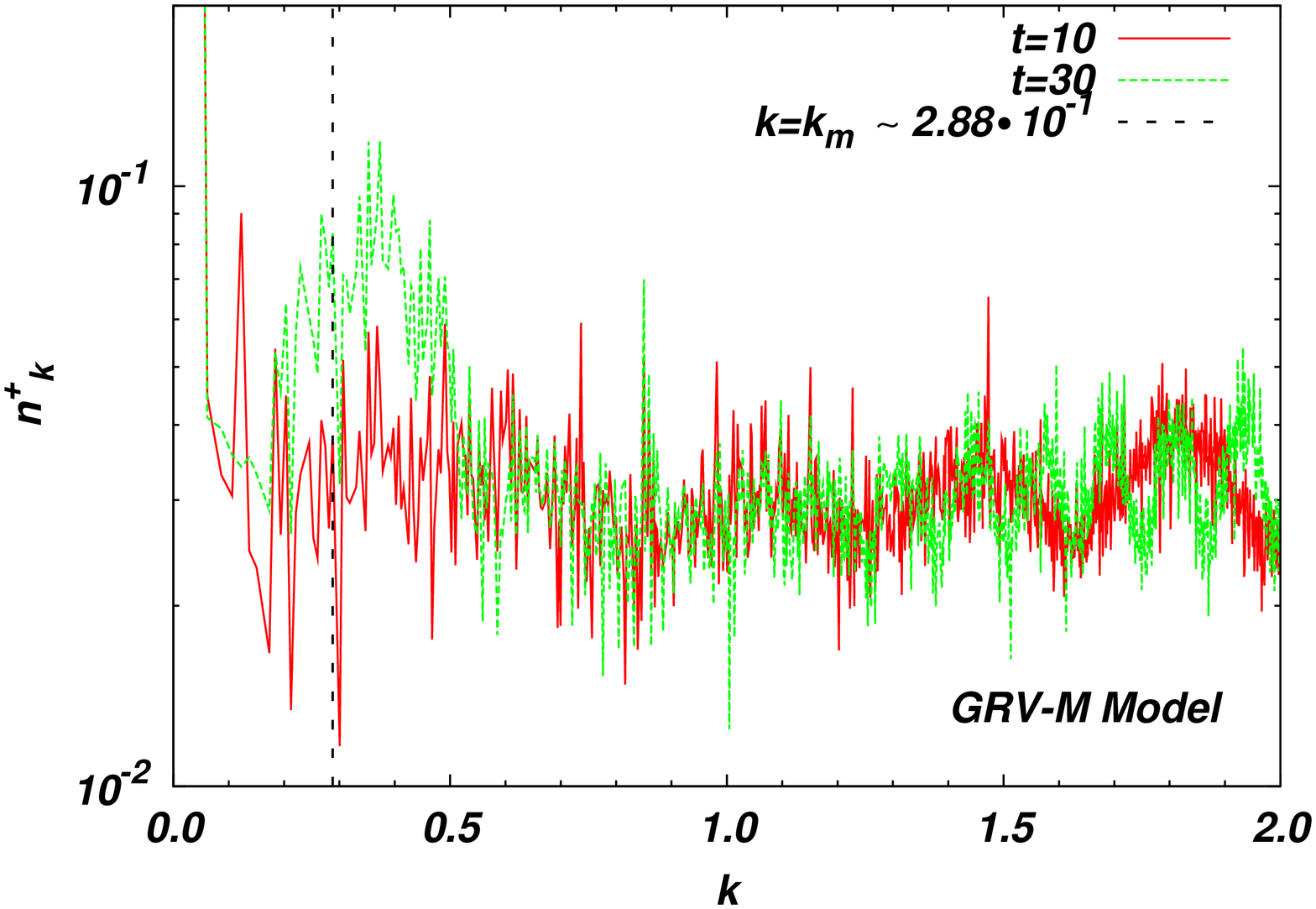}
	\includegraphics[angle=-90, scale=0.31]{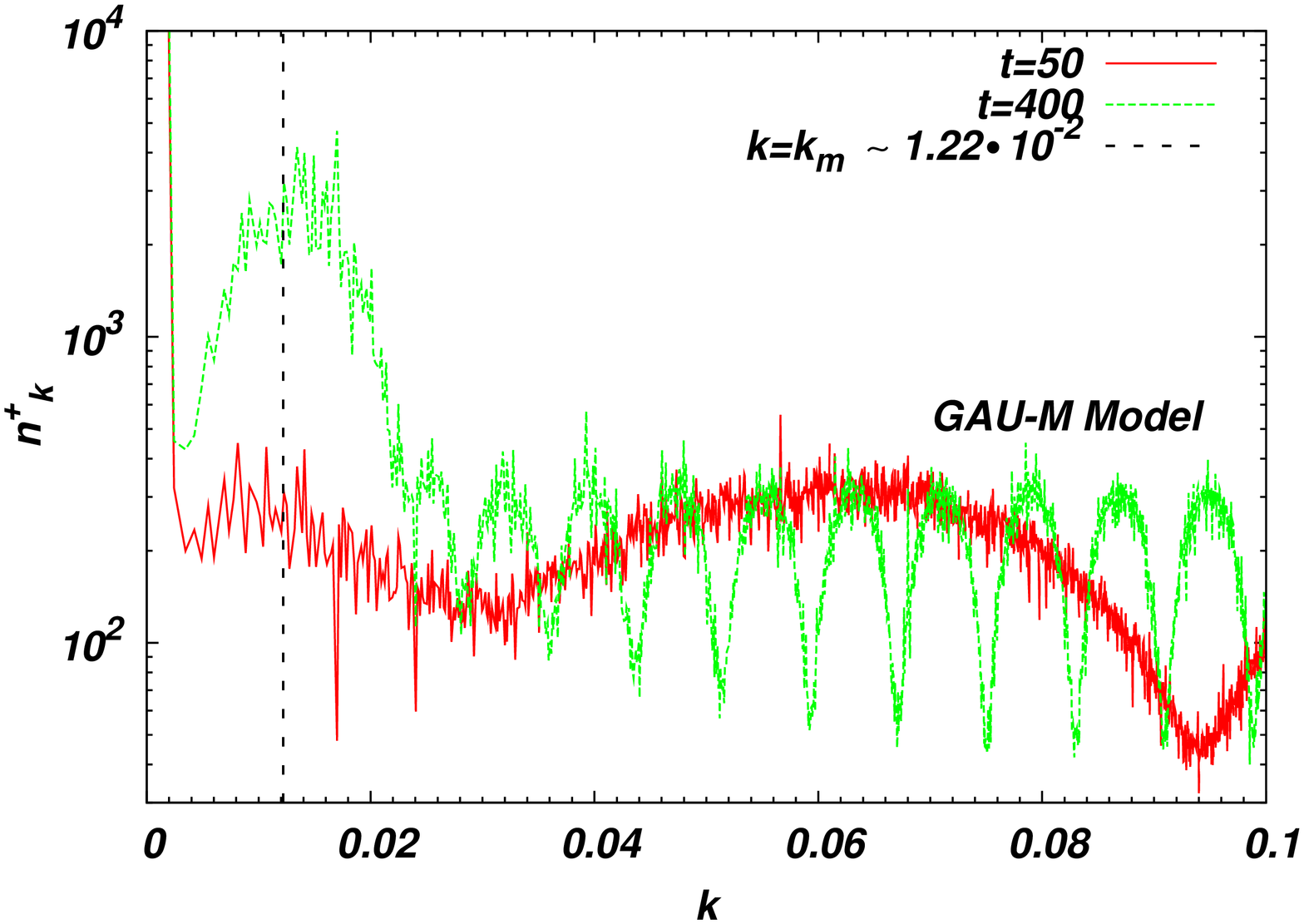}\\
	\includegraphics[angle=-90, scale=0.31]{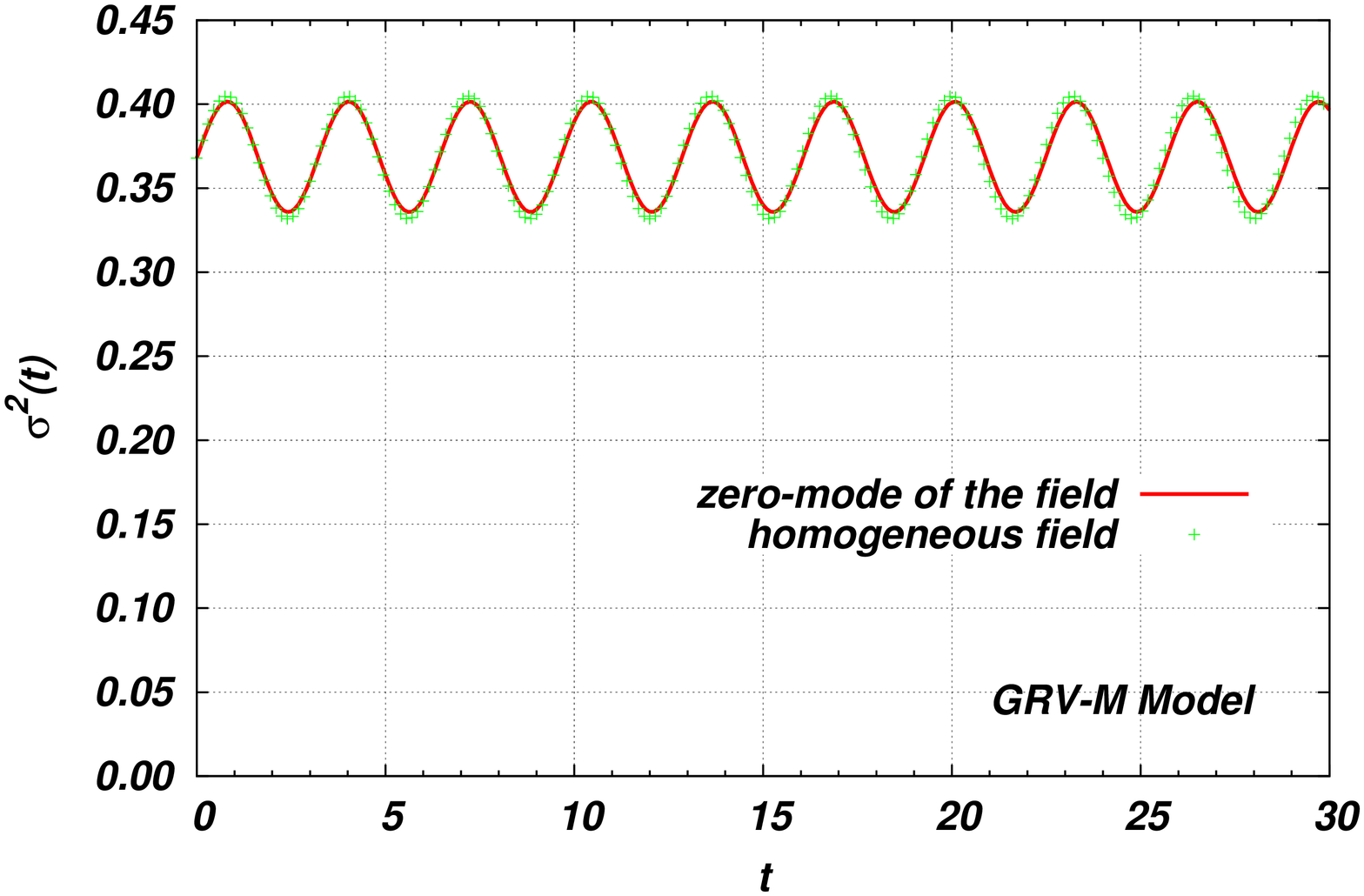}
	\includegraphics[angle=-90, scale=0.31]{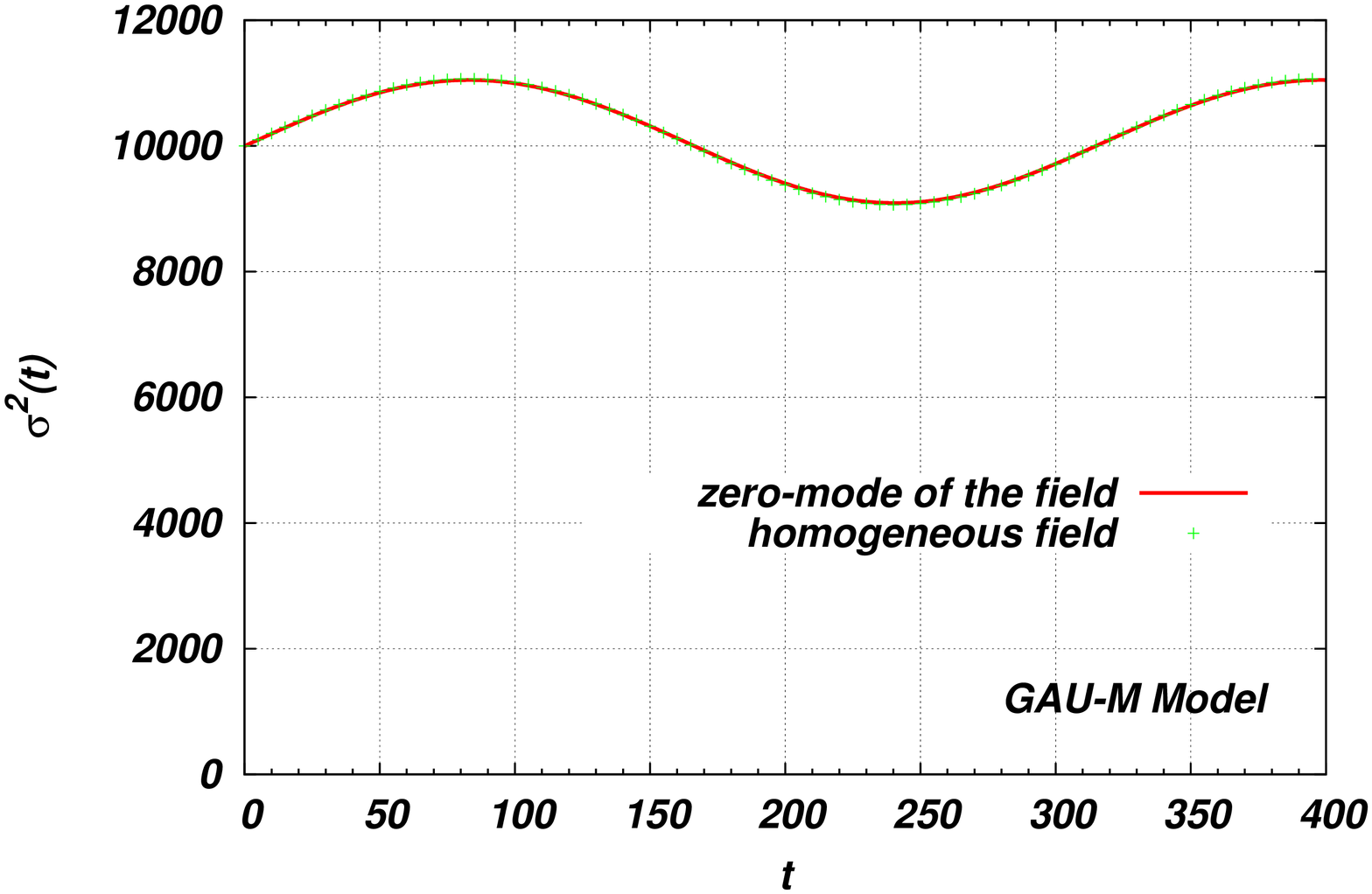}\\
	\includegraphics[angle=-90, scale=0.31]{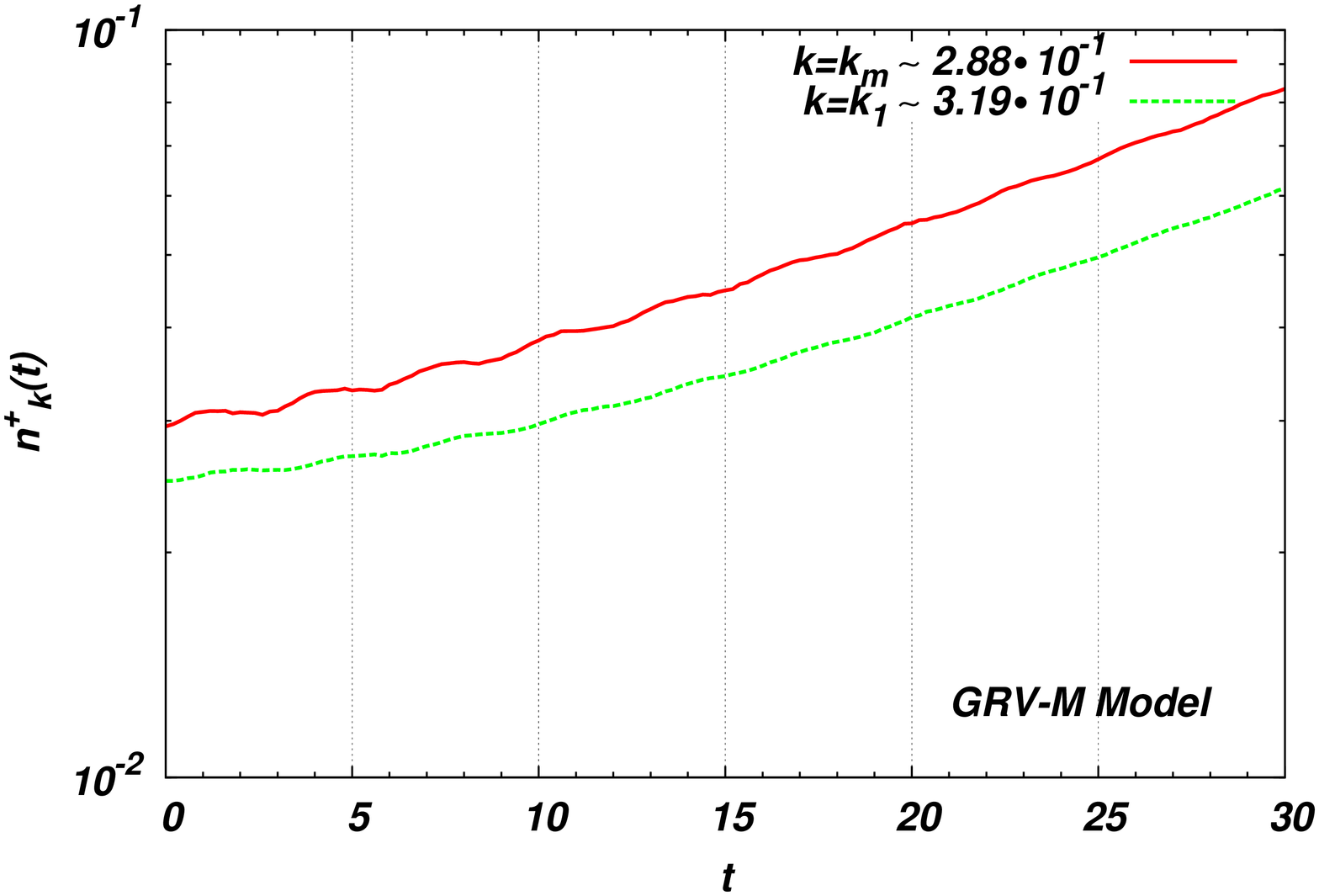}
	\includegraphics[angle=-90, scale=0.31]{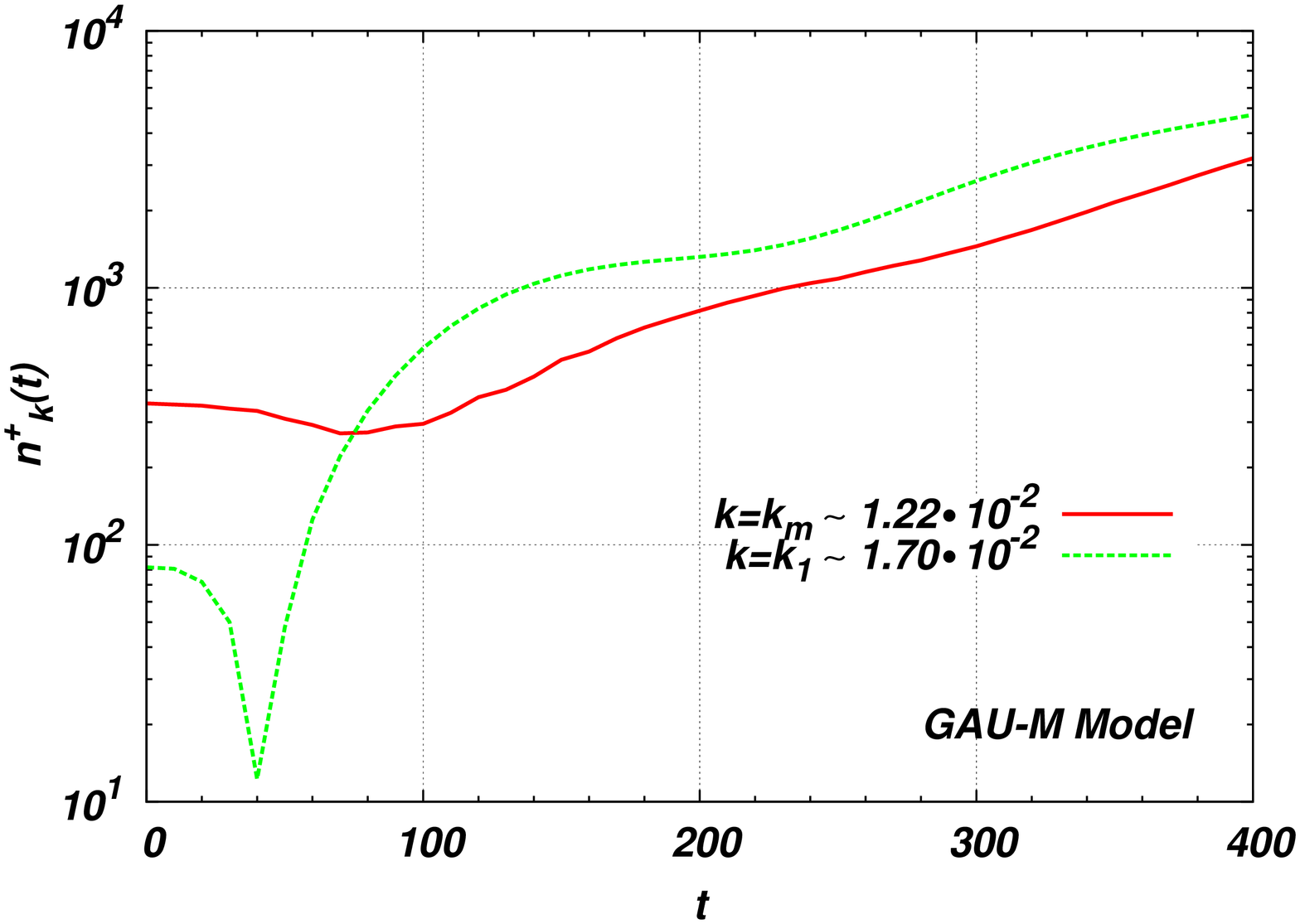}
  \end{center}
  \caption{ \textbf{(Color online)}  In the top two panels, we plot the amplitude of $n^+_k(t)$ with two different time steps for GRV-M Model in the left panel and GAU-M Model in the right panel, where we took the average of $n^+_\mathbf{k}(t)$ over the axes of $\mathbf{k}$. The black-dashed vertical lines indicate the analytical values of the most amplified modes $k_m$ obtained from \eqs{ext13}{ext12}. In the middle panels (GRV-M Model on left and GAU-M Model on right), we compare the zero-mode $\overline{\sigma^2}$ (red-solid lines) and the homogeneous field $\sigma^2$ (green-plus dots) obtained in the bottom panels of \fig{fig:sigsq}. In the bottom panels of \fig{fig:nklin}, we plot the evolution of $n^+_k(t)$ for both analytic values $k_m$ (red-solid lines) and numerical values $k_1$ (green-dashed lines) of $n^+_k$ shown in the top two panels.}
  \label{fig:nklin}
\end{figure}

In the top two panels of \fig{fig:nklin}, we plot the amplitude of $n^+_k(t)$, where we took the average of $n^+_\mathbf{k}(t)$ over the axes of $\mathbf{k}$. We show the amplitudes of $n^+_k(t)$ for GRV-M Model in the left panel and for GAU-M Model in the right panel with two different time steps. In the panels, we indicate the analytical values of the most amplified modes $k_m$ obtained from \eqs{ext13}{ext12} with black-dashed vertical lines. In GRV-M Model, the amplitude with $t=30$ (green-dashed line) is a little noisy to see the first peak $k_1$ in terms of $k$. Our analytical estimate, $k_m\sim 2.88\times 10^{-1}$, is located at a more infrared region than the point $k=k_1\sim 0.34$, and the periodic structure can be seen in the higher-momentum space. In GAU-M Model, on the other hand, we can confirm that our analytical value, $k=k_m\sim 1.22\times 10^{-2}$, agrees with the numerical value, $k_1\sim 1.70\times 10^{-2}$, in the green-dashed line; however, the analytical value appears in a slightly more infrared region. We also observe the periodic structure in the higher-momentum modes as it was reported in \cite{Kasuya:2001hg}. In the middle panels (GRV-M Model on left and GAU-M Model on right), we compare both the zero-mode, $\overline{\sigma^2}$ (red-solid lines), and the homogeneous field, $\sigma^2$ (green-plus dots), shown in the bottom panels of \fig{fig:sigsq}. The middle panels in both cases show that the zero-mode does not decay quickly, and it oscillates around $\sigma^2=\sigma^2_{cr}$. We can also check that our numerical parameters are appropriate, minimising numerical errors. In the bottom panels of \fig{fig:nklin}, we plot the evolution of $n_k(t)$ for the modes both $k_m$ (red-solid lines) and $k_1$ (green-dashed lines). In the left bottom panel, we can see the exponential growth of the amplitude in GRV-M Model for both modes, and step-like particle production exists at the beginning of the evolution as broad resonant preheating \cite{Kofman:1997yn} (c.f. \eq{evochrg}), and it begins to create the particles exponentially afterwards. The particles are produced quickly when the zero-mode $\sigma^2(t)$ increases in time at the beginning, see the middle panels. This is the different feature of the evolution compared to the case of resonant preheating, where particle production for the broad resonance occurs nonadiabatically when the zero mode (inflaton field) crosses the zero axis. In the right bottom panel, we can see more clearly the step-like particle creations for both modes, and then this step-like evolution smoothes out, which leads to the exponential particle production as in the GRV-M case. We believe that the adiabatic condition, $\ddot{S}\ll \dot{S}^2$, is ``softly'' violated only in this initial stage since we can not see the clear exponential growth at the beginning of this evolution. In the next paragraph, we discuss the late linear evolution when this nonadiabatic evolution ceases, and show that our analytical results agree with numerical ones more nicely.

\vspace*{10pt}
\paragraph*{\underline{\bf Up to the nonlinear time:}}

\begin{figure}[!ht]
  \begin{center}
	\includegraphics[angle=-90, scale=0.31]{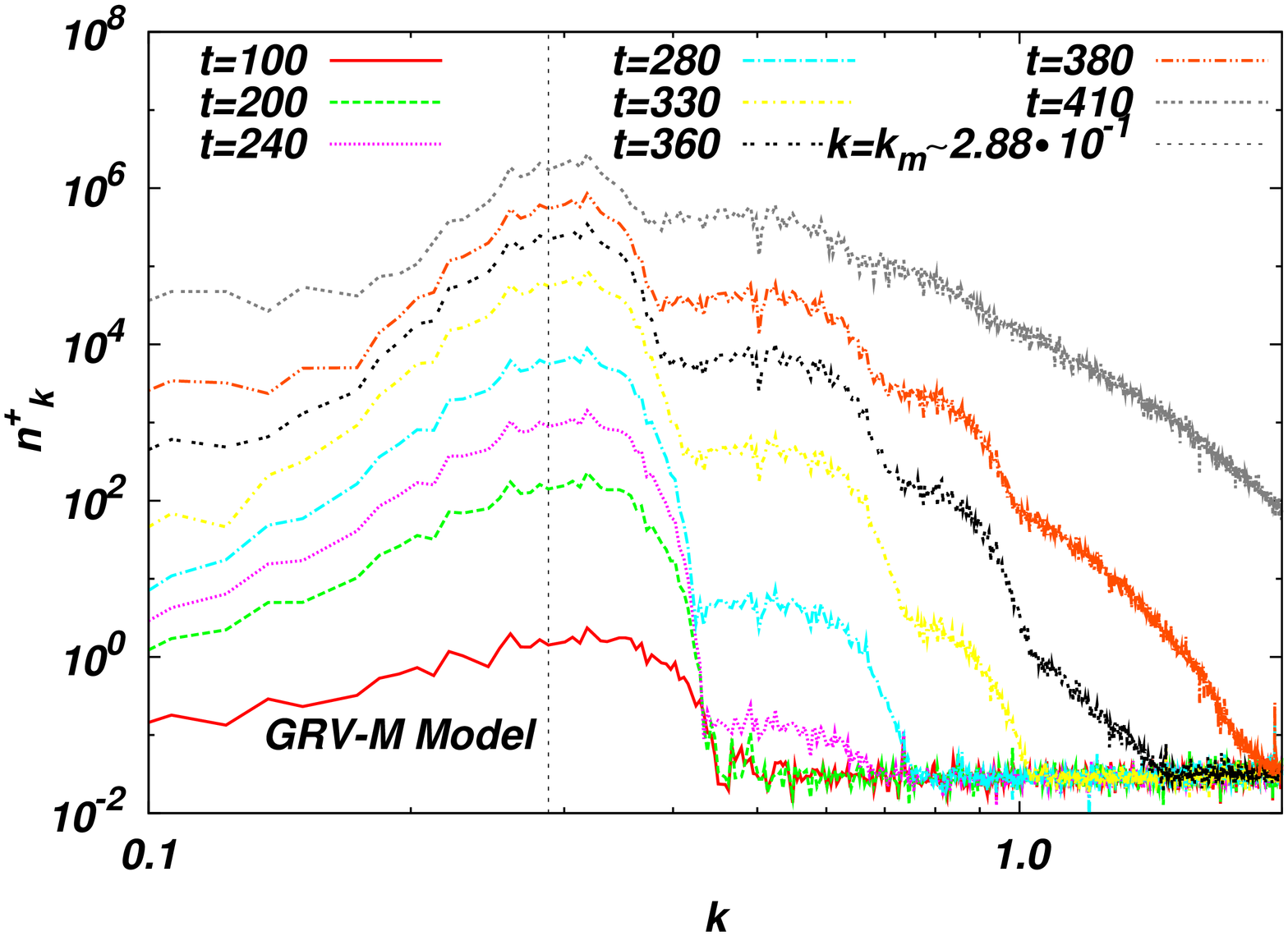}
	\includegraphics[angle=-90, scale=0.31]{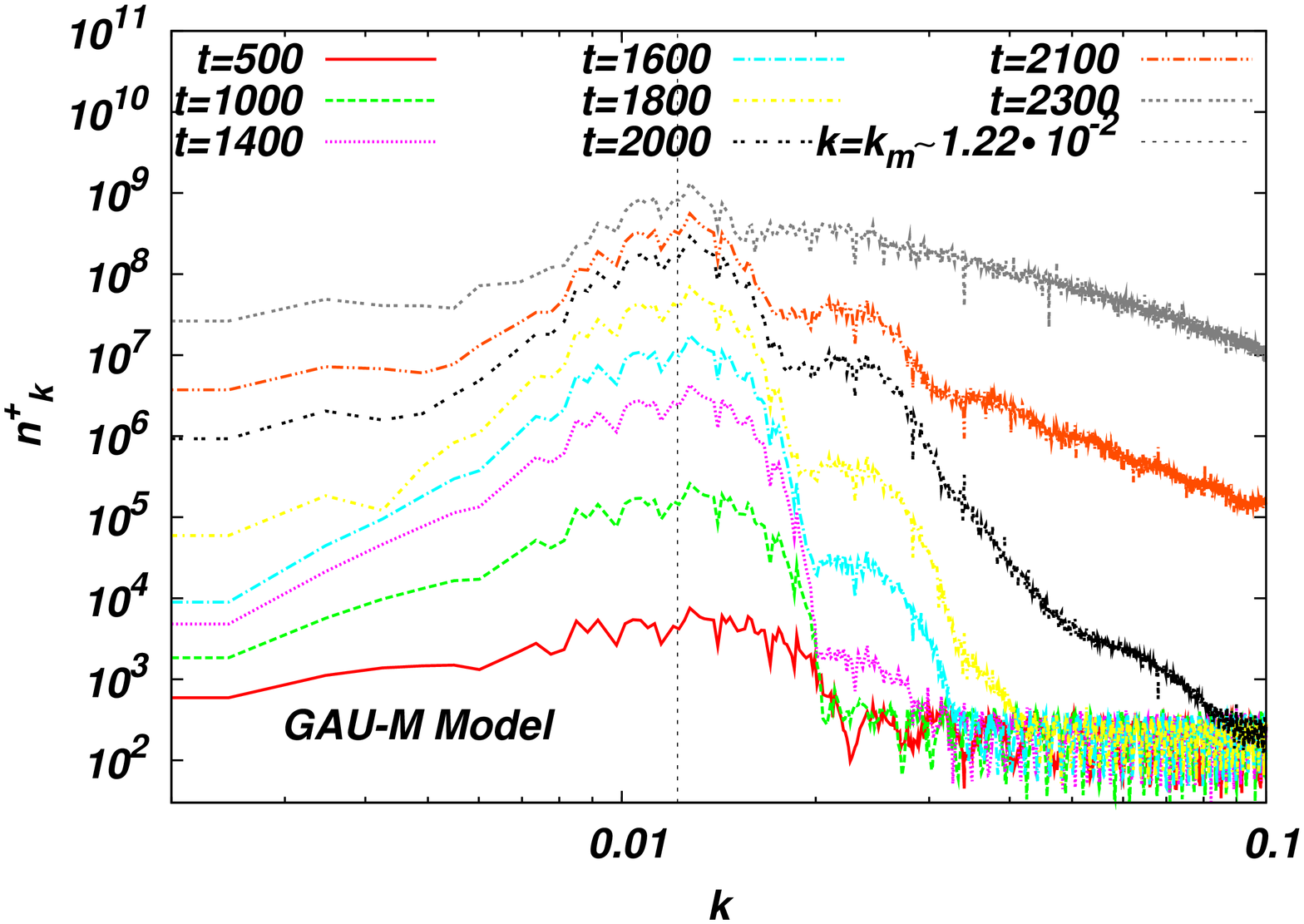}\\
	\includegraphics[angle=-90, scale=0.31]{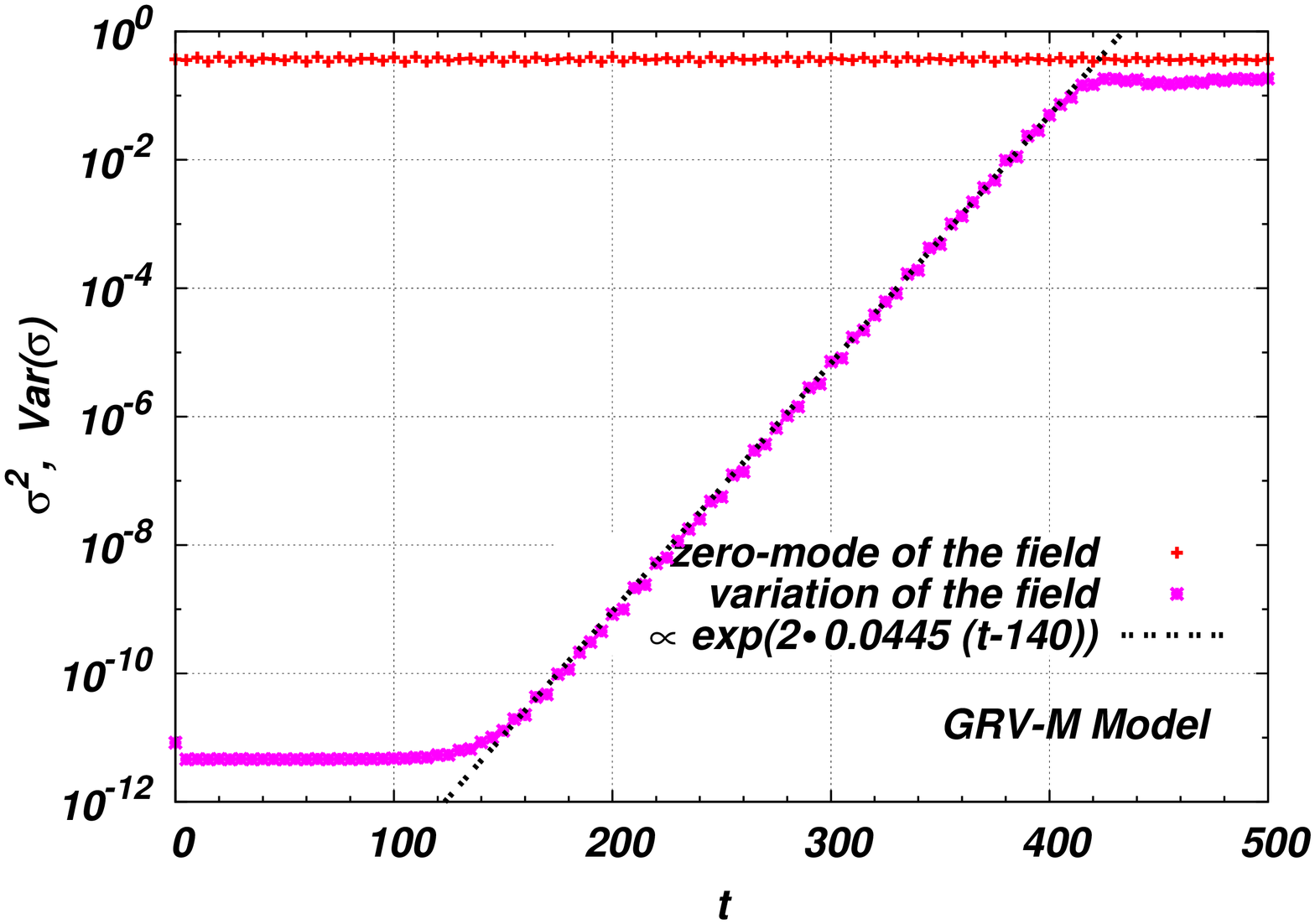}
	\includegraphics[angle=-90, scale=0.31]{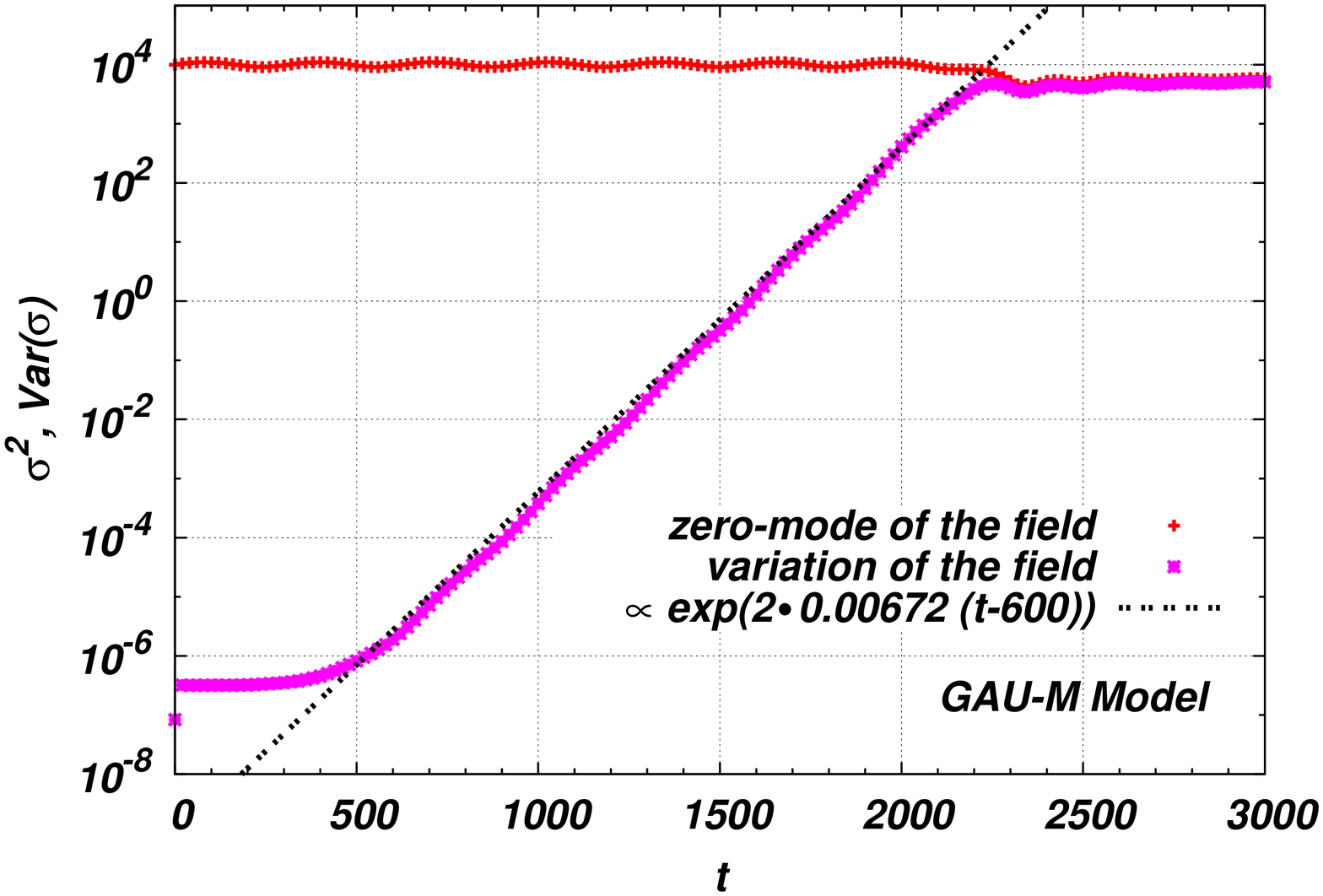}\\
	\includegraphics[angle=-90, scale=0.31]{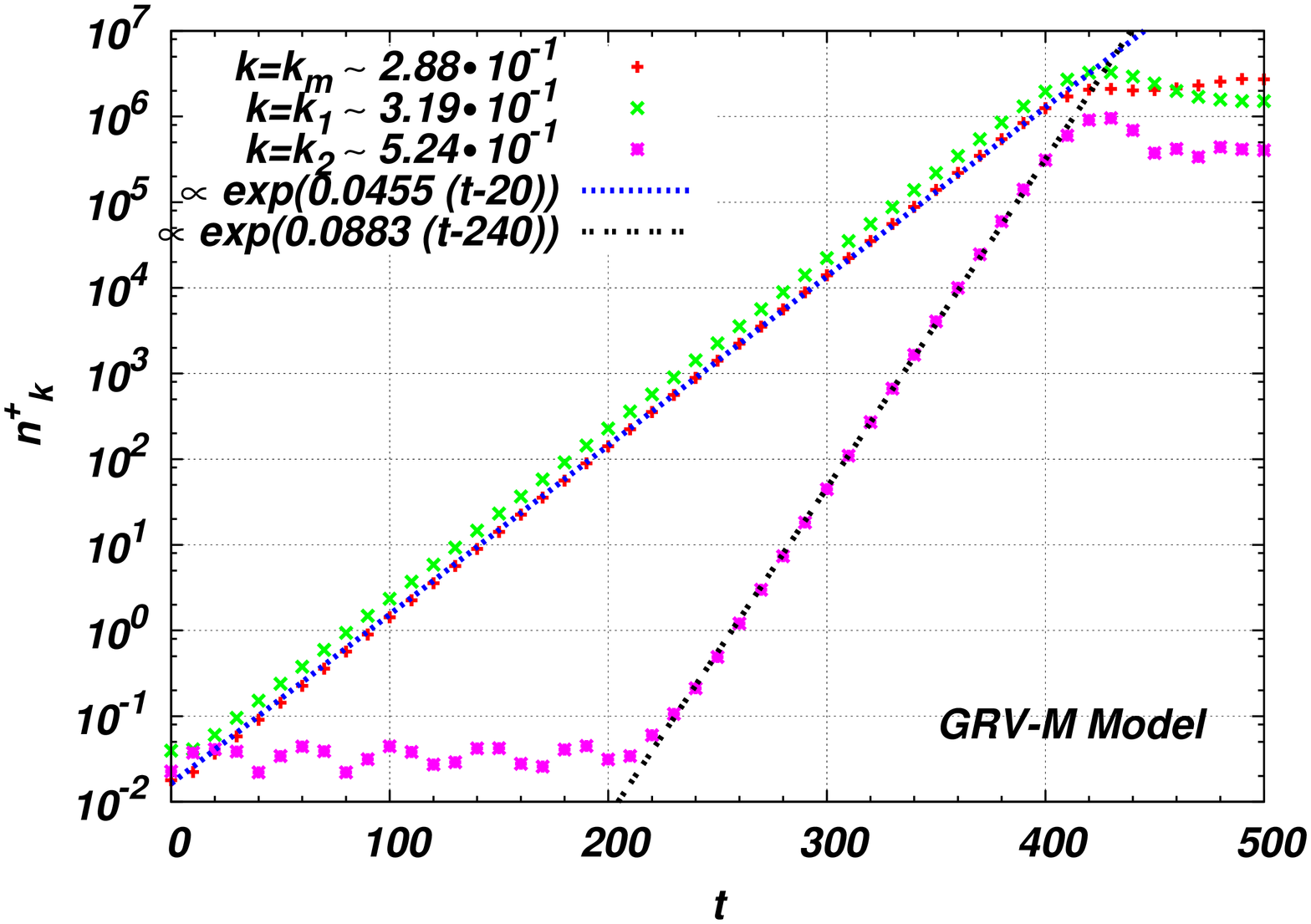}
	\includegraphics[angle=-90, scale=0.31]{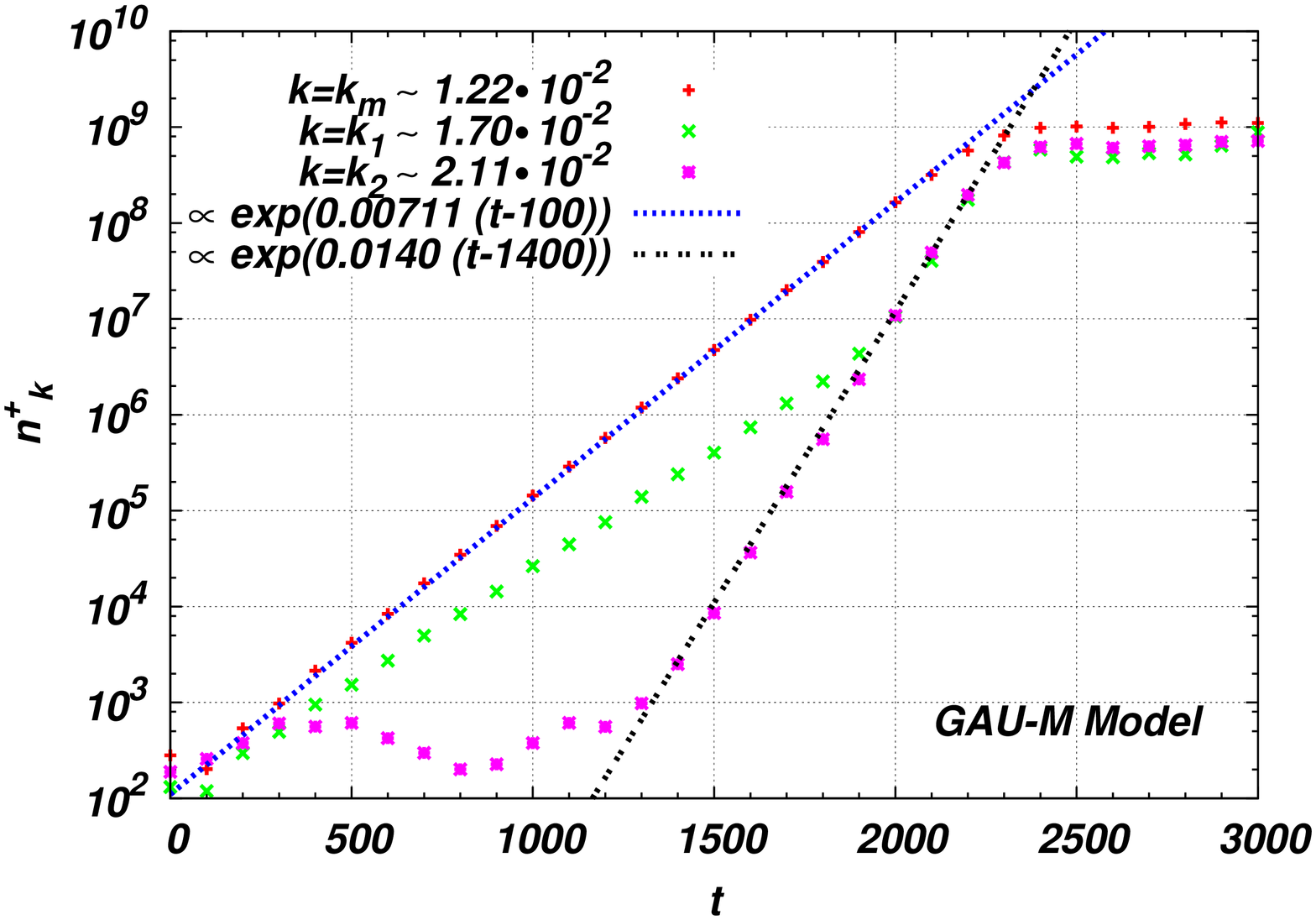}
  \end{center}
  \caption{ \textbf{(Color online)} The top panels plot the amplitude of $n^+_k$ with various times in both GRV-M Model (left) and GAU-M Model (right). The analytical value of the most amplified mode $k_m$ in black-dashed vertical lines agrees with the first peak, $k_1$, of the spectra in both cases. The middle panels show the evolution of zero-mode field, $\overline{\sigma^2}$ (red-plus dots), and the variance of the field, Var$(\sigma)$ (green-cross dots), up to the nonlinear time $t=t_{NL}$, where we can set $t_{NL}\sim 420$ in GRV-M Model and $t_{NL}\sim 2200$ in GAU-M Model. In the bottom panels, we plot the evolution of the amplitude $n^+_k$ for the first $k_1$ (red-plus dots), second peak $k_2$ (green-cross dots) modes and the analytical most amplified modes $k_m$ (purple squared-cross dots).}
  \label{fig:nknext}
\end{figure}

In \fig{fig:nknext}, we show the evolution of the various physical quantities in the late stage of linear perturbations: $n^+_k,\; \overline{\sigma^2}$ and Var$(\sigma)$. The top panels plot the amplitude of $n^+_k$ with various times in both GRV-M Model (left) and GAU-M Model (right). Notice that we plot it against the logarithmic scale of $k$ as opposed to the linear scale seen in the top panels of \fig{fig:nklin}. For all time steps shown there, our analytical values of $k_m$ agree well with the first peak mode $k_1$, at which the amplitudes are most amplified. Notice that the zero-momentum mode does not decay in both cases. After the first peak of the amplitude is well developed, the second peak appears in the spectra, and later the third peak can be barely observed. Roughly speaking, the $n^{th}$ peaks appear around the values which are $k_m$ multiplied by $n$. These higher peaks are suppressed by rescattering processes in which a particle from the first peak transfers some of its momentum to a particle from the zero-momentum modes (AD condensates) \cite{Khlebnikov:1996mc}. Later, all modes of the particle spectra, $n^+_k$, develop quickly, but the first peak is still visible. The middle panels illustrate the evolution of a zero-mode field $\overline{\sigma^2}$ and the variance of the field Var$(\sigma)$ up to the nonlinear time $t=t_{NL}$. As we saw in the top panels, the zero mode does not decay even after the nonlinearity comes in, whilst the variance of the field develops exponentially from $t\sim 140$ in GRV-M Model (left) and from $t\sim 600$ in GAU-M Model (right). This delay of the exponential growth comes from the fact that the other modes do not evolve initially except the mode $k_m$; thus, we can set these times as $t_*$ defined in \eq{nlt}. We fit a function, $\propto exp\bset{2 \dot{S}_{num}(t-t_*)}$, against the exponential evolution for the variations, where $\dot{S}_{num}$ is a numerical value, and we obtained $\dot{S}_{num} \sim 4.45\times 10^{-2}$ in GRV-M Model and $\dot{S}_{num} \sim 6.72\times 10^{-3}$ in GAU-M Model, which match satisfactorily with the analytical ones in \eqs{ext11}{ext12}, where we computed as $\state{\dot{S}_m}\sim 4.20\times 10^{-2}$ in GRV-M Model and $\state{\dot{S}_m}\sim 7.07\times 10^{-3}$ in GAU-M Model. From the middle panels, the nonlinear time is approximately both $t_{NL}\sim 420$ in GRV-M Model and $t_{NL}\sim 2200$ in GAU-M Model, and these values agree well with the analytical estimates in \eq{nlt}, where the analytical values are $t_{NL}\sim 262+140\sim 422$ in GRV-M Model and $t_{NL}\sim 1628+600\sim 2228$. In the bottom panels, we plot the evolution of the amplitude $n^+_k$ for the first peak mode (red-plus dots), second peak mode (green-cross dots) and the analytical most amplified modes (purple squared-cross dots). The numerical values of the exponents for the most amplified modes $k_m$ in blue long-dotted lines, ($\dot{S}_{num} \sim 4.55\times 10^{-2}$ in GRV-M Model and $\dot{S}_{num} \sim 7.11\times 10^{-3}$ in GAU-M Model) match with the analytical ones in \eqs{ext11}{ext12}. The second peaks $k_2$ in black short-dotted lines start to grow at $t\sim 220$ in GRV-M Model and at $t\sim 1300$ in GAU-M Model, and we can set these values as $t_0$ defined in \eq{n+k}. The initial behaviour of the amplitude of second peak of $n^+_k$ seems to be quasi-periodic, which implies that $\state{S}$ for the mode, $k_2$, is pure imaginary, see \eq{scdsig} (c.f. the bottom panels of FIG. 5 in \cite{Rosa:2007dr}). Surprisingly, the growth rates are about twice larger than the values of both $\state{\dot{S}_m}$ and $\dot{S}_{num}$ for Var($\sigma$) and $k_1$. Note that the initial evolution is not adiabatic, so that the growth rates are not strictly exponential as we have seen in the bottom panels of \fig{fig:nklin}. For example, the growth of the first peaks, $k_m$ (or $k_1$), in GAU-M Model is not exponential initially, but it becomes exponential as the growth of the second peak mode $k_2$.

\vspace*{10pt}
\paragraph*{\underline{\bf Bubbles pinched out of filaments:}}

\begin{figure}[!ht]
  \begin{center}
	\includegraphics[scale=0.4]{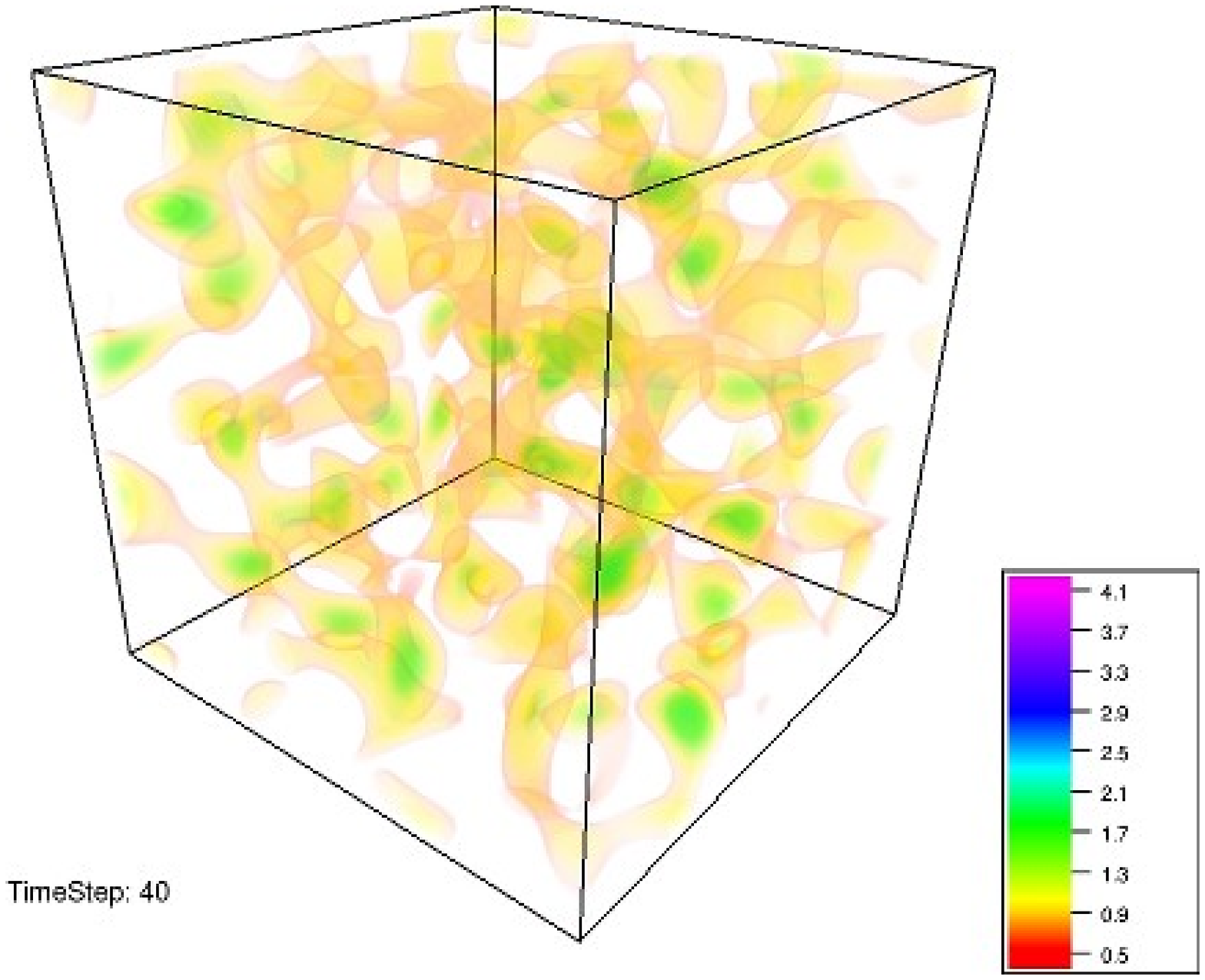}
	\includegraphics[scale=0.4]{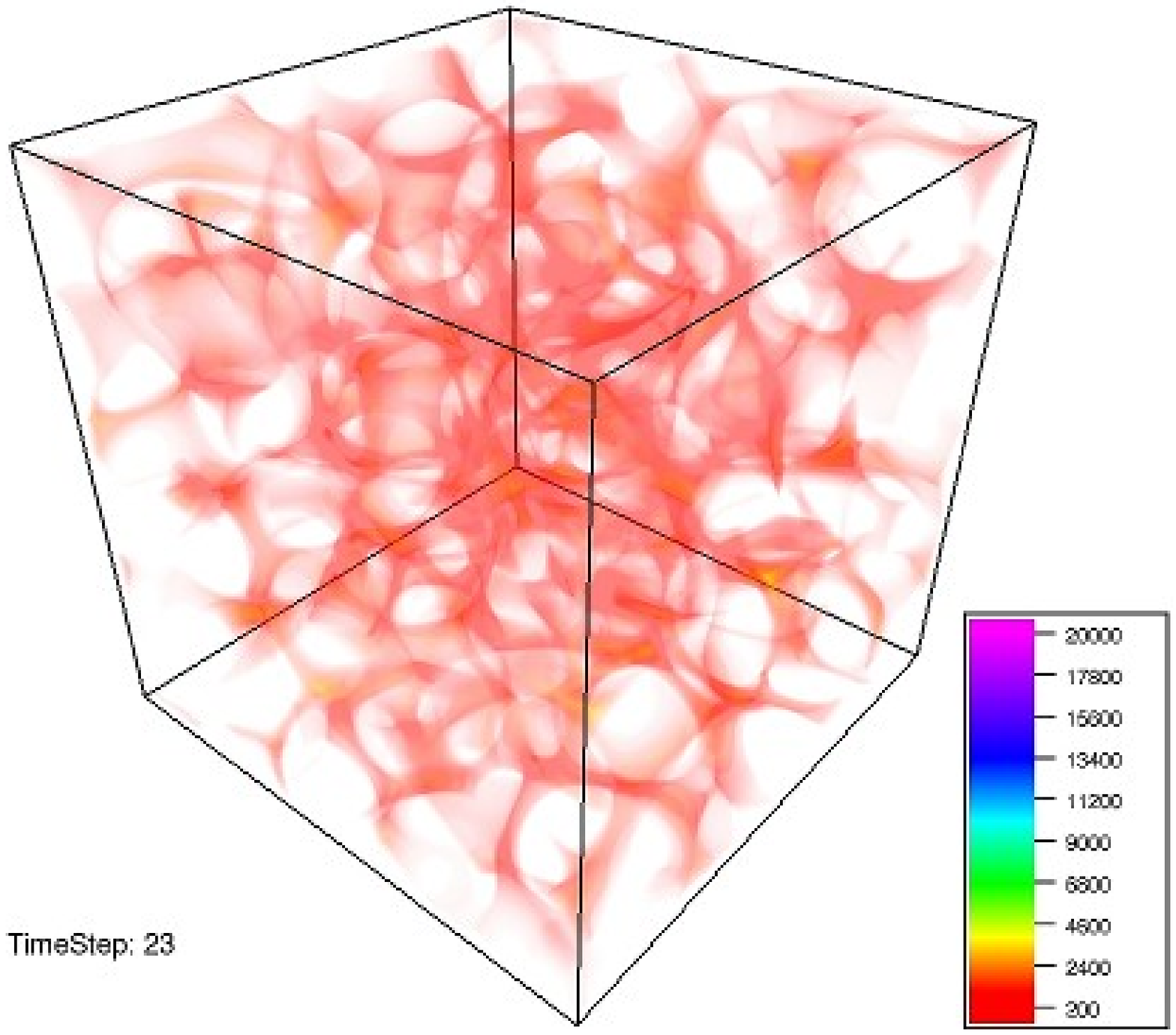}\\
	\includegraphics[scale=0.4]{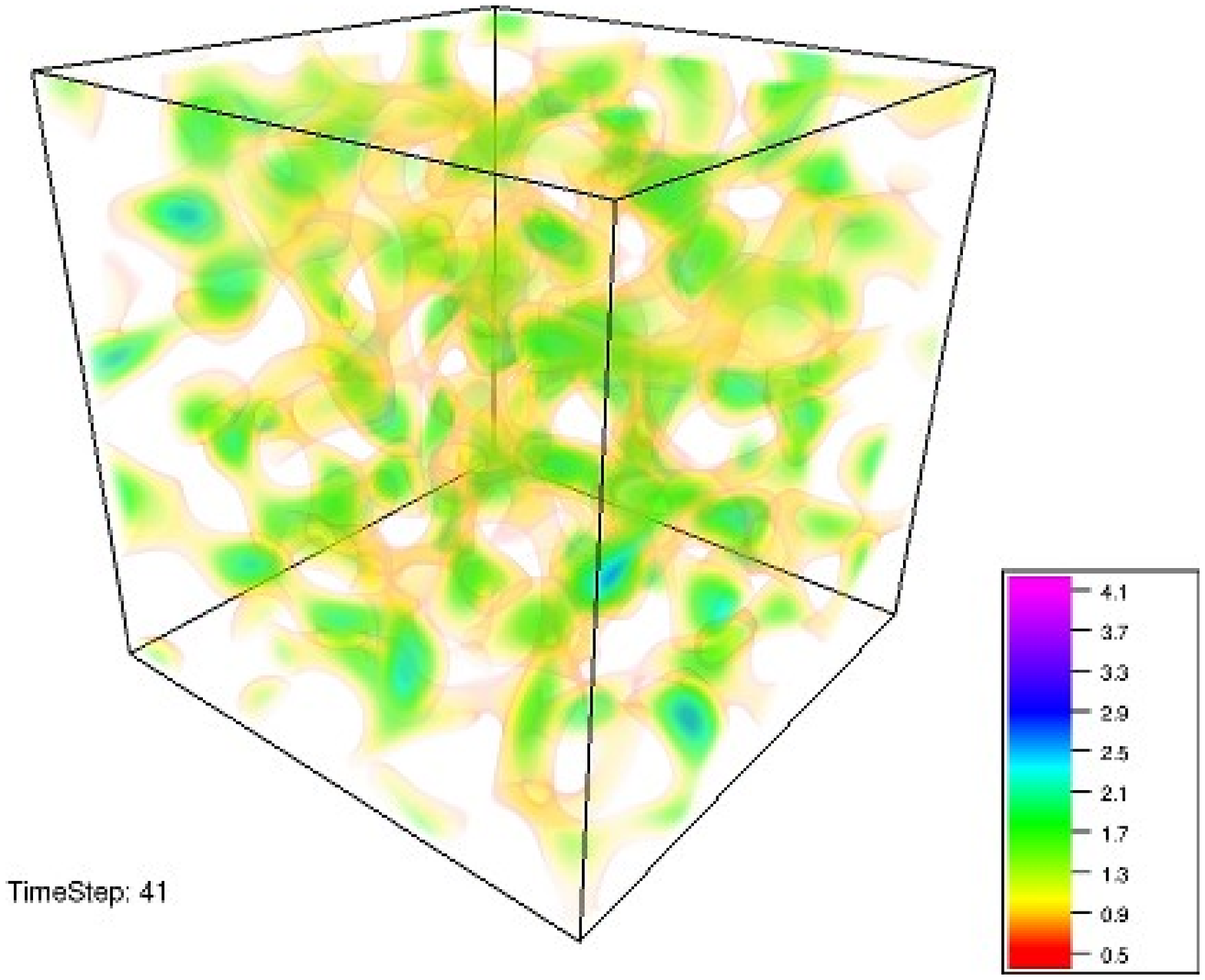}
	\includegraphics[scale=0.4]{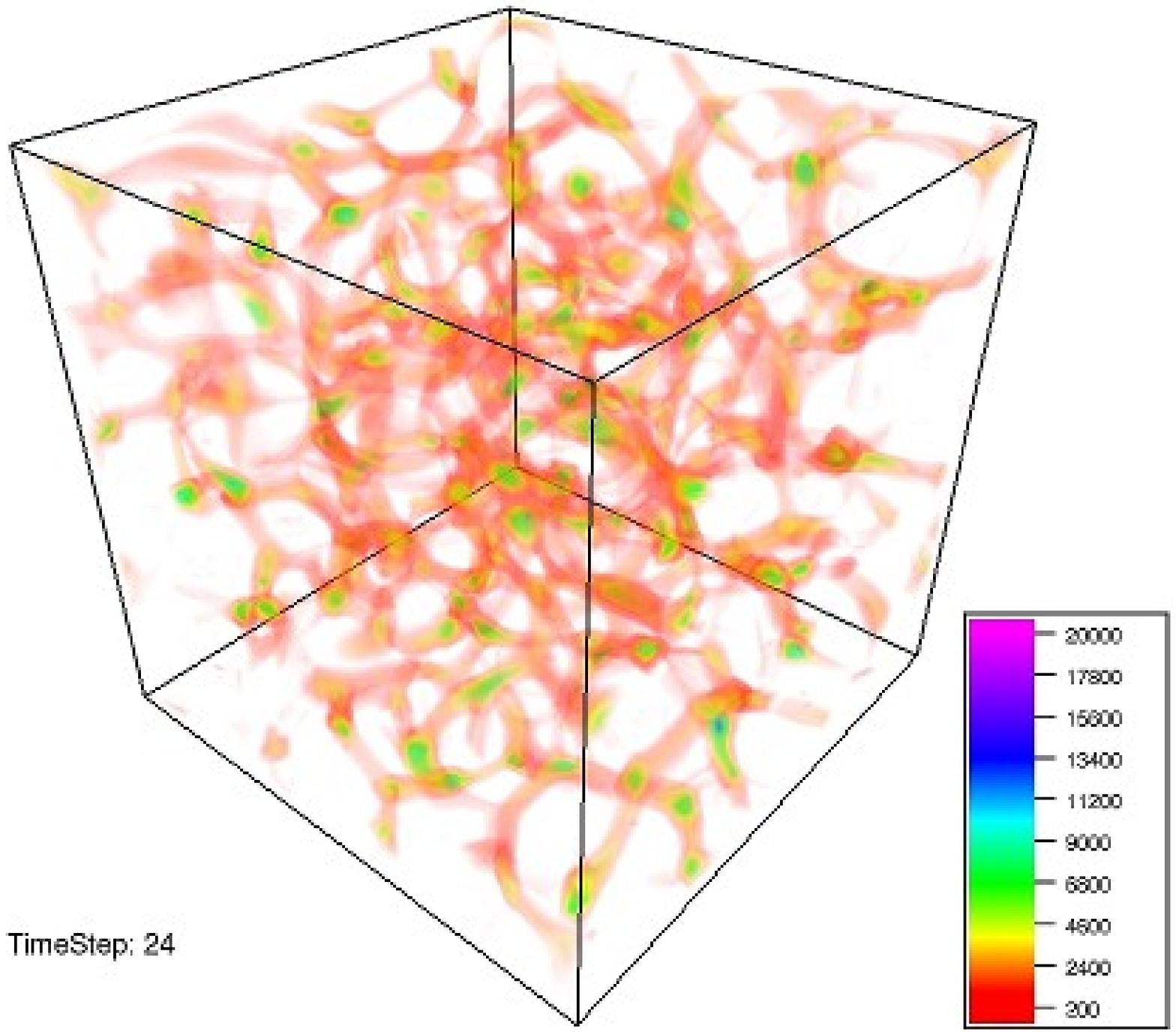}
  \end{center}
  \caption{ \textbf{(Color online)} In the top and bottom panels, we show the snapshots of the positive charge density $n^+(\mathbf{x})$ for GRV-M Model (left panels) and GAU-M Model (right panels) around $t\sim t_{NL}$, where 'Timestep' in the panels denotes the actual time divided by $10$ in GRV-M Model and the actual time divided by $10^2$ in GAU-M Model, and the colour bars illustrate the values of the positive charge density. After the nonlinearity is fully developed, many bubbles form, which are pinched out of ``highly'' charged filaments.}
  \label{fig:qb}
\end{figure}

In \fig{fig:qb}, we show the snapshots of the positive charge density $n^+(\mathbf{x})$ for GRV-M Model (left panels) and GAU-M Model (right panels) around $t\sim t_{NL}$, where 'Timestep' in the panels denotes the actual time divided by $10$ in GRV-M Model and the actual time divided by $10^2$ in GAU-M Model. The colour bars illustrate the values of the positive charge density. We can see long-wavelength objects (sometimes called 'filaments') in both cases, and the charge in some regions is compactified into spheres, see bottom panels. These filaments and bubbles correspond to the nonlinear solutions, which may be nontopological strings \cite{Copeland:1988ra} and the excited states of $Q$-balls, respectively. The radii of these bubbles are of the same order as the wave-length which corresponds to the most amplified modes, $k_m$. As we will see in the next subsection, these bubbles grow by colliding and merging each other. Note that this bubble creation is nothing to do with bubble nucleation in first-order phase transition as opposed to the case in \cite{Lee:1994qb}, in which case the AD condensate is classically stable against special perturbation, but not quantum mechanically.

\subsubsection{Nonlinear evolution}

\paragraph*{\underline{\bf Bubble collisions and mergers:}}

\begin{figure}[!ht]
  \begin{center}
	\includegraphics[scale=0.26]{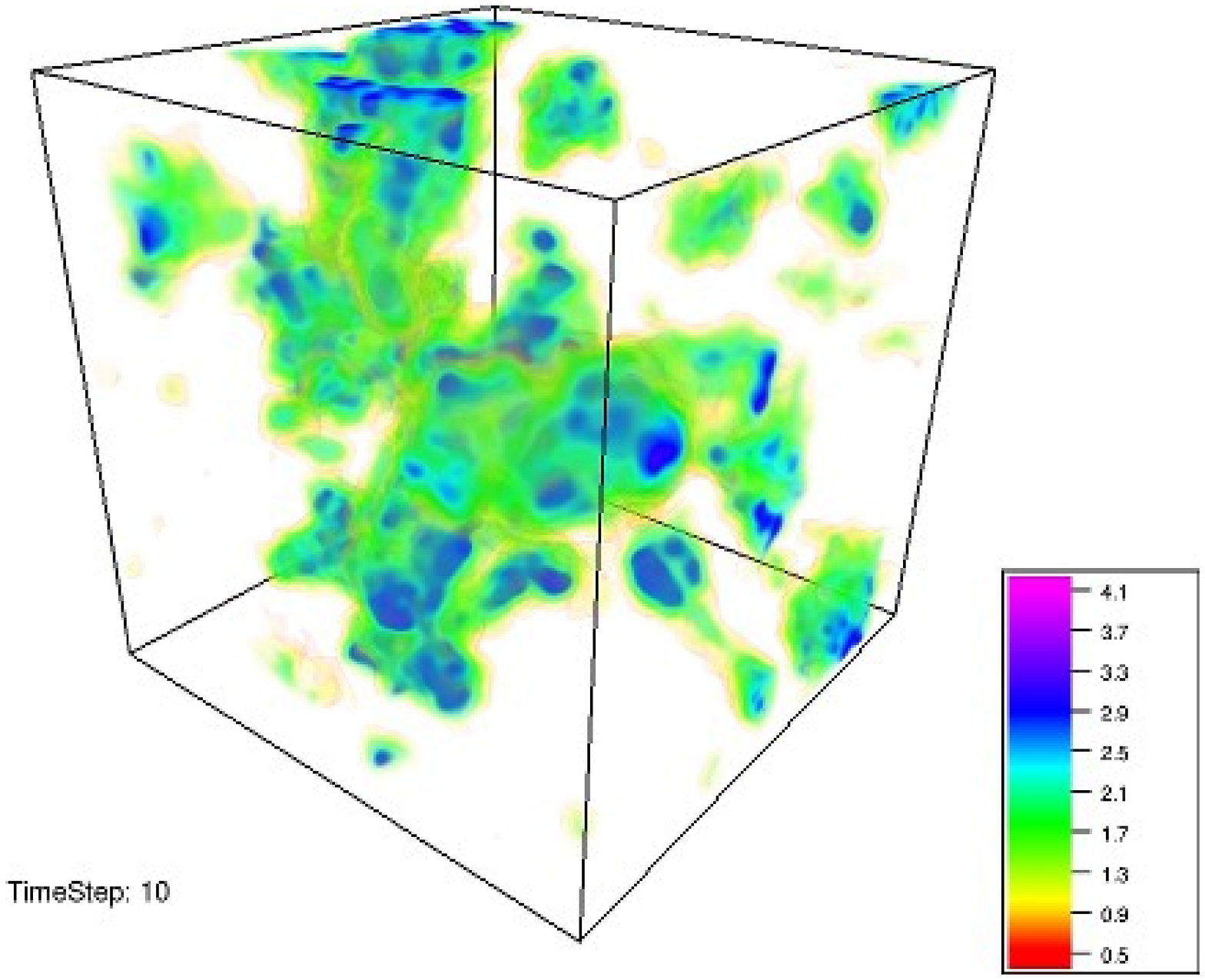}
	\includegraphics[scale=0.26]{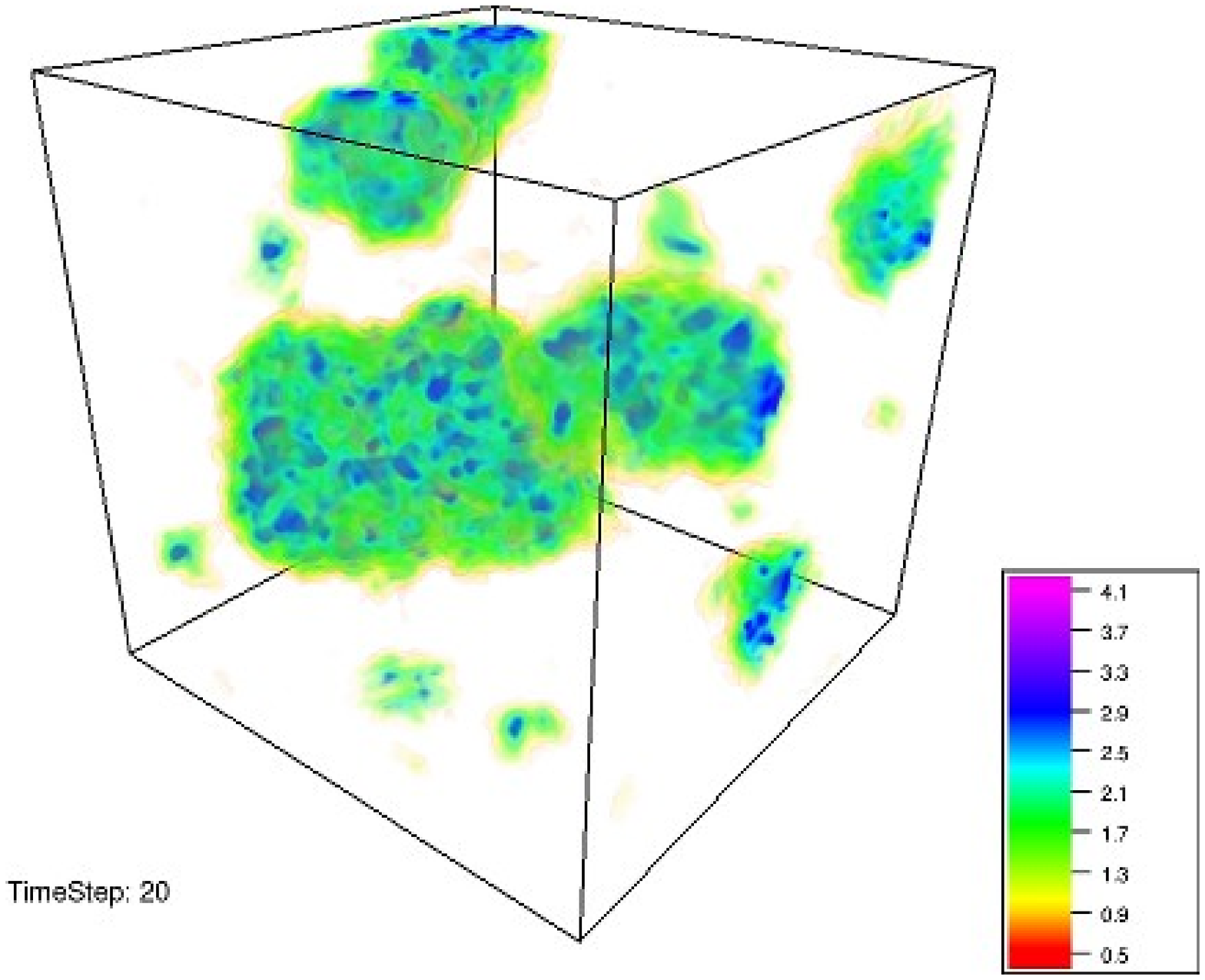}
	\includegraphics[scale=0.26]{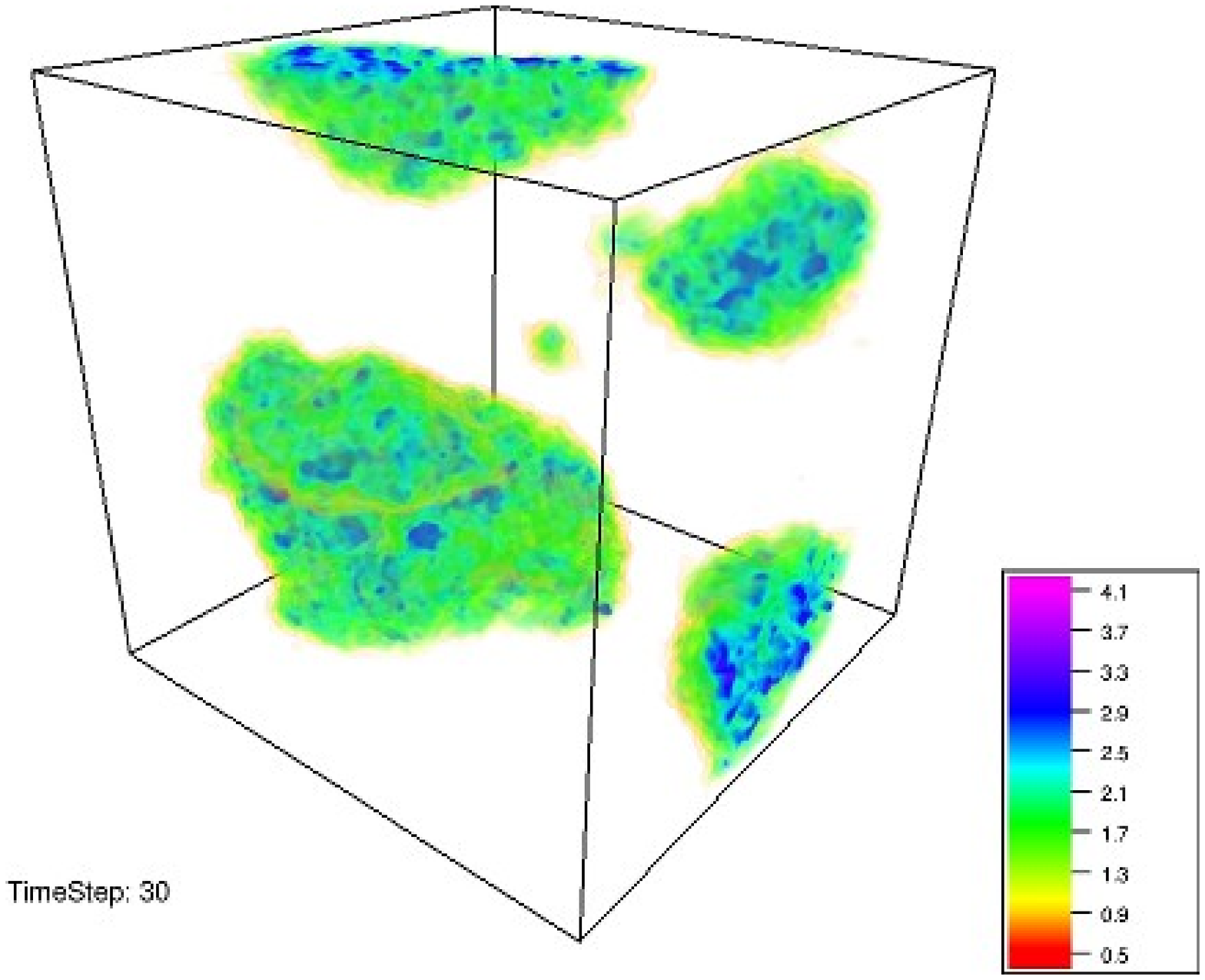}\\
	\includegraphics[scale=0.26]{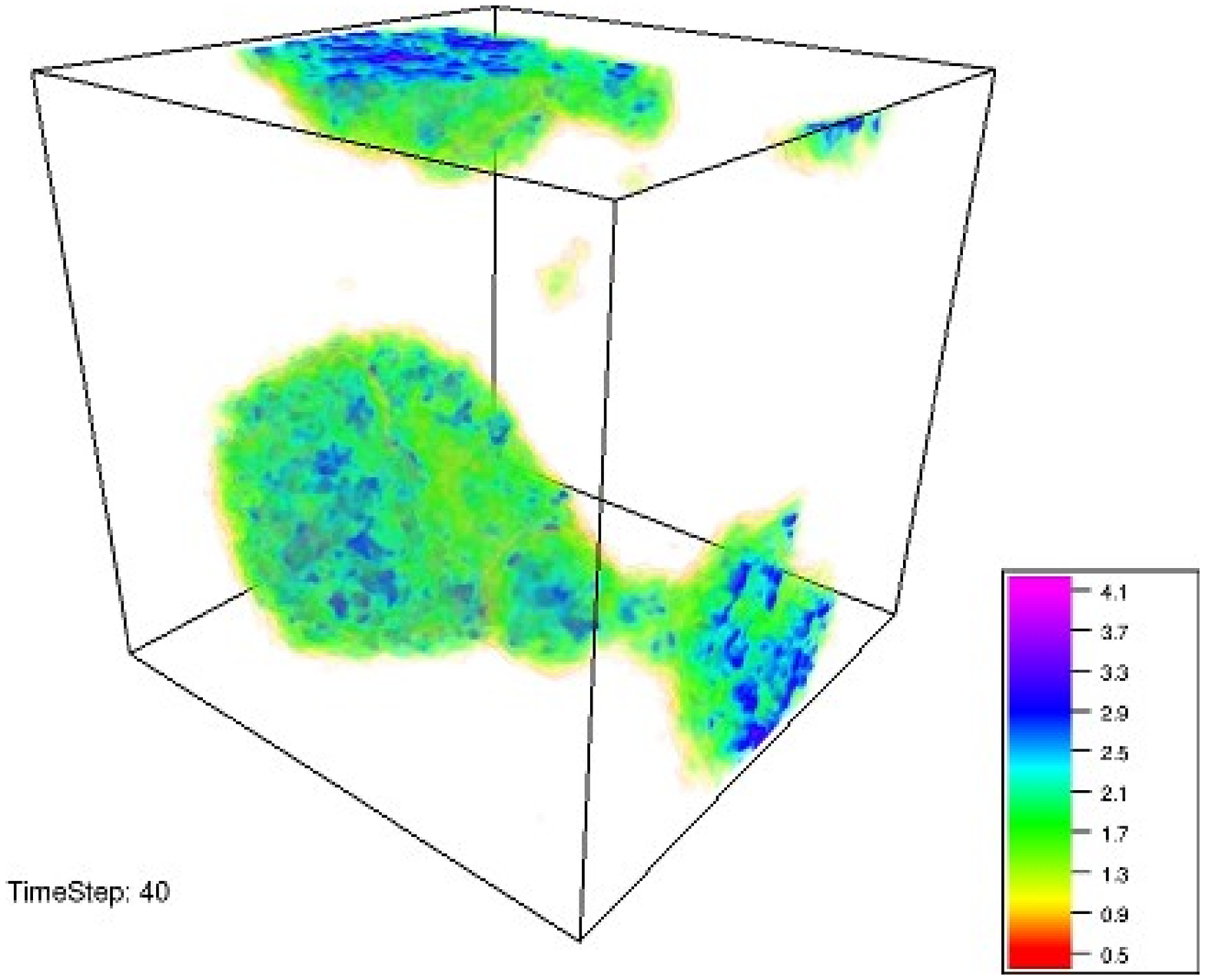}
	\includegraphics[scale=0.26]{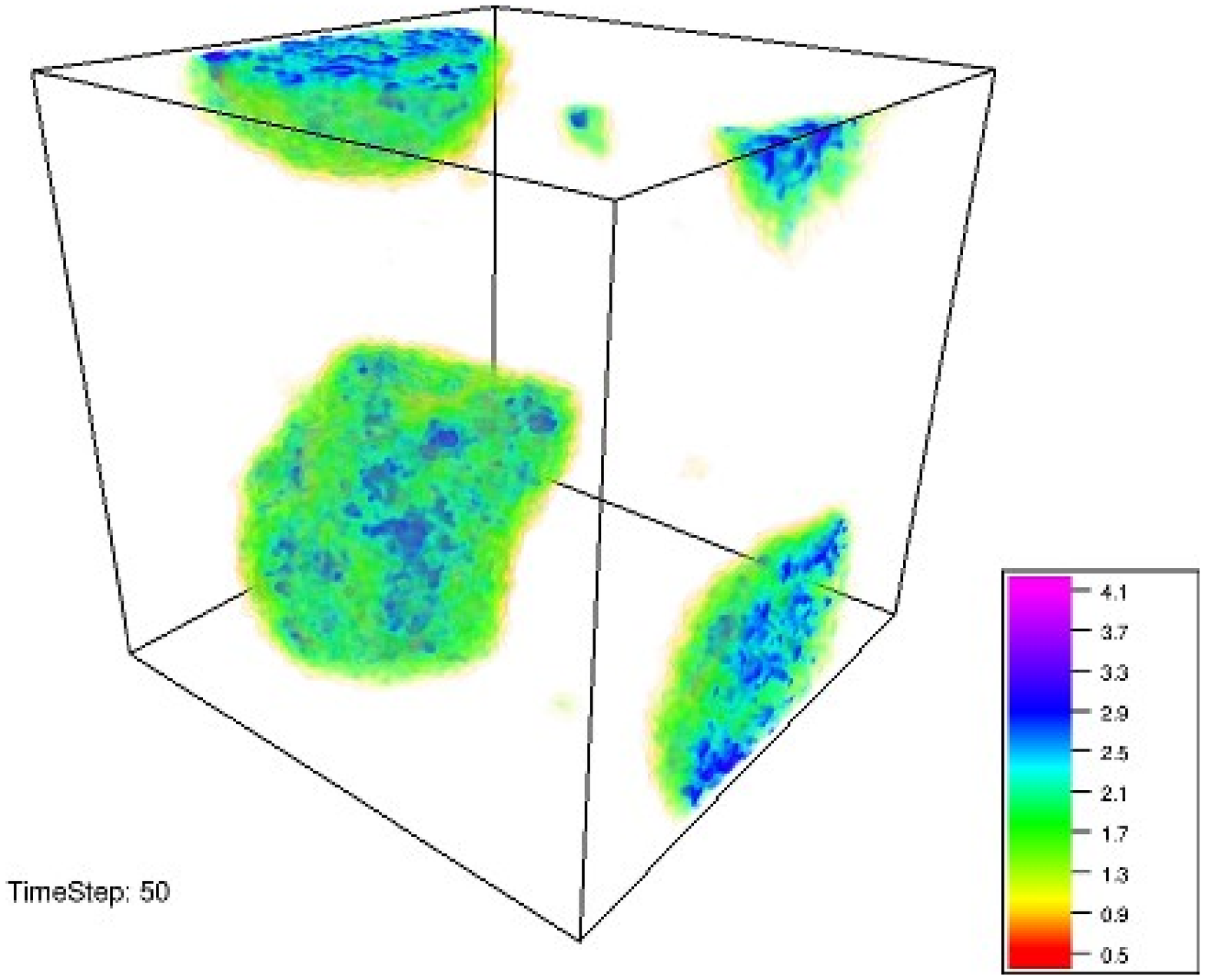}
	\includegraphics[scale=0.26]{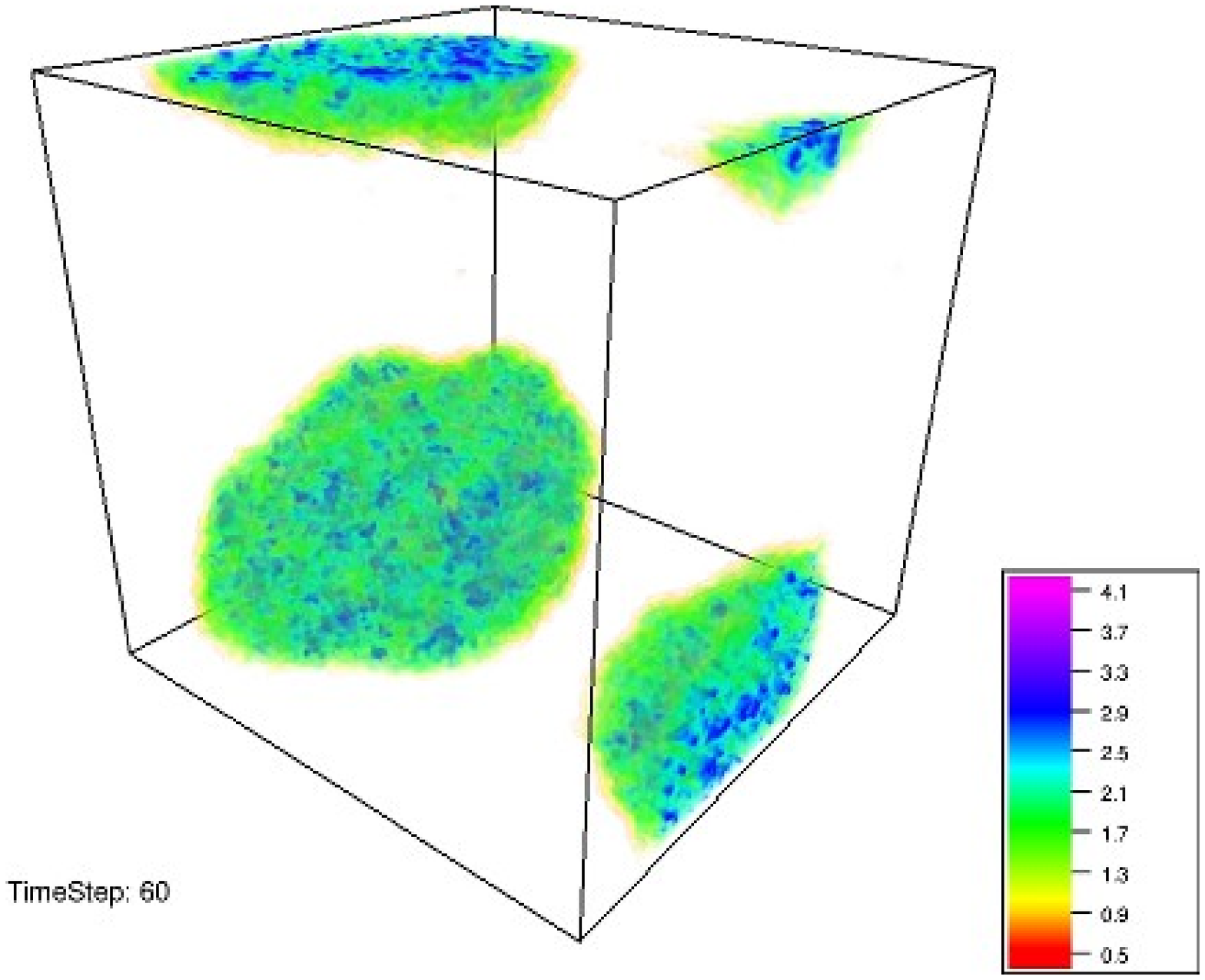}
  \end{center}
  \caption{ \textbf{(Color online)} We show the snapshots of the positive charge density for GRV-M Model in different time steps ($t=1000,\ 2000,\ 3000,\ 4000,\ 5000$ and $6000$), where 'Timestep' in the figure denotes the actual simulation time divided by $10^2$ and the colour bars illustrate the values of the positive charge density. A few created lumps collide and merge into a large cluster in the end.}
  \label{fig:qbformgrv}
\end{figure}

In \fig{fig:qbformgrv}, we show the snapshots of the positive charge density for GRV-M Model in different time steps up to $t=6000$, where 'Timestep' in the figure denotes the actual simulation time divided by $10^2$ and the colour bars illustrate the values of the positive charge density. After the system goes into a nonlinear regime, we can see a few lumps in the first few panels of the snapshots, and those lumps merge into larger lumpy objects. Finally, we can see a large cluster which consists of a complicated inner structure, see the last snapshot. Recall that we are using the periodic boundary condition.

\vspace*{5pt}

\begin{figure}[!ht]
  \begin{center}
	\includegraphics[scale=0.26]{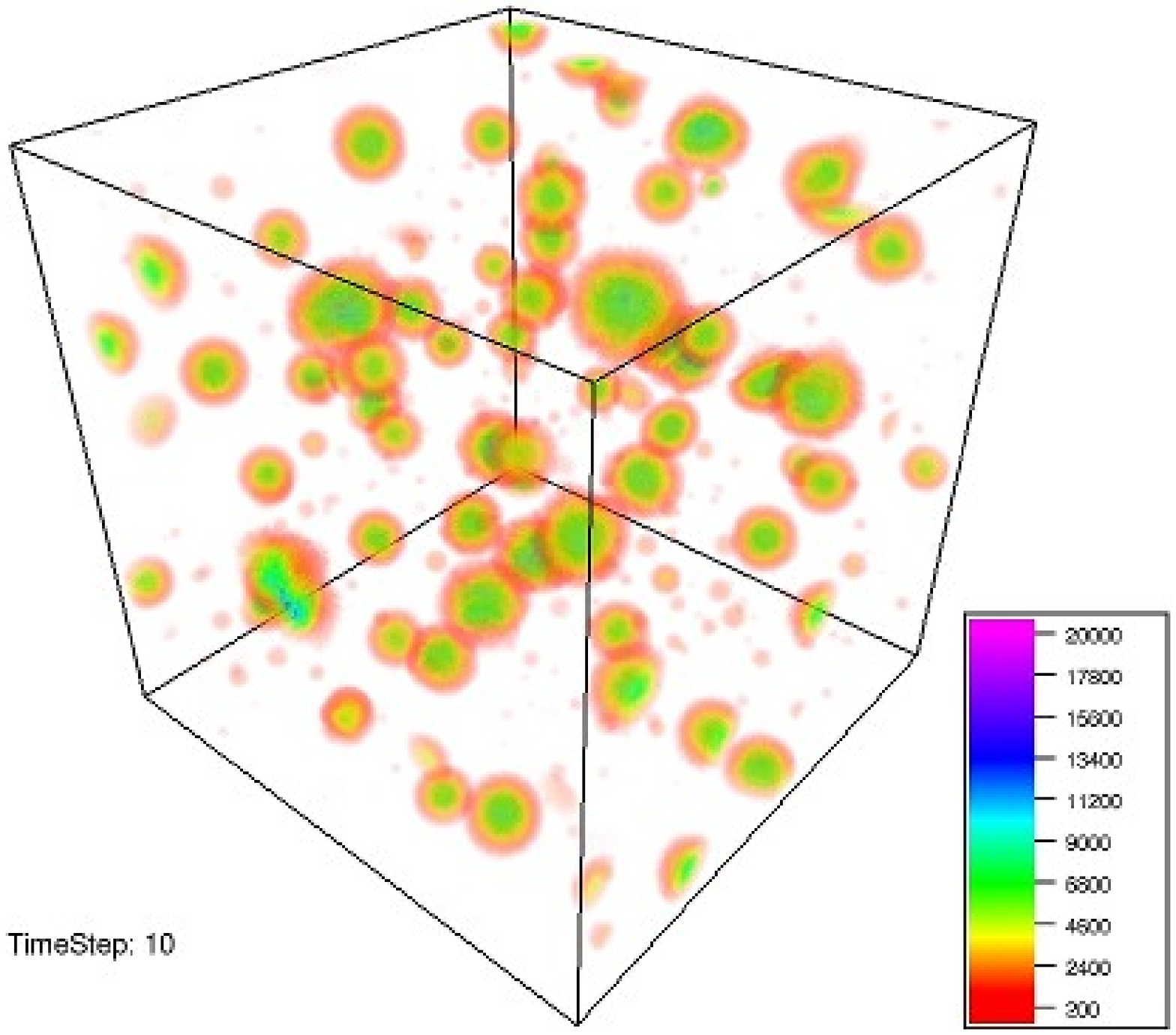}
	\includegraphics[scale=0.26]{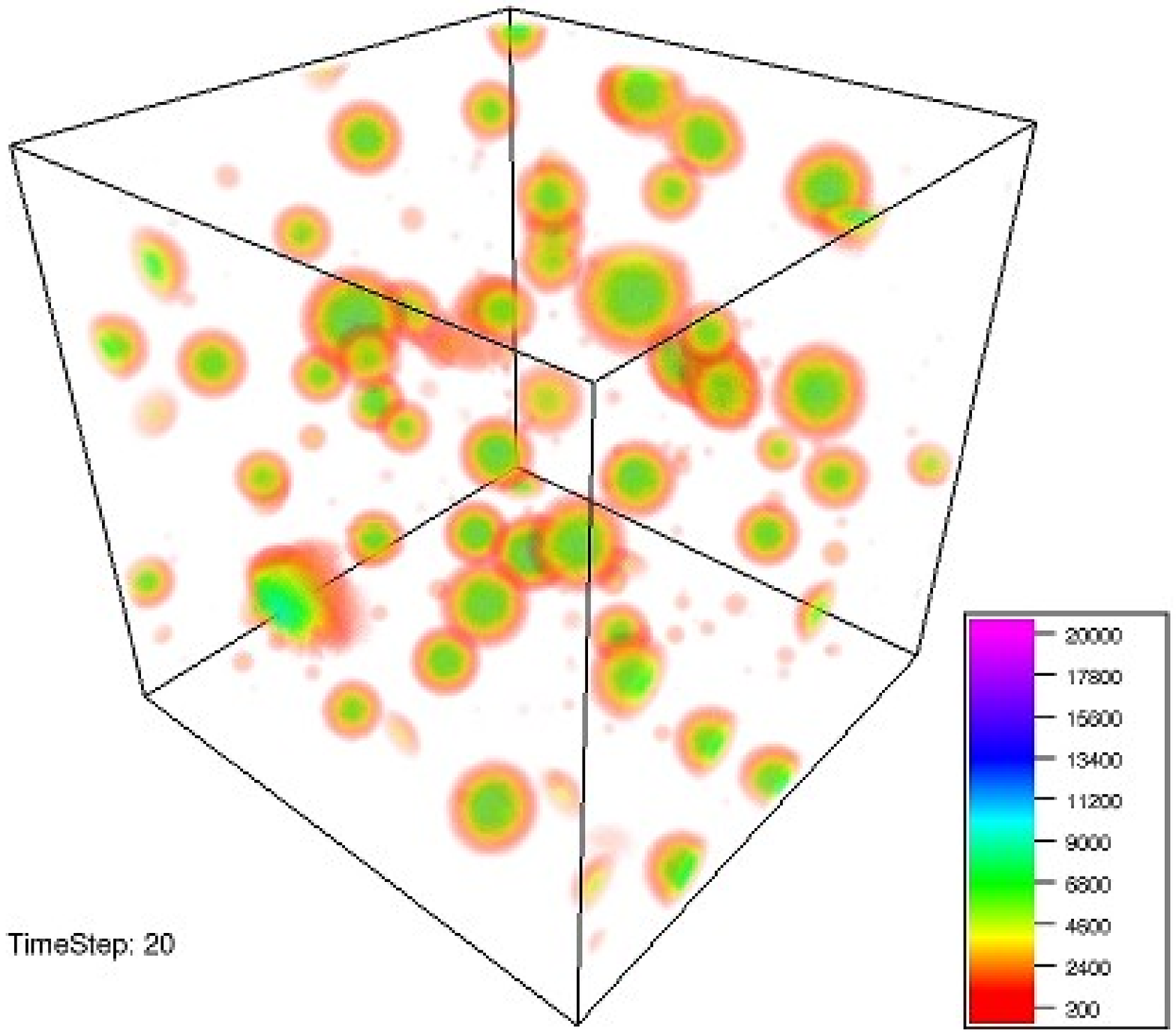}
	\includegraphics[scale=0.26]{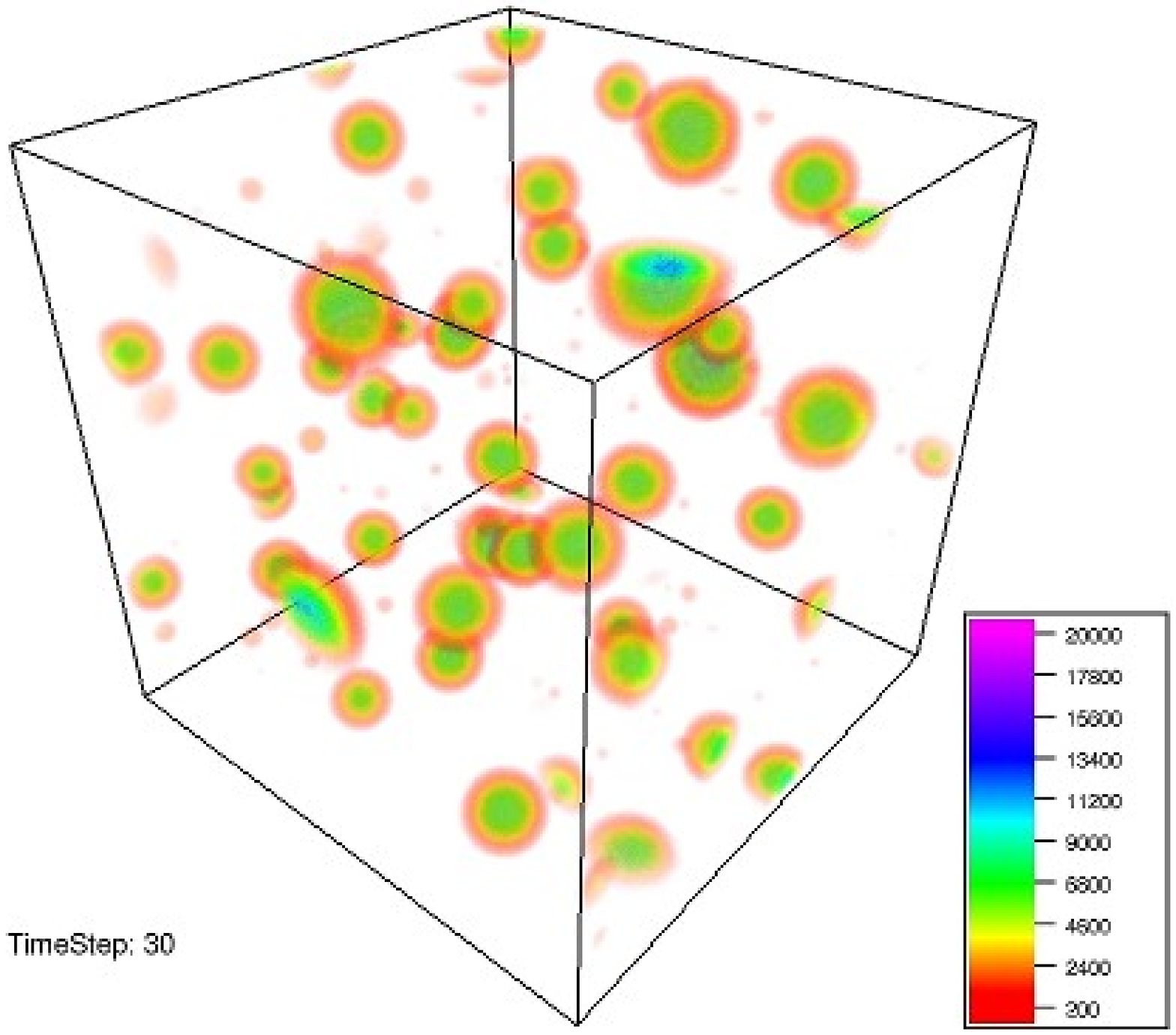}\\
	\includegraphics[scale=0.26]{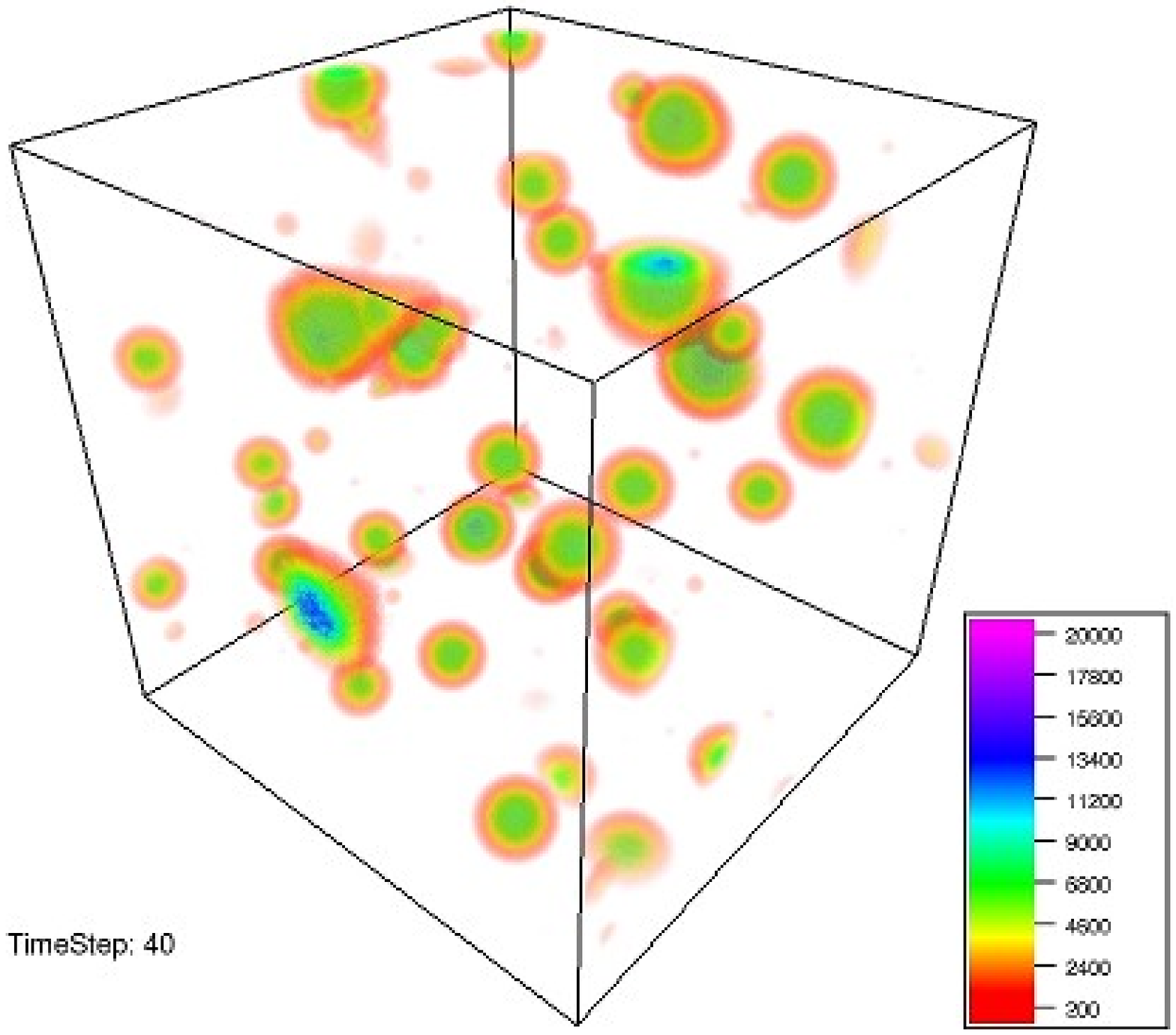}
	\includegraphics[scale=0.26]{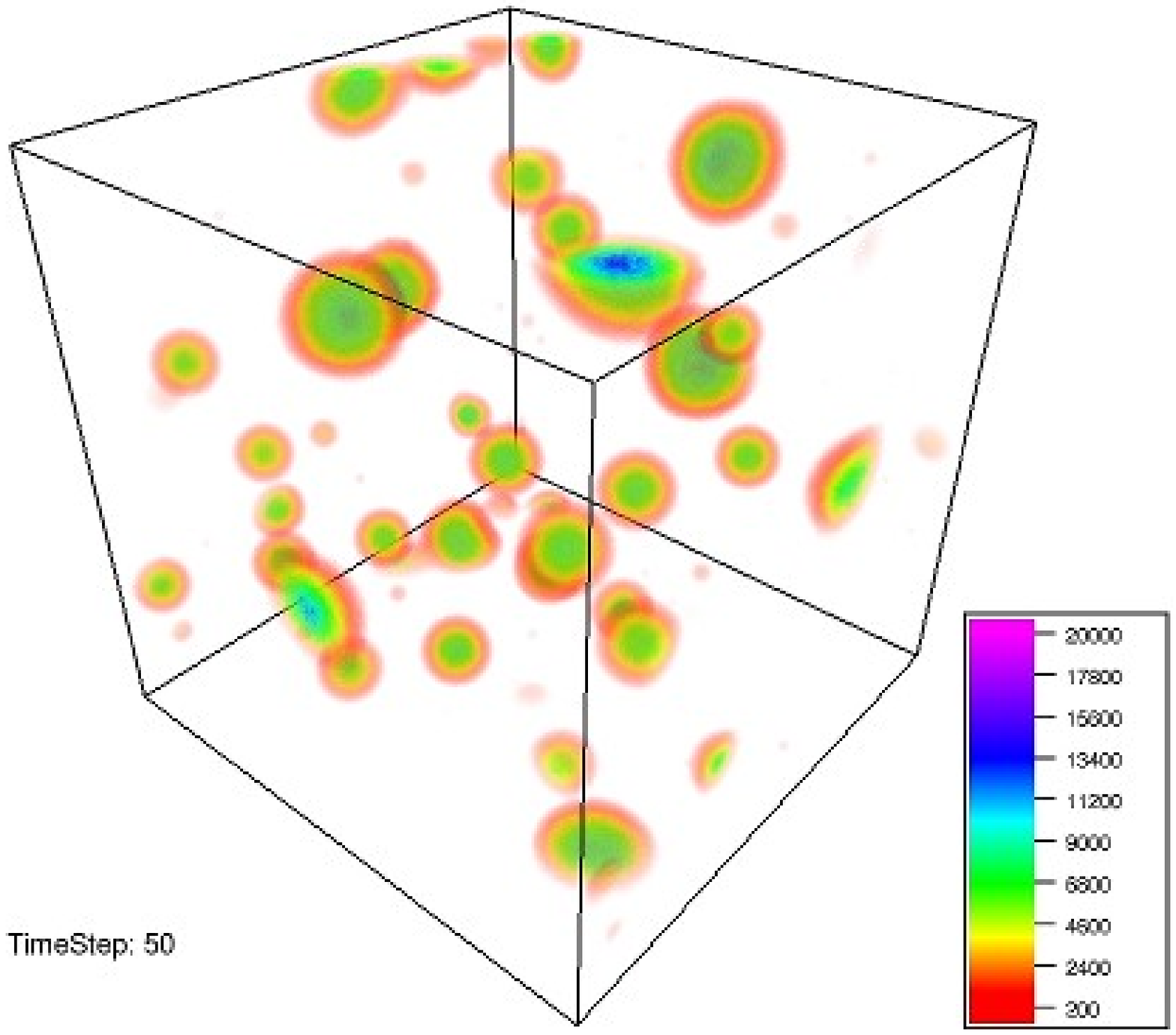}
	\includegraphics[scale=0.26]{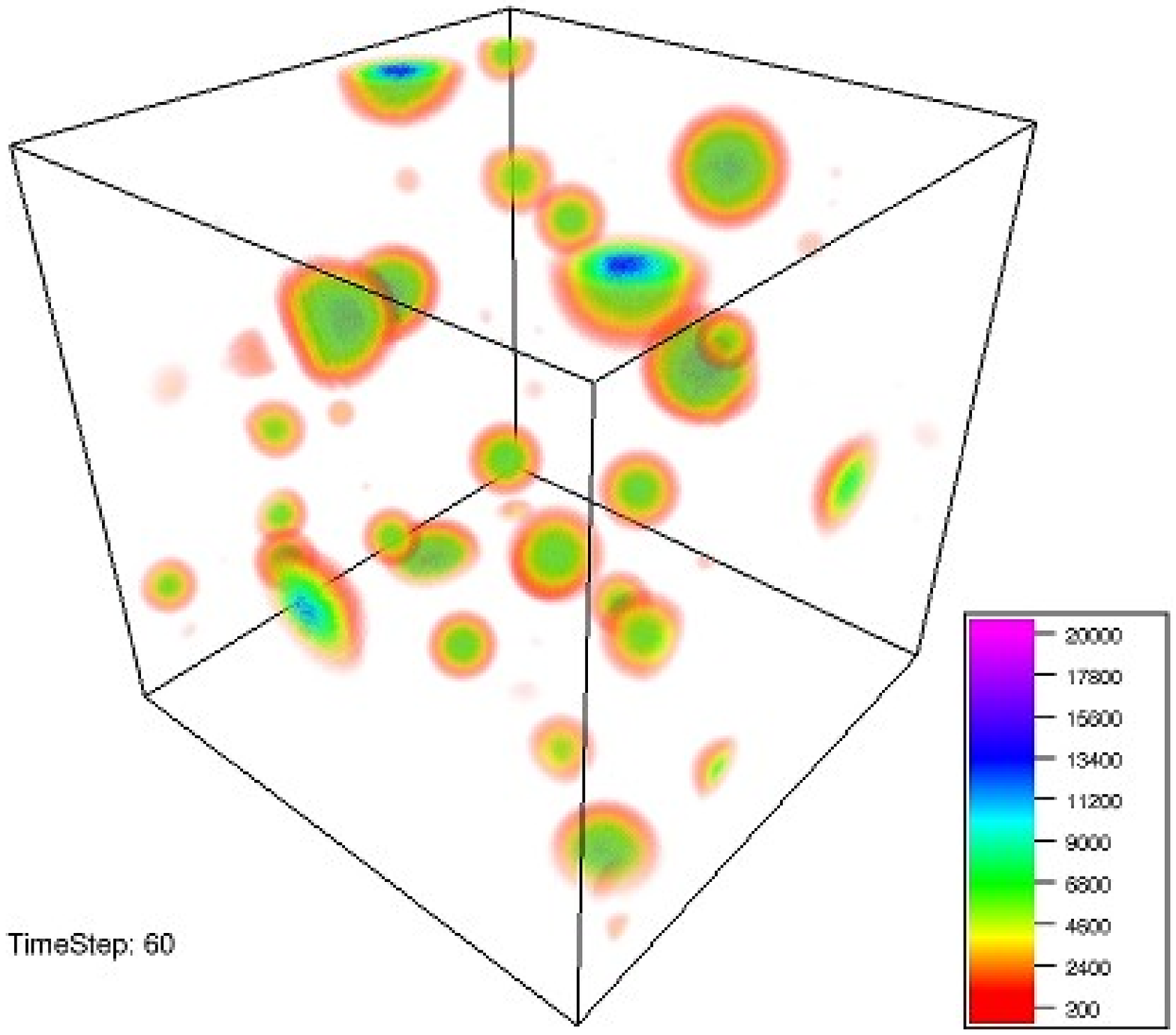}
  \end{center}
  \caption{ \textbf{(Color online)} We illustrate the detailed evolution of the positive charge density for GAU-M Model in different time steps ($t=10000,\ 20000,\ 30000,\ 40000,\ 50000$ and $60000$), where 'Timestep' in the figure denotes the actual simulation time divided by $10^3$ and the colour bars illustrate the values of the positive charge density. There are smaller number of bubbles left in the end.}
  \label{fig:qbformgau}
\end{figure}

\fig{fig:qbformgau} shows the detailed evolution of the positive charge density for GAU-M Model in different time steps up to $t=60000$, where 'Timestep' in the figure denotes the actual simulation time divided by $10^3$ and the colour bars illustrate the values of the positive charge density. A large number of small bubbles can be observed, and nearby bubbles collide and merge into larger bubbles. In the final panel, there are smaller number of bubbles left (compare to the first panel). We believe that this time arrow is followed by the fact that the total energy of large bubbles is smaller than the total energy of smaller bubbles, c.f. fission stability of $Q$-balls in \cite{Tsumagari:2008bv}. These large bubbles are able to carry a large amount of charge inside of them as we saw in the left panel of FIG. 9 in \cite{Copeland:2009as} in the ``thin-wall'' $Q$-ball limit. 

\vspace*{5pt}

The differences of the evolution between GRV-M and GAU-M models come from a number of facts, e.g. different initial conditions, stability conditions and momentum fluxes due to asymptotic profiles at a large distance from the cores, see \cite{Copeland:2009as}.

\vspace*{10pt}

\paragraph*{\underline{\bf Distributions of the negative charge density:}}

\begin{figure}[!ht]
  \begin{center}
	\includegraphics[scale=0.4]{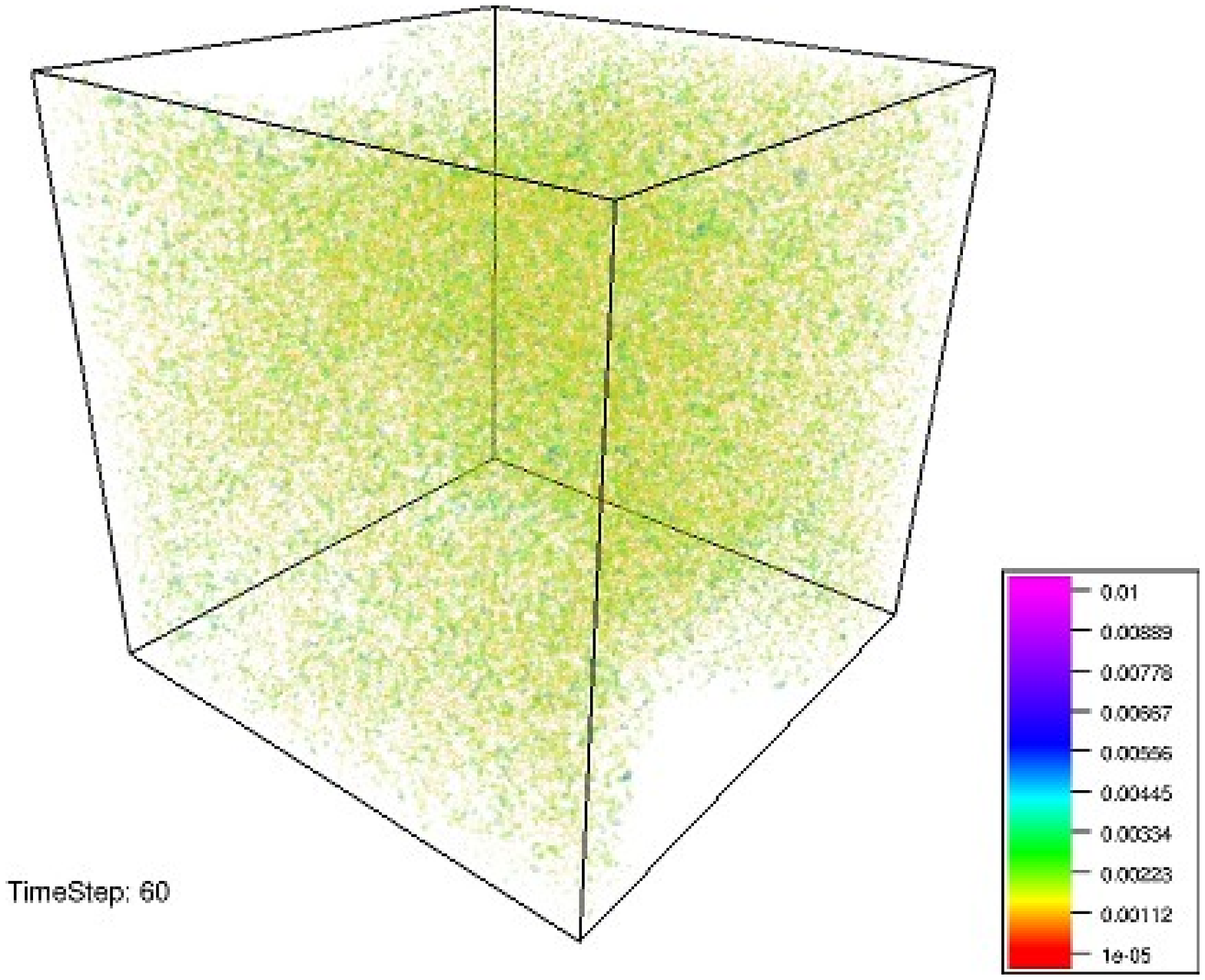}
	\includegraphics[scale=0.4]{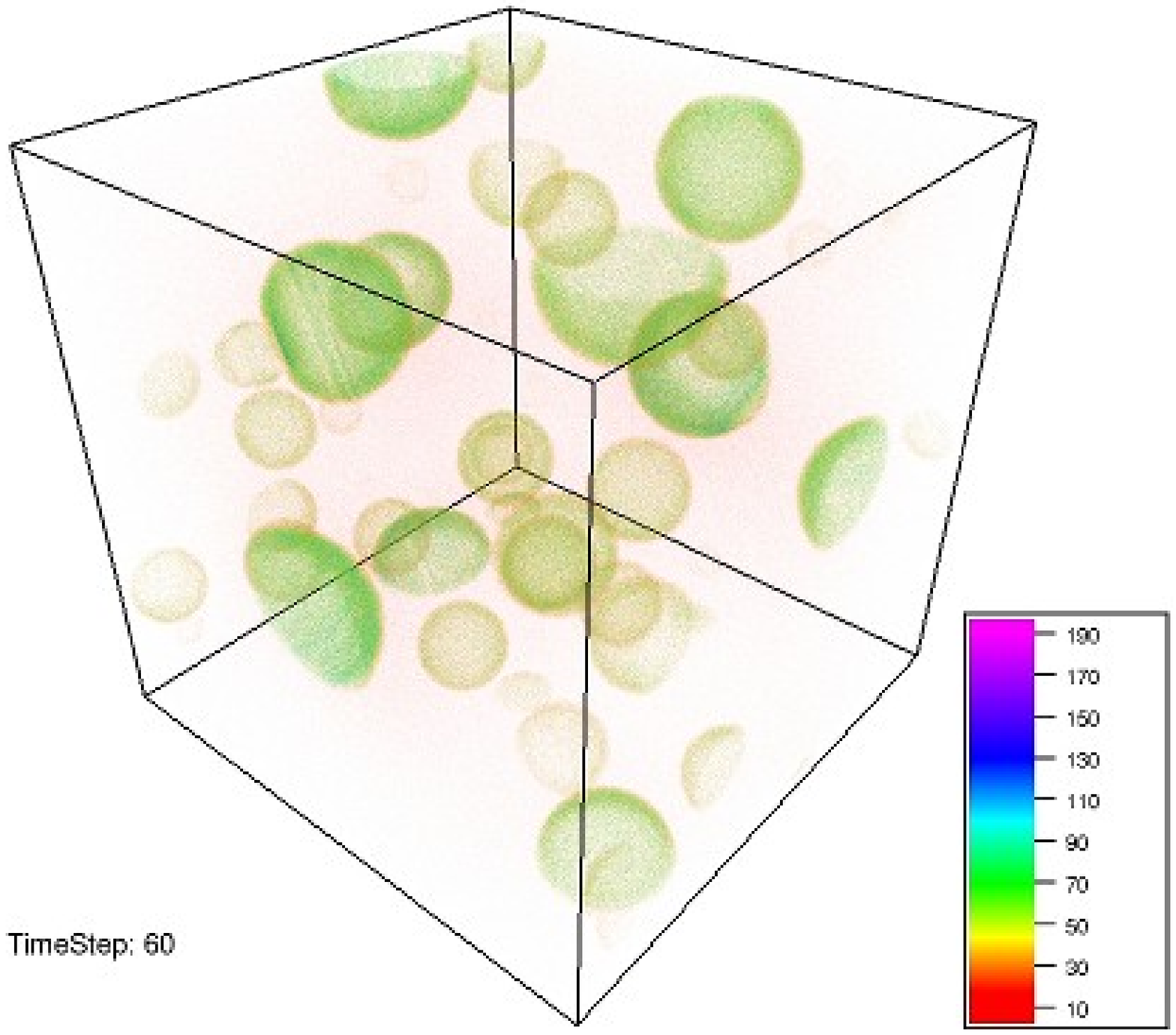}
  \end{center}
  \caption{ \textbf{(Color online)} We present the snapshots of the negative charge density for GRV-M Model (left panel) at $t=6.0\times 10^{3}$ and GAU-M Model (right panel) at $t=6.0\times 10^{4}$, where the colour bars illustrate the values of the negative charge density. The negative charge density for GRV-M Model is surrounded by the large positive charged cluster; however, the distribution spreads out over the lattice space, whereas the negative charge density for GAU-M Model is concentrated around the positive charged lumps (compare them to the last panels of \figs{fig:qbformgrv}{fig:qbformgau}).}
  \label{fig:qbnm}
\end{figure}

We show the snapshots of the negative charge density for GRV-M Model (left panel) at $t=6000$ and GAU-M Model (right panel) at $t=1.0\times 10^{5}$ in \fig{fig:qbnm}, where the colour bars illustrate the values of the negative charge density. These times correspond to the same times as in the final snapshots of \figs{fig:qbformgrv}{fig:qbformgau}. The values of charge density in both models are much smaller than the values of positive charge density in \figs{fig:qbformgrv}{fig:qbformgau}. This implies that we are observing the plots of thermal plasma rather than the charged (nonlinear) lumps. Their distributions are quite different each other. The negative charge density for GRV-M Model is surrounded by the large positive charged cluster seen in the last panel of \fig{fig:qbformgrv}, and it is distributed all over the lattice; whereas, for GAU-M Model the distributions of the negative charged plasma are highly concentrated only around the surface of the lumps (compare the last panel of \fig{fig:qbformgau}).

\vspace*{10pt}

\paragraph*{\underline{\bf Driven turbulence:}}

\begin{figure}[!ht]
  \begin{center}
	\includegraphics[angle=-90, scale=0.31]{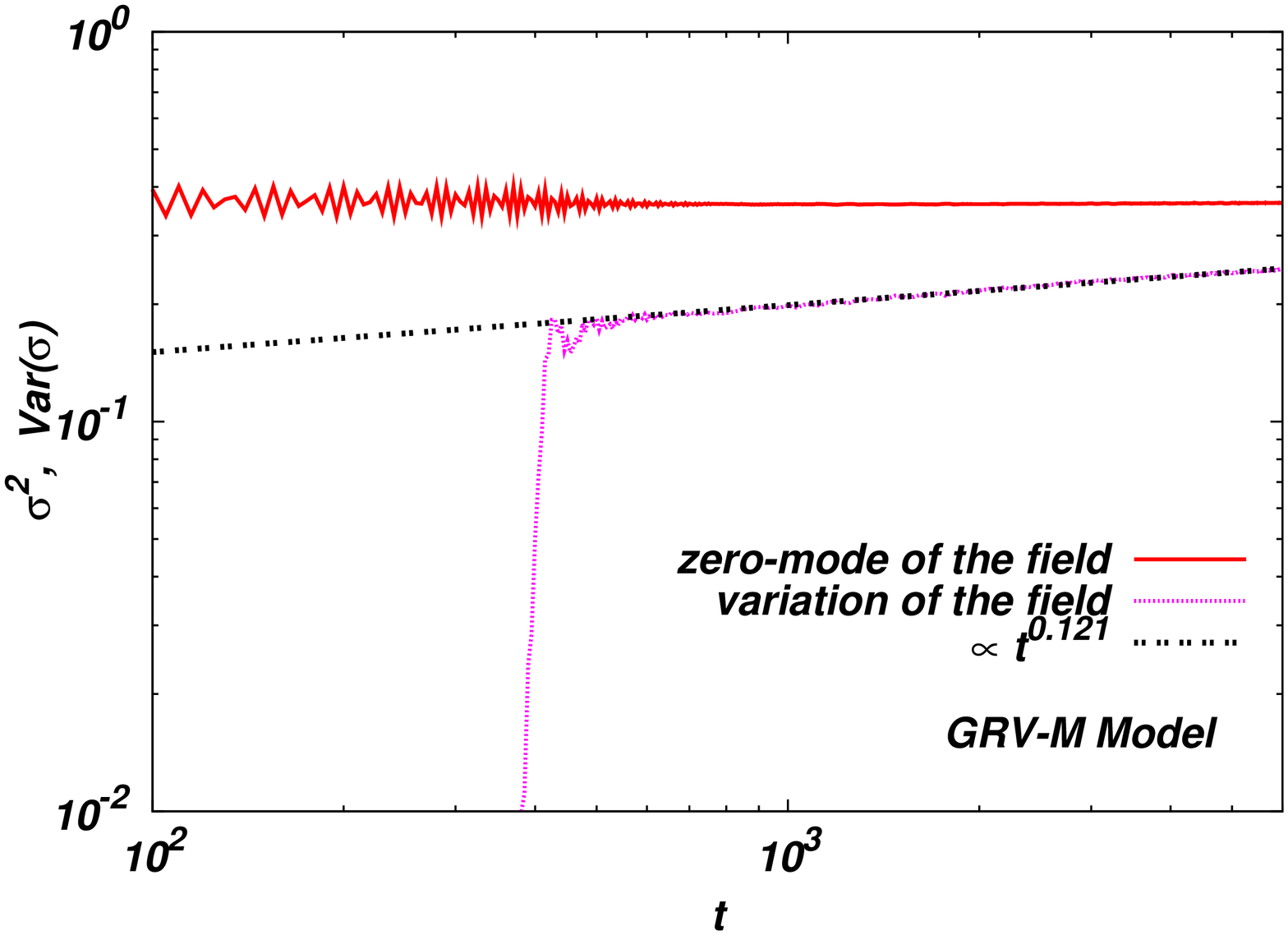}
	\includegraphics[angle=-90, scale=0.31]{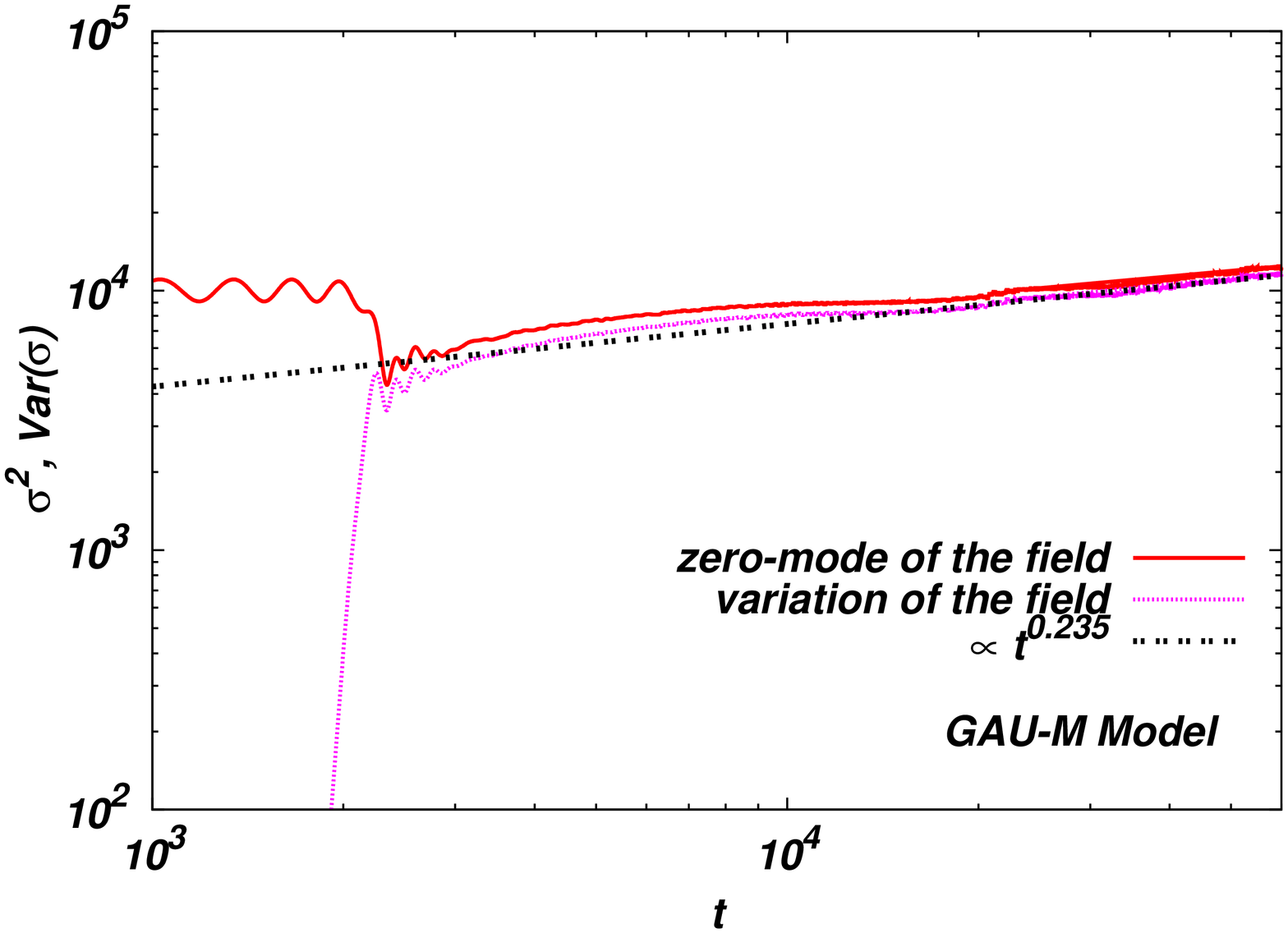}\\
	\includegraphics[angle=-90, scale=0.31]{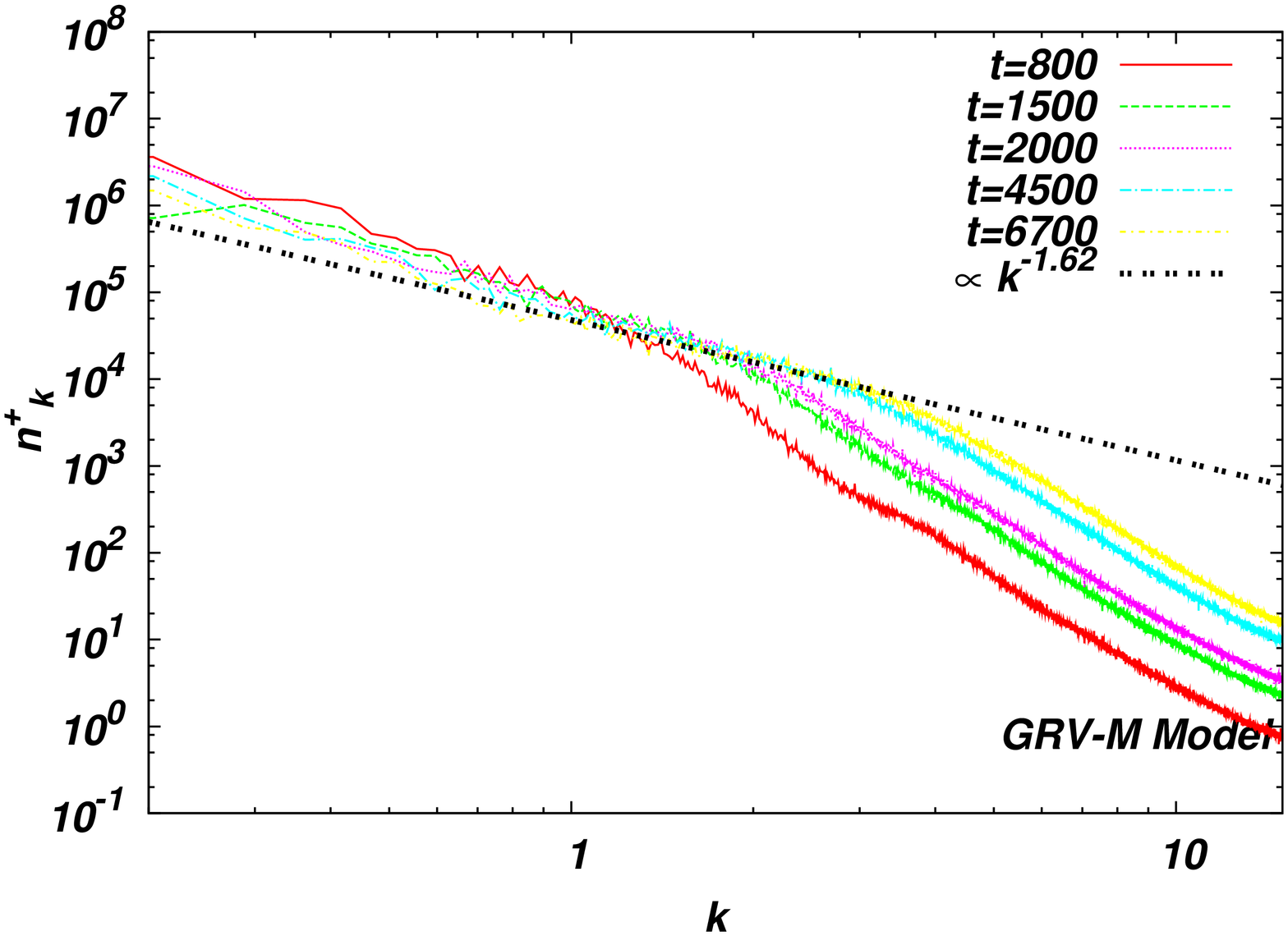}
	\includegraphics[angle=-90, scale=0.31]{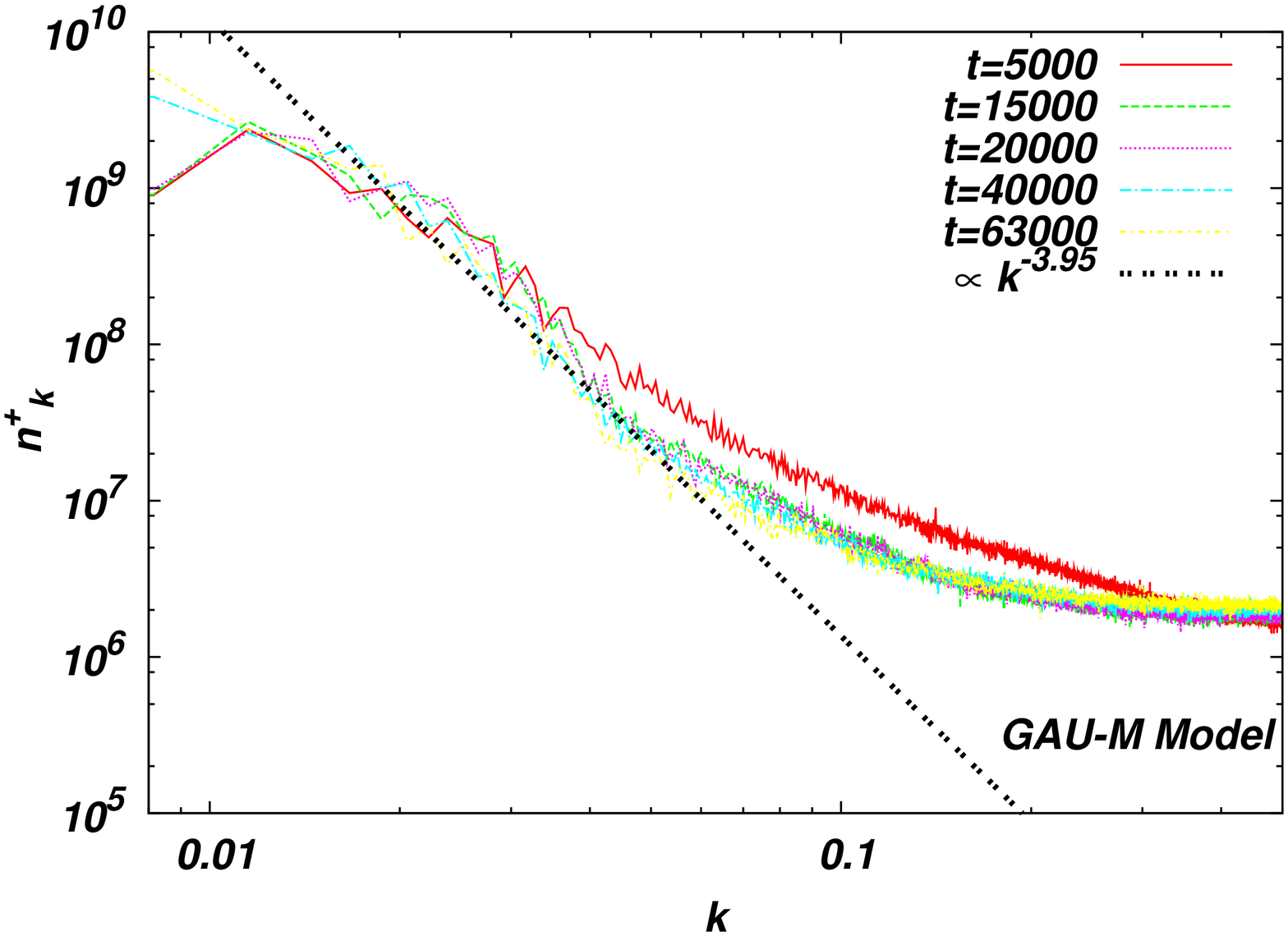}\\
	\includegraphics[angle=-90, scale=0.31]{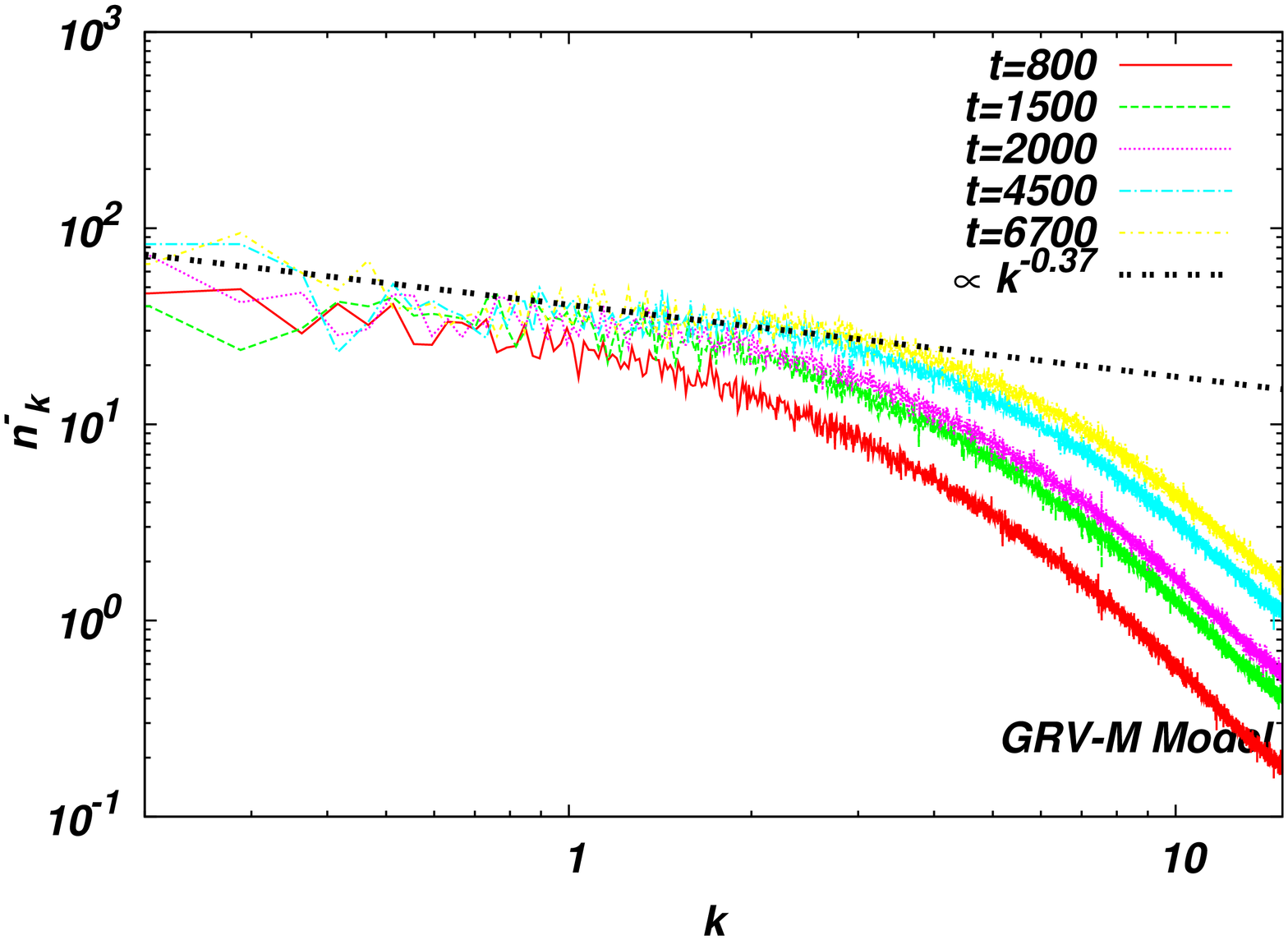}
	\includegraphics[angle=-90, scale=0.31]{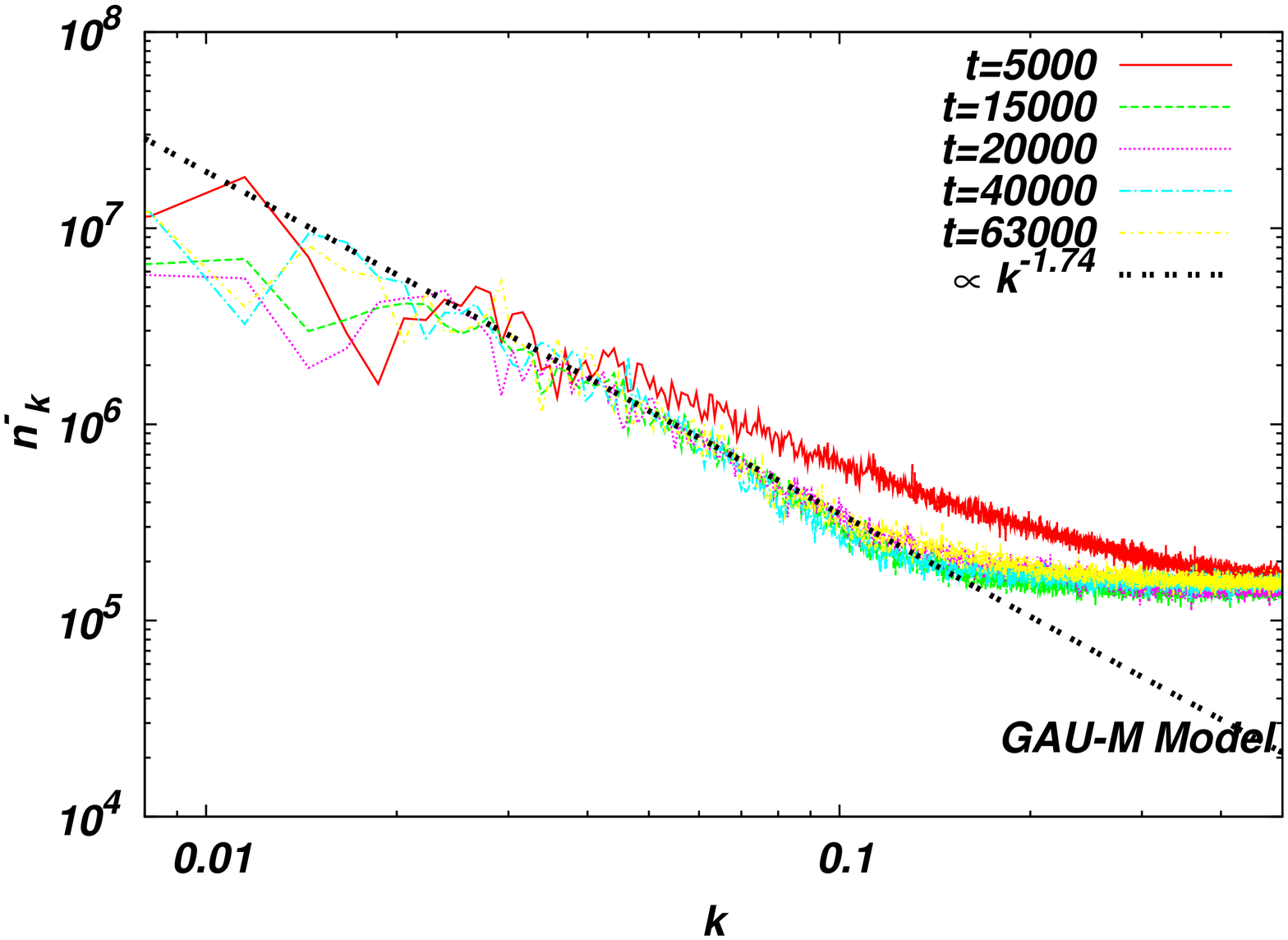}
  \end{center}
  \caption{ \textbf{(Color online)} Left panels (GRV-M Model) and right panels (GAU-M Model): the top panels show the evolution of zero-mode (red-solid lines) and the variations for $\sigma$ (dotted-dashed purple lines), whose latter evolution are fitted by a function, $\propto t^{\gamma_1}$, (black dashed lines) where $\gamma_1$ is a numerical value as the power of \eq{variation}. In the middle and bottom panels, we plot, respectively, the amplitudes of $n^+_k$ and $n^-_k$ in different times for both models, and we fit them by a function of $\propto k^{-\gamma_2}$ where $\gamma_2$ is a numerical value.}
  \label{fig:nknonlin}
\end{figure}

The top panels of \fig{fig:nknonlin} show the evolution of zero-mode (red-solid lines) and the variations for $\sigma$ (dotted-dashed purple lines), whose latter evolution are fitted by a function, $\propto t^{\gamma_1}$, (black dashed lines), where $\gamma_1$ is a numerical value as the power of \eq{variation}. For both models (GRV-M Model on the left panel and GAU-M Model on the right panel), the asymptotic evolution after the linear perturbation regime is overlapped by the function, where $\gamma_1 \sim 0.121$ for GRV-M Model and $\gamma_1\sim 0.235$ for GAU-M Model. Our analytic values can be matched by setting $p\sim 0.111$ with $m=5$ in GRV-M Model and $p\sim 0.250$ with $m=3$ in GAU-M Model, see \eq{variation}. Hence, we could identify this regime as driven (stationary) turbulence, and the main dynamics in each model is caused by either a five-particle interaction or three particle interaction, respectively. Note that our nonrenormalisation term has a $\phi^6$ term in both models. In the middle and bottom panels of \fig{fig:nknonlin}, we plot, respectively, the amplitudes of $n^+_k$ and $n^-_k$ in different times for GRV-M Model (left panels) and GAU-M Model (right panels). For $n^{\pm}_k$ of GRV-M Model, the amplitudes of the high momentum modes grow in time, whilst the lower momentum modes do not decay completely and stay for a long time. We fit a function, $\propto k^{-\gamma_2}$, (yellow dotted lines) where $\gamma_2$ is a numerical value onto the spectra at $t=6700$ for the region where the function is fitted as shown in black dashed lines. We find that $\gamma_2\sim 1.62$ for the $n^+_k$ case and $\gamma_2\sim 0.37$ for the $n^-_k$ case. In the right middle and bottom panels, we plot the amplitudes of $n^{\pm}_k$ for GAU-M Model in various times. The amplitudes of the high momentum modes decrease as opposed to the GRV-M case, and the slopes of the spectra for $n^{\pm}_k$ at $t=63000$ in yellow-dotted lines are steeper than the GRV-M case, where we fit the numerical spectra by the following values shown in black dashed lines: $\gamma_2\sim 3.95$ for the $n^+_k$ case and $\gamma_2\sim 1.74$ for the $n^-_k$ case.

\subsubsection{From driven turbulence to near equilibrium -- Thermalisation:}

\begin{figure}[!ht]
  \begin{center}
	\includegraphics[angle=-90, scale=0.31]{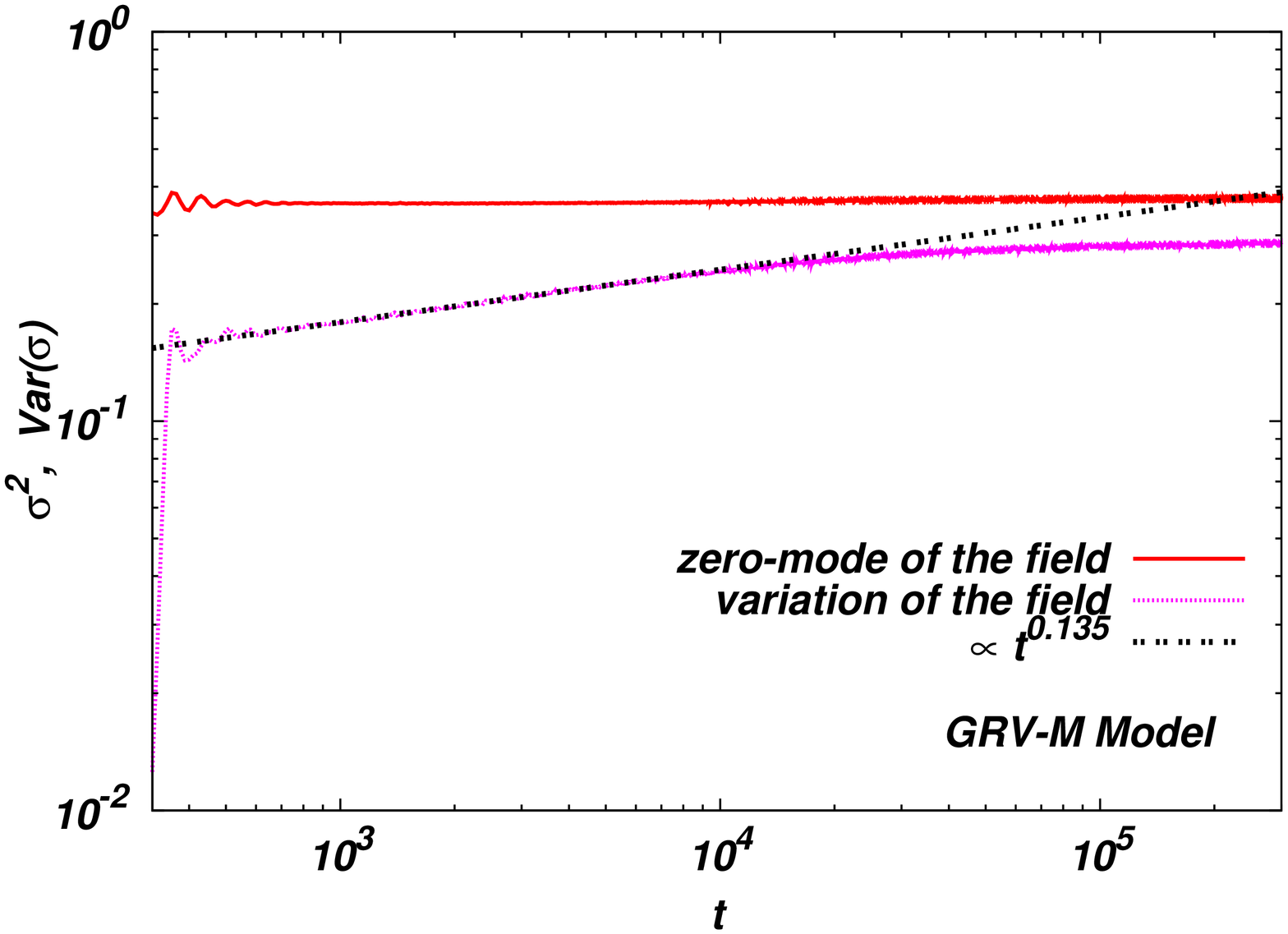}
	\includegraphics[angle=-90, scale=0.31]{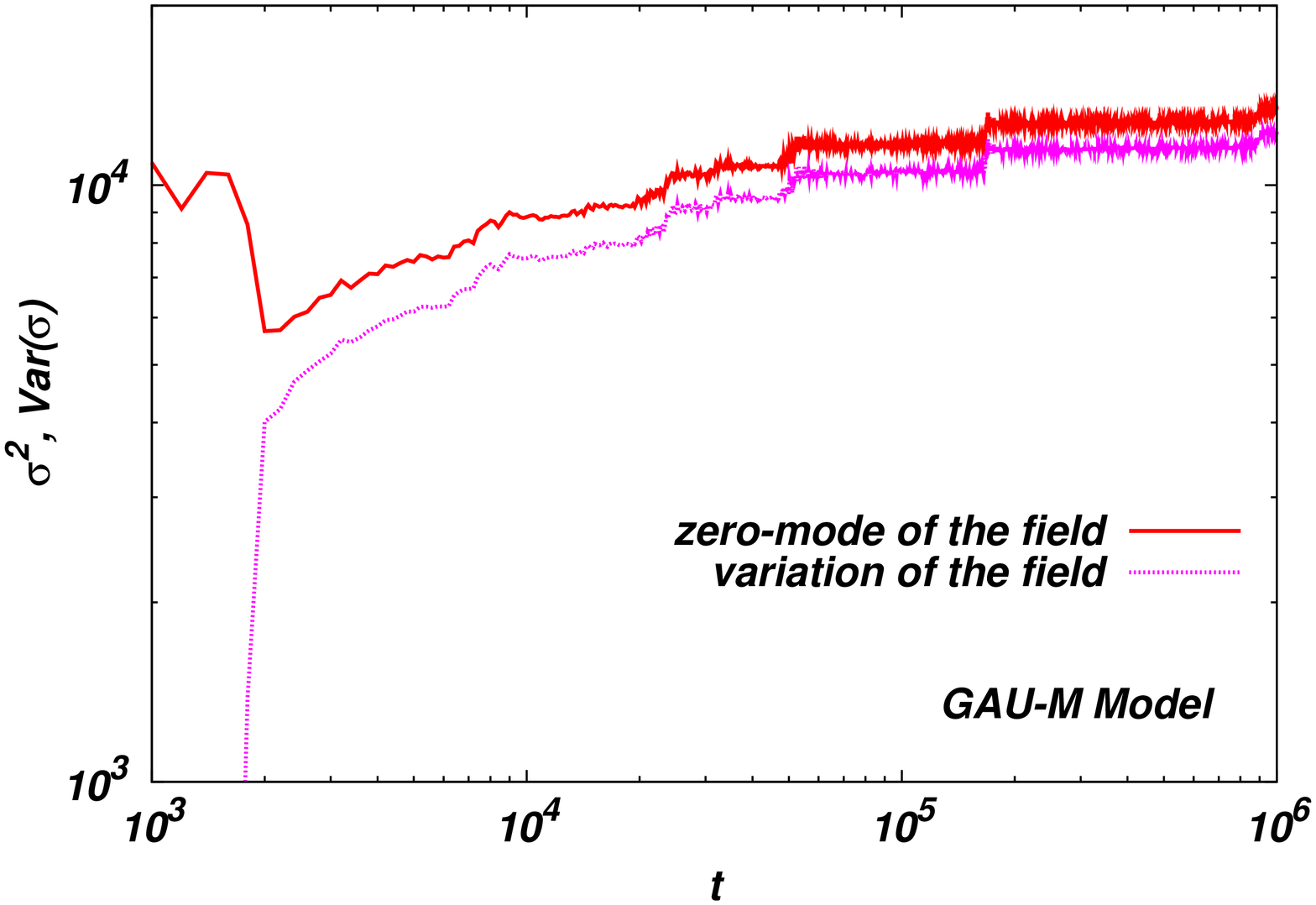}\\
	\includegraphics[scale=0.40]{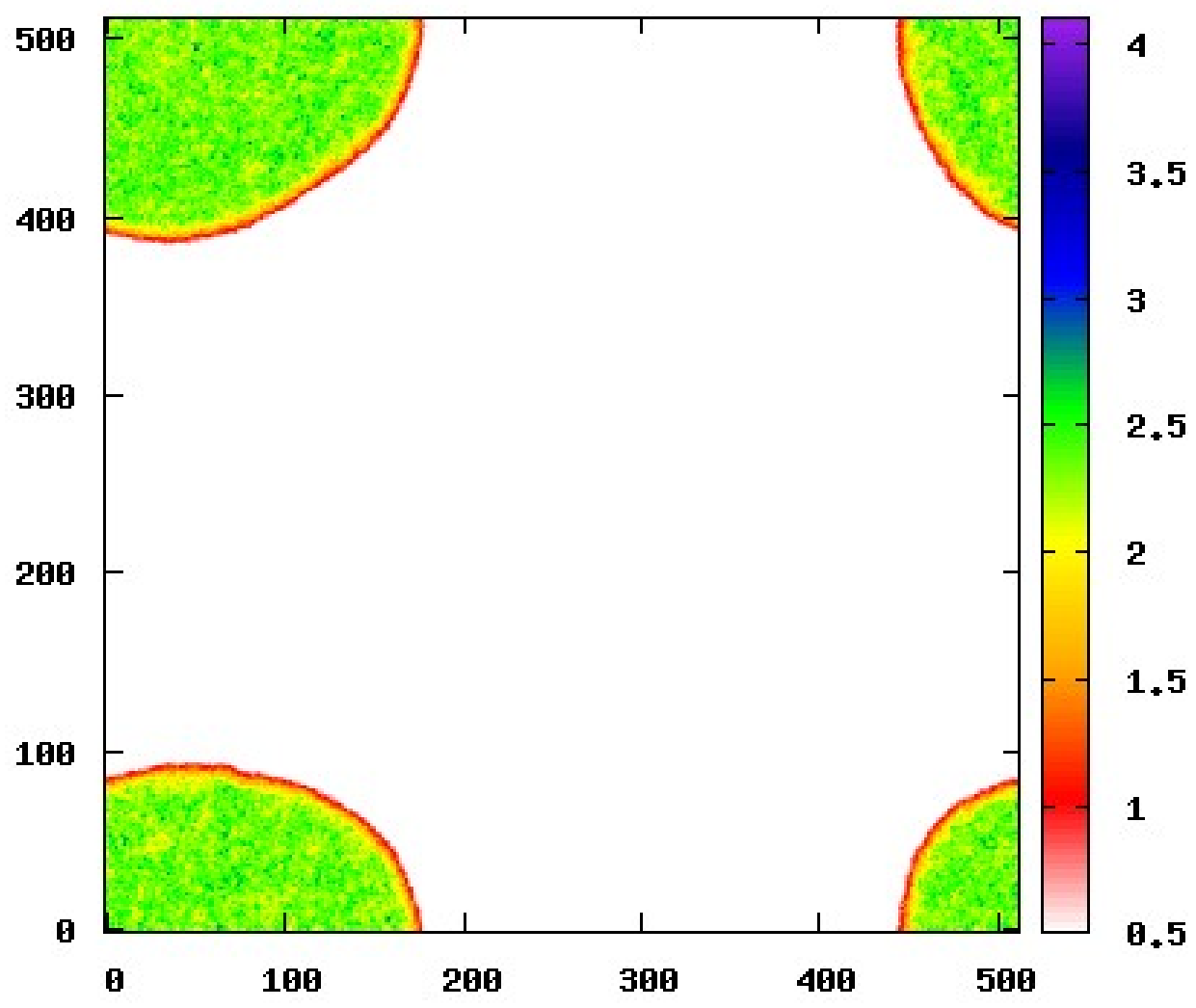}
	\includegraphics[scale=0.40]{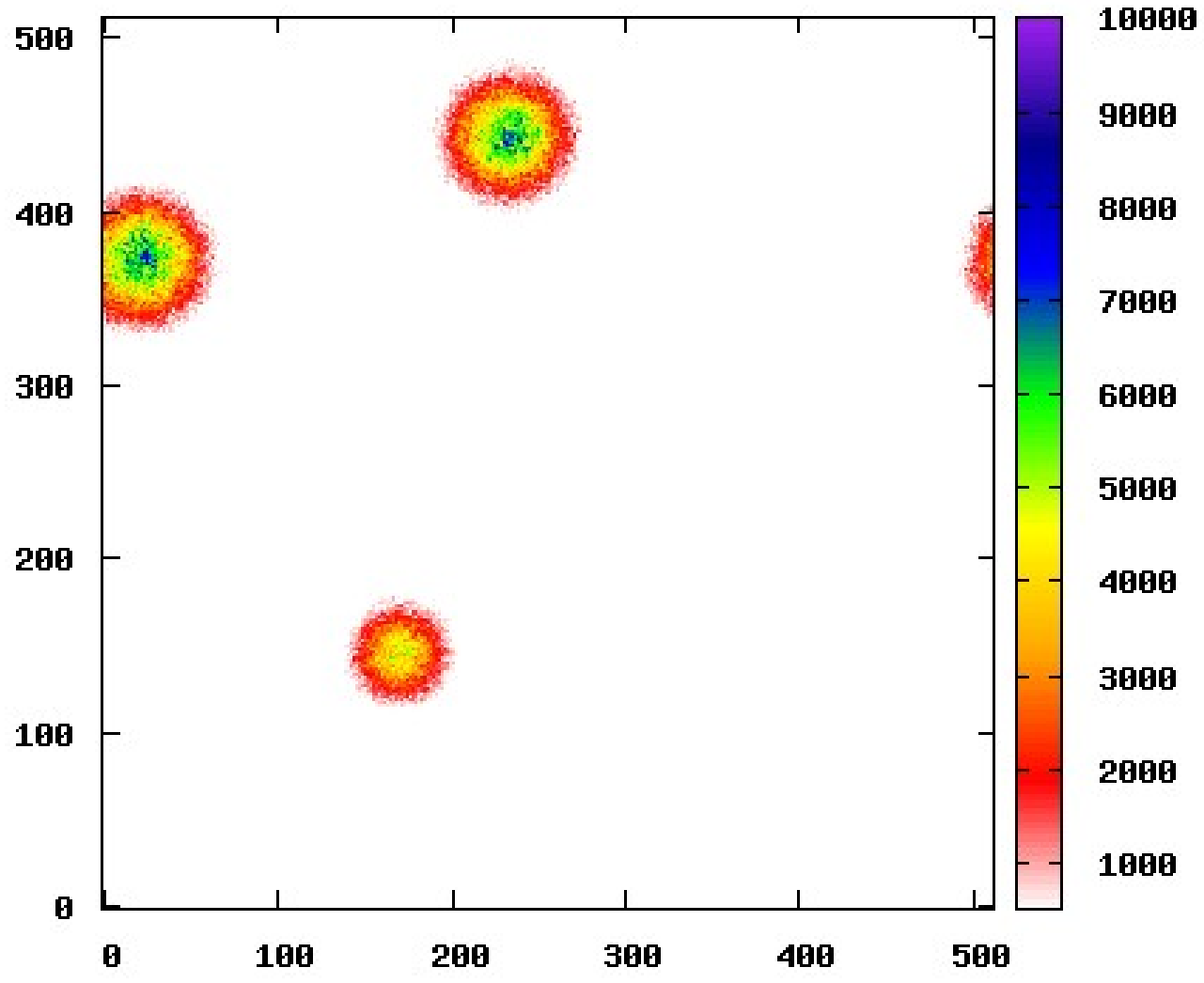}
  \end{center}
  \caption{ \textbf{(Color online)} Left panels (GRV-M Model) and right panels (GAU-M Model) in $2+1$ dimensions: the top panels show the evolution of zero-mode (red-solid lines) and the variations for $\sigma$ (dotted-dashed purple lines). In the bottom panels, we plot the energy density (at $t=3.5\times 10^5$ in the left-bottom panel and at $t=1.7\times 10^7$ in the right-bottom panel) instead of the charge densty to compare the $Q$-ball profiles seen in FIG. 3 and FIG. 7 of \cite{Copeland:2009as} where the colour bars illustrate the values of energy density. We can see that almost all of the charge is trapped into bubbles which may be ``thin-wall'' $Q$-balls, recall that we are imposing a periodic boundary condition.}
  \label{fig:neq}
\end{figure}

In order to significantly reduce the simulation time, we carry out $2+1$-dimensional lattice simulations with the same initial conditions used in $3+1$-dimensional cases, where our lattice units are reduced from $512^3$ to $512^2$. In the top panels of \fig{fig:neq} (GRV-M Model in the left panels and GAU-M Model in the right panels), we illustrate the evolution of zero-mode and the variances of $\sigma$, and in the bottom panels we plot the energy density (at $t=3.5\times 10^5$ in the left-bottom panel and at $t=1.7\times 10^7$ in the right-bottom panel) instead of the charge density to compare with the $Q$-ball profiles at zero-temperature, which we obtained in FIG. 3 and FIG. 7 in \cite{Copeland:2009as}. The colour bars in the bottom panels of \fig{fig:neq} illustrate the values of energy density. Note that we are using the same parameters for GRV-M Model as the ones used in \cite{Copeland:2009as}, whilst the potential for GAU-M Model used there is a generalised version of our present potential \eq{gauge-pot}, so the profiles in GAU-M Model should look similar only qualitatively, but not quantitatively. From the top panels, we can also see, in particular GRV-M Model, the scaling exponent evolution during the driven turbulence stage after the pre-thermalisation ends as confirmed in the top panels of \fig{fig:nknonlin}. The subsequent evolution, however, is different each other and also unique apart from a characteristic free turbulence stage. These features of thermalisation process are caused by stable nonlinear solutions, namely ``$Q$-balls''; in GRV-M Model (left panels), the variance does not evolve that much after the driven turbulence stage ends and we can see thin walled like charged lumps in the end, see the left-bottom panel. In GAU-M Model (right panels) the variance has a step-like evolution, at which stage we confirmed that two (or sometimes more) charged lumps collide and merge into a larger lump. The collision rate is very low since the motions of these ``heavy'' bubbles are nonrelativistic, but we expect that there will be only one single $Q$-ball left ultimately as similar as the GRV-M case. Generally, we observe that almost all of the total energy is trapped into these lumps, where we also confirm that the total charge is absorbed into these lumps, as reported in \cite{Kasuya:2000wx, Kasuya:1999wu}. As the ``thin-wall'' $Q$-balls in GAU-M Model do not have an extremely thin-wall thickness \cite{Copeland:2009as}, the profiles seen in the right bottom panel do not have such a thin-shell thickness. Note that the ``thick-wall'' $Q$-balls in GAU-M Model may suffer from classical instability and fission against spatial perturbations around the $Q$-ball solutions, and decay into smaller $Q$-balls as opposed to the case of ``thick-wall'' $Q$-balls in GRV-M Model. The reader should also notice that the potential for GAU-M Model in the present case is different from Eqs. (40) and (41) in \cite{Copeland:2009as}, which may change the classical stability of the $Q$-balls in the ``thick-wall'' limit. Furthermore, the stability of $Q$-balls is related to their own charge $Q$ so that the initial ratio, $E/(mQ)$, can also cause the different evolution. Therefore, we believe that the evolution is very sensitive to the parameters of models used and the initial conditions. It is worth mentioning, in the left-bottom panel, that the value of charge density within the charged cluster is slightly larger than the value of the thin-wall $Q$-balls in the zero-temperature case (compare to right bottom panel of FIG. 3 in \cite{Copeland:2009as}). We believe that this is because this charged cluster appears in the thermal background, in which the thermal effects change their profiles.

\vspace*{10pt}

Let us recap our findings in this section. We have shown in both GRV-M and GAU-M models that the AD condensate that has a negative pressure is generally unstable against linear fluctuations, and the fluctuations evolve exponentially. The condition for the presence of the negative pressure corresponds to the existence condition of $Q$-balls, and under our initial conditions shown in \tbl{parameterSET}, we observed that almost all of the total charge is trapped into a single (and a few) spherical lump(s) (``thermal $Q$-balls'') in the end of our numerical simulations. In the intermidiate regions between the initial exponential amplification stage and thermalisation stage in the presence of the nonlinear solutions, we identified that the driven turbulence is active; we then found the scaling exponent evolution for the variance of $\sigma$, and we saw that this stage lasts much longer than the case of tachyonic reheating.

\section{Conclusion and discussion}\label{concl}

In this paper, we discussed both analytically and numerically two main issues: Affleck-Dine (AD) dynamics and their subsequent nonequilibrium dynamics in the presence of nonlinear solutions. We showed that the AD dynamics has the same features as orbital motions of planets, replacing the gravitation force by an isotropic harmonic oscillator force. As the relativistic correction on the Newtonian potential gives a precession for the planetary orbit motion, the orbits of AD fields are disturbed by the nonrenormalisable and quantum correction terms. Note that the essential origin of these corrections is physically different. In the presence of a negative pressure of the AD condensate, we have shown that the condensate is classically unstable, and the evolution of the system is similar to the dynamics of reheating the Universe, i.e. \emph{pre-thermalisation}, \emph{bubble collisions} and \emph{thermalisation}. We found that the thermalisation process occurs in the presence of charged lumps, which merge into a single (or a few) ``thin-walled $Q$-ball(s)'' absorbing most of the homogeneous charge distributed initially in a lattice.

\vspace*{10pt}

In Sec. \ref{sectorbit}, we introduced two phenomenological models which are motivated by the MSSM, i.e. gravity-mediated (GRV-M) model and gauge-mediated (GAU-M) model. We obtained the frequencies of the rotation for the nearly circular orbits, and showed that the condensate can have a negative pressure in both cases, see Sec. \ref{MODEL-ABC}. Furthermore, we checked numerically our analytic results with the various cases in both a non-expanding and expanding universe.

Our analytic expressions have a number of advantages. In the existing literature on preheating for complex scalar fields \cite{Chacko:2002wr, Postma:2003gc}, the motion of the complex scalar field is assumed to be of an elliptical form, but their ansatz does not hold (compare our expressions in \eqs{tisig}{sigsol} and \eq{sigantz} and their ansatz). In the multi-flat direction cases, our analytic expressions of the AD field give the exact Mathieu equation when the interaction term between the AD field $\phi$ and another field $\chi$, which parametrises another flat direction, is given by $g^2 |\phi|^2|\chi|^2$, where $g$ is a coupling constant between them. The previous literature \cite{Chacko:2002wr} suggested that the resonant SUSY preheating for nearly circular orbits is not effective since the characteristic dimensionless quantity $q$ is much less than unity, recalling that broad resonant preheating (nonadiabatic evolution) occurs for $q\gg 1$. This statement also holds for our case when the orbit of the AD field is nearly circular since $q\propto \varepsilon^2$ where $\varepsilon^2$ is the third eccentricity of the orbits, recalling that nearly circular orbits correspond to the case of $\varepsilon^2 \ll 1$. 

We obtained the successful ans\"{a}tze, \eq{sigantz}, for nearly circular orbits in an expanding universe (see also the top panels in \fig{fig:sigexp}), but our analytical expressions could be improved by the action variable technique as a real scalar field case \cite{Berkooz:2005sf}. These issues on understanding the analytic orbit forms are related to the dynamics of spinning scalar fields, which can be responsible for the early- and late- time exponential expansions of the Universe (spinflation \cite{Easson:2007dh} and spintessence \cite{Boyle:2001du}) since the AD condensate can possess a negative pressure, which can satisfy the condition of slow-roll inflation, $w<-1/3$. It has been discussed in \cite{Liddle:1998xm} about an oscillating field responsible for dark energy (see a recent review \cite{Copeland:2006wr}), and it gives a constraint on the power of a power-law potential in order to obtain the attractor solutions \cite{Copeland:1997et}. As in real scalar fields, a complex scalar field has been investigated, see for example \cite{Kasuya:2001pr}. Following our analytical work, one can investigate the further analysis on dark energy for a complex scalar field and their late evolution in order to place constraints on parameters of the models, avoiding $Q$-ball formation.
\vspace*{10pt}

In Sec. \ref{sectinst}, we explored the late evolution of AD fields in Minkowski spacetime in both GRV-M and GAU-M models. As a usual nonequilibrium dynamics, we proposed that the dynamics of the $Q$-ball formation goes through three distinct regimes: \emph{pre-thermalisation}, \emph{bubble collision} (driven turbulence) and \emph{thermalisation}. We showed analytically that the AD condensate is unstable against spatial perturbations when the condensate has a negative pressure, and the perturbations grow exponentially. The presence of the negative pressure satisfies the existence of $Q$-balls as well as the fact that the sound wave of the perturbation has an imaginary value of the sound speed. Assuming the adiabatic linear evolution, we have analytically presented that the perturbations for the most amplified mode $k=k_m$ in \eq{kmost} grow with the exponent $\dot{S}_m$ in \eq{Sevo}, which we obtained by taking the average over one rotation of the orbits of the AD field. In the previous literature \cite{Enqvist:2002si, Kasuya:2001hg}, these values were obtained by ignoring the nonrenormalisable term and by assuming that the orbit is circular. By including the nonrenormalisable term and considering more general elliptic orbits, we recovered their results as the leading order of our solutions in Sec. \ref{NCO}. We also showed that the nonlinear time is delayed compared to the time which the authors \cite{Enqvist:1998en} obtained, since the other modes are not well developed when the most amplified mode starts to grow exponentially. With our $3+1$-dimensional numerical lattice simulations that are run for much longer time with much larger simulation sizes than the past lattice simulations in \cite{Kasuya:2000wx, Kasuya:1999wu, Kasuya:2001hg, Kusenko:2008zm, Enqvist:2002si}, our analytic results are well checked. In addition, we found that the adiabatic condition is violated at the beginning stage of the linear perturbations as seen in broad resonant preheating. In the driven turbulence stage, we observed that many bubbles form and collide/merge into larger bubbles in both GRV-M and GAU-M models. Note that these bubbles are nothing to do with bubbles due to first order phase transition. By concerning the variance of the radial field $\sigma$, we have seen that the evolution follows a scaling exponent law as a signature of the driven turbulence \cite{Micha:2004bv}. As opposed to the case of tachyonic preheating, this driven turbulence stage, in our case, lasts for longer time, which may be caused by the presence of classical nonlinear solutions, i.e. ``$Q$-balls''. We saw in $2+1$-dimensional numerical results that a thermalisation stage actually exists where the evolution for the variance of a field has a different scaling law from the one which appears in the driven (first) turbulence stage. We believe that quantum effects should be non-negligible in this late turbulence stage, and the classical thermalisation process, in our case, should be different from the corresponding quantum-mechanical thermalisation. Since the thermalisation process is generally extremely long, a lattice simulation in an expanding background encounters serious problems in the ultra-violet limits; thus, we ignored Hubble expansion in this paper. By considering the quantum-mechanical effects as well as Hubble expansion, it is worth to investigate the cosmological consequences as our future work.

In the context of a (p)reheating scenario, it has been suggested \cite{GarciaBellido:2007af} that the collision of bubbles during the driven turbulence stage can be an effective source of gravitational waves, which can be detected by LIGO \cite{ligo-web-page} and LISA \cite{lisa-web-page} in the near future. We noticed that this analysis should be applicable to the same driven turbulence stage of the $Q$-ball formation, which was initially proposed in \cite{Tsumagari:2008bv}. The problem of gravitational waves emitted in the fragmentation stage has been discussed \cite{Kusenko:2008zm}, while the analysis of the gravitational wave emissions in the driven turbulence stage still remains.

\section*{ACKNOWLEDGMENTS}

This work was partially done when the author visited the Cosmology group, University of Helsinki. This visit was financially supported by the BESTS scholarship, University of Nottingham. Our numerical code is developed from the code, 'LATfield' \cite{latfield-web-page}. The numerical simulations were carried out by UK National Cosmology Supercomputer, Cosmos, funded by STFC, HEFCE and silicon Graphics, and carried out by Nottingham HPC facility. The author would like to thank E.~Copeland for continuous encouragement and positive comments on the final draft of this manuscript. The author is also grateful to N.~Bevis, D.~Buck, K.~Enqvist, M.~Kawasaki, K.~Maeda, P.~Saffin and O.~Seto for valuable discussions, A.~Mazumdar for leaving useful comments, and A.~Josan for carefully reading this manuscript.

\appendix

\section{Perturbations on multiple scalar fields}\label{MULTI}

In this appendix, we obtain Euler-Lagrange equations for multiple scalar fields $\hat\varphi^a$ with a symmetric field space metric $G_{ab}(\hat\varphi)=G_{ba}(\hat\varphi)$, following the notations \cite{Sasaki:1995aw, Weinberg:2008zzc}. Our aim is to obtain equations of motion for the background homogeneous (zero-mode) fields $\varphi^{a}(t)$ and the perturbed fields $\delta\varphi^a(t,\mathbf{x})$ in a fixed unperturbed background (Friedmann-Robertson-Walker) metric, $g_{\mu\nu}=\textrm{diag}(-1,\ a(t),\ a(t),\ a(t))$ where $a(t)$ is scale factor of the Universe and $H=\dot{a}/a$ is Hubble parameter. Here, an over-dot denotes the time-derivative. As the simplest nontrivial example of the multiple scalar fields, we find equations of motion for a complex scalar field $\hat{\phi}\equiv\hat{\sigma} e^{i\hat{\theta}}$ where $\hat{\sigma}$ and $\hat{\theta}$ are real scalar fields and the system possesses a U(1) symmetry.

Let us start off with the action
\be{ext2}
S=\int d^4x \sqrt{-g}\bset{-\half g^{\mu\nu}G_{ab}(\hat{\varphi})\partial_\mu \hat\varphi^a \partial_\nu, \hat{\varphi}^b-V(\hat\varphi)},
\ee
where $g\equiv \textrm{det}\bset{g_{\mu\nu}}$ and $V(\hat\varphi)$ is a potential for $\varphi$.
Because of the action principle, we obtain the Euler-Lagrange equation for $\hat{\varphi}$
\be{ELmulti}
\frac{1}{\sqrt{-g}}\partial_\rho\bset{\sqrt{-g}g^{\rho\nu}G_{cb}\partial_\nu \hat\varphi^b}=\half g^{\mu\nu}G_{ab,c}\partial_\mu \hat\varphi^a \partial_\nu \hat\varphi^b + V_{,c},
\ee
and the energy momentum tensor
\be{EGMmulti}
T_{\mu\nu}=G_{ab}\partial_\mu \hat\varphi^a \partial_\nu \hat\varphi^b + g_{\mu\nu}\sbset{-\half g^{\rho\sigma}G_{ab}\partial_\rho\hat\varphi^a\partial_\sigma\hat\varphi^b - V(\hat\varphi)}.
\ee
Here, we defined $G_{ab,c}\equiv \frac{dG_{ab}}{d\hat\varphi^c}$, and so on. The energy density and pressure can be given by $T_{\mu\nu}$ \cite{Weinberg:2008zzc}
\bea{energy-multi}
\rho_E&=&-\half g^{\mu\nu} G_{ab} \partial_\mu \varphi^a \partial_\nu \varphi^b + V(\varphi),\\
\label{press-multi} p&=& -\half g^{\mu\nu} G_{ab} \partial_\mu \varphi^a \partial_\nu \varphi^b - V(\varphi).
\eea
By pertubing the fields as $\hat\varphi^a=\varphi^a(t)+\delta\varphi^a(t,\mathbf{x})$ where $|\varphi|\gg |\delta\varphi|$, the homogeneous part gives, from \eq{ELmulti},
\be{homovp}
\frac{D}{dt}\dot{\varphi}^a+3H \dot{\varphi}^a +G^{ab}V_{,b}=0,
\ee
where the covariant derivative, $D/dt$, can be defined by the ``Christoffel symbols'' $\gamma^{a}_{bc}\equiv\half G^{ad}\times$ $\bset{G_{dc,b}+G_{db,c}-G_{bc,d}}$; thus, $\frac{D}{dt}\dot{\varphi}^a\equiv \frac{d}{dt}\dot{\varphi}^a + \gamma^{a}_{bc}\dot{\varphi}^b\dot{\varphi}^c$.
On the other hand, we can obtain the equations of motion for the pertubed fields $\delta \varphi$ from \eq{ELmulti}
\be{perturb}
\frac{D^2}{dt^2}\delta\varphi^a+3H\frac{D}{dt}\delta\varphi^a-\bset{\frac{\nabla}{a}}^2\delta\varphi^a-\gamma^{a}_{bcd} \dot{\varphi}^b\dot{\varphi}^c\delta\varphi^d+(V^{;a})_{;d}\delta\varphi^d=G^{ab}G_{bc,d}G^{ce}V_{,e}\delta\varphi^d,
\ee
where we used $\frac{D}{dt}\delta\varphi^a = \delta\dot{\varphi}^a+\gamma^a_{bc}\dot\varphi^b\delta\varphi^c$, defined the ``Riemann tensors'' as $\gamma^a_{bcd}\equiv \gamma^a_{bd,c}-\gamma^a_{bc,d}+\gamma^a_{ce}\gamma^e_{bd}-\gamma^a_{de}\gamma^e_{bc}$, and denoted the covariant derivative as a usual notion $';'$. Notice that we used $V_{,b}\equiv \frac{\partial V}{\partial\hat\varphi^b}(\hat\varphi)\simeq \left.\frac{\partial V}{\partial \hat\varphi^b}(\hat\varphi)\right|_{\varphi}+\delta\varphi^c \left.\frac{\partial^2 V}{\partial \hat\varphi^b\partial \hat\varphi^c}\right|_{\varphi}+\dots$.

When the system has a $O(2)\sim U(1)$ symmetry as $\hat\varphi_a=\bset{\hat\sigma, \hspace*{5pt} \hat\theta}$ and a flat field metric is $G_{ab}=\textrm{diag}(1,\; \hat\sigma^2)$, we can obtain $\gamma^1_{22}=-\hat\sigma;\, \gamma^2_{12}=\gamma^2_{21}=1/\hat\sigma$. Then, \eq{homovp} with a potential $V(\sigma)$ gives
\bea{sigeom}
\ddot{\sigma}+3H\dot{\sigma}-\sigma\dot{\theta}^2+\frac{dV}{d\sigma}&=&0,\\
\label{thetaeom} \ddot{\theta}+3H\dot{\theta}+\frac{2}{\sigma}\dot{\sigma}\dot{\theta}&=&0.
\eea
Here, the third term in \eq{sigeom} corresponds to the ``centrifugal force'' due to the spin in a field space, and the third term in \eq{thetaeom} corresponds to the ``Colliori force''. In addition, the energy density and pressure can be given by from \eqs{energy-multi}{press-multi}
\be{epcom}
\rho_E=\half\bset{\dot{\sigma}^2+\sigma^2\dot{\theta}^2}+V,\spc p=\half\bset{\dot{\sigma}^2+\sigma^2\dot{\theta}^2}-V.
\ee
Furthermore, \eq{perturb} gives
\bea{pertsig}
&&\ddot{\dsig}+3H\dot{\dsig}-\bset{\bset{\frac{\nabla}{a}}^2+\dot{\theta}^2-\frac{d^2V}{d\sigma^2}}\dsig-2\sigma\dot{\theta}\dot{\dtheta}=0,\\
\label{pertth}&&\ddot{\dtheta}+\bset{3H+\frac{2\dot{\sigma}}{\sigma}}\dot{\dtheta}-\bset{\frac{\nabla}{a}}^2\dtheta+\frac{2\dot{\theta}}{\sigma^2}\bset{\sigma\dot{\dsig}-\dot{\sigma}\dsig}=0.
\eea
We use \eqs{sigeom}{thetaeom} to concern with the orbits of AD condensates in Sec. \ref{sectorbit}, and use \eqs{pertsig}{pertth} to investigate the linear spatial instability of the condensates in Sec. \ref{sectinst}.

\section{The orbit of an Affleck-Dine ``planet''}\label{App2}

In this appendix, we obtain an exact orbit form in a quadratic potential case when Hubble expansion is assumed to be small and adiabatic. The orbit of a AD field, or more precisely an eccentricity of the elliptic motion in the complex field-space, is determined by the initial charge and energy density. In order to obtain analytic expressions of the orbit in more general potential cases in which we are more interested, we restrict ourself to work in Minkowski spacetime and on the orbit which should be nearly circular. We then obtain the perturbed orbit equation and necessary conditions for closed orbits, where the orbits come back to their original positions after some rotations around the minimum of the effective potential. By including the effects of Hubble expansion, we shall introduce ans\"{a}tze in Sec. \ref{numAD}, which are inspired by our solutions obtained in Minkowski spacetime, and our numerical results support the ans\"{a}tze, assuming that the rotation frequency $W$ is always much greater than Hubble expansion $H$ \cite{Turner:1983he}.

\subsection{The exact orbit in an expanding universe}\label{sec2}
The exact orbit expressions of an AD field in an expanding universe can be obtained with a quadratic potential,
\be{quad}
V=\frac{M^2}{2}\sigma^2=\frac{M^2}{2}\bset{\frac{a_0}{a}}^3\tisig^2,
\ee
where $M$ is a mass of the field $\phi$ and we rescaled the field $\sigma$, $\sigma(t)=\bset{\frac{a_0}{a(t)}}^{3/2}\tisig(t)$. From now on, we solve the orbit equations, \eq{modrad}, for $\tisig(t)$ at first, and then solve them for $\tiu(\theta)$, replacing the time-dependence by a phase variable $\theta$. We then show that the orbits are of usual elliptic forms with a third eccentricity $\varepsilon^2$ for $\tisig(t)$ and $\tiu(\theta)$.

\subsubsection{The orbit for $\tisig(t)$}

In this subsection we obtain an expression for the orbit $\tisig(t)$ by solving \eq{modrad}. Substituting \eq{quad} into \eq{modrad} and ignoring the terms involving $H^2$ and $\ddot{a}/a$, we obtain
\be{radquad}
\ddot{\tisig} -\frac{\tirhoQ^2}{\tisig^3}+M^2\tisig=0 \spc \lr \spc \frac{d\tirhoE}{dt}=0,
\ee
where $\tirhoE\equiv \half \bset{\frac{d\tisig}{dt}}^2+\half M^2 \tisig^2+\frac{\tirhoQ^2}{2\tisig^2}\neq a^{-3}_0 \rho_E$, which is approximately conserved. Since $\half \frac{d^2}{dt^2}(\tisig^2)=\dot{\tisig}^2+\tisig\ddot{\tisig}=2\tirhoE-2M^2\tisig^2$, \eq{radquad} leads to a harmonic oscillator form,
\be{ext6}
\frac{d^2}{dt^2}(\tisig^2)=-4M^2\bset{\tisig^2-\frac{\tirhoE}{M^2}}
\ee
whose solution is
\bea{tisigsol}
\tisig^2(t)&=&\frac{\tirhoE}{M^2}+ A \cos\sbset{2M(t+t_0)},\\
\label{tisig}&=&\frac{\tirhoE}{M^2}\bset{1+\varepsilon^2 \cos\sbset{2M(t+t_0)} }.
\eea
Here, $B$ is some constant value and we set $t_0$ as a time when the AD field starts to rotate. We also defined a third eccentricity $\varepsilon^2 \equiv \frac{AM^2}{\tirhoE}= \frac{\tisig^2_{max}-\tisig^2_{min}}{\tisig^2_{max}+\tisig^2_{min}}$, where the apocentral and pericentral points are, respectively, given by $\tisig^2_{max}\equiv\frac{\tirhoE}{M^2}+A$ and $\tisig^2_{min}\equiv\frac{\tirhoE}{M^2}-A$.
Notice that the circular orbit case corresponds to $\varepsilon^2=0$, which implies that $\tisig^2_{max}=\tisig^2_{min}$, and also note that the eccentricity is real and has a value between 0 and 1 in the present quadratic potential \footnote{In an inverse-squared central force, the first eccentricity can be larger than equal 1, which corresponds to the cases where the orbits are parabola or hyperbola.}.

We can obtain a period $\tau$ of this orbit,
\be{period}
\tau=\frac{\pi}{M}.
\ee
Substituting \eq{tisigsol} into $\tirhoE$, we obtain $A=\frac{\sqrt{\tirhoE^2-M^2\tirhoQ^2}}{M^2}$. From the above expressions for $\varepsilon^2$ and $A$, we can obtain $\frac{M\tirhoQ}{\tirhoE}=\sqrt{1-\varepsilon^4}$. Using this and \eq{tisigsol}, it ends up with
\be{dotth}
\dot{\theta}(t)=\frac{\tirhoQ}{\tisig^2}=\frac{M\sqrt{1-\varepsilon^4}}{1+\varepsilon^2\cos\sbset{2M(t+t_0)}}.
\ee
For the circular orbits $\varepsilon^2=0$, $\dot{\theta}$ is time-independent as we can expect, and the ratio, $\tirhoE/(M\tirhoQ)$, is unity. While for the radial orbits $\varepsilon^2=1$, $\dot{\theta}=0$ as we can also expect, and we can find $\tirhoE/(M\tirhoQ)\gg 1$.
\subsubsection{The orbit for $\tiu(\theta)=\tisig^{-1}(\theta)$}

What follows is that we express $\tisig(t)$ as a function of $\theta$ by using the second expression in \eqs{modrad}{tisigsol}. We then obtain
\be{theta}
\tan(\theta-\theta_0)=\frac{\tisig_{min}}{\tisig_{max}}\tan\bset{M(t+t_0)},
\ee
where $\theta_0$ is an integration constant and we used the following integral formula, $\int \frac{dx}{a_1+a_2\cos{x}}=\frac{2}{\sqrt{a^2_1-a^2_2}} \textrm{Arctan}\bset{\frac{(a_1-a_2)\tan(\frac{x}{2})}{\sqrt{a^2_1-a^2_2}}}$ with some real values $a_1$ and $a_2$. Without loss of generality, we can choose $t_0=\theta_0=0$, which imply that the orbit at $t=0,\; \tau/2$ gives, respectively, $\theta=0,\; \pi/2$, recalling \eq{period}. By comparing \eq{tisigsol} to \eq{theta}, we obtain 
\bea{tisigsol2}
\tisig^2(\theta)&=&\frac{\tisig^2_{max}\tisig^2_{min}}{\tisig^2_{min}\cos^2\theta+\tisig^2_{max}\sin^2\theta},\\
\label{uel1}\lr \tiu^2(\theta)&=&\frac{1}{\tisig^2}=\frac{\cos^2\theta}{\tisig^2_{max}}+\frac{\sin^2\theta}{\tisig^2_{min}},\\
\label{uel} &=& \frac{\tisig^2_{max}+\tisig^2_{min}}{2\tisig^2_{max}\tisig^2_{min}}\bset{1-\varepsilon^2\cos(2\theta)}.
\eea
Hence, we can see that $\theta=0$ when $\tisig=\tisig_{max}$ and $\theta=\pi/2$ when $\tisig=\tisig_{min}$. Finally, we obtained the expressions for the orbits as usual elliptic forms in \eqs{tisig}{uel}. For the circular orbits, $\varepsilon^2=0$, we can obtain $\tiu^2=const.$ from \eq{uel} as we can expect.
\subsection{The nearly circular orbits in Minkowski spacetime}
Without Hubble expansion, we can investigate a nearly circular bounded orbit of an AD field in general potentials which satisfy \eq{condbound}. Because of the above reasons, we concentrate on non-expanding background in this subsection, and obtain a time-dependent expression for the nearly circular orbits as \eq{tisig}. We then find the expression that depends on the phase $\theta$ as \eq{uel1}. Moreover we obtain conditions for closed orbits, in which the perturbations are expanded up to 1st order (for the complete proof of the condition up to 4th order, see Bertrand's theorem \cite{Bertrand}).

\subsubsection{The orbit for $\sigma(t)$}\label{sigorbitMIN}
In Minkowski spacetime, we can find an expression for the orbit $\sigma(t)$ in a general potential $V(\sigma)$ as in \eq{tisig}. Notice that the tilde variables are the same as un-tilde ones in the present non-expanding background. Recall the equation of motion \eq{radhomo} in Minkowski spacetime,
\be{rad}
\ddot{\sigma}+\frac{dV_+}{d\sigma}=0.
\ee
Suppose that the orbit that is nearly circular as $\sigma(t)=\sigma_{cr}+\delta(t)$, where $\sigma_{cr}\gg |\delta|$, recalling $\sigma_{cr}$ is defined by \eq{sigcr}. Substituting this expression of $\sigma$ into \eq{rad} and keeping $\delta$ terms up to 1st order, we obtain a harmonic oscillator form
\be{del}
\ddot{\delta}+W^2 \delta=0,
\ee
where the reader should recall the condition, \eq{condbound}, for the bound orbits, and $W$ is constant since we are working in Minkowski spacetime. 

Thus, the solution of \eq{del} is 
\be{deltaeq}
\delta(t)=\sigma_{cr} B\cos(Wt),
\ee 
where $B$ is a small positive dimensionless constant, i.e. $B\ll 1$ due to $\sigma_{cr}\gg |\delta|$, and we set the differentiation constant as $0$ to ensure that $\sigma(0)=\sigma_{max}$. We can find that $\sigma_{max}=\sigma_{cr}(1+B)$, $\sigma_{min}=\sigma_{cr}(1-B)$, and $\sigma_{max}\sigma_{min}\simeq \sigma^2_{cr}\bset{1+\order{B^2}}$ ,which give $B=\frac{\sigma_{max}-\sigma_{min}}{\sigma_{max}+\sigma_{min}}$, $\sigma_{cr}=\frac{\sigma_{max}+\sigma_{min}}{2}$, and $2B\simeq \frac{\sigma^2_{max}-\sigma^2_{min}}{\sigma^2_{max}+\sigma^2_{min}}= \varepsilon^2$, where we used the definition of the third eccentricity. We can check that the condition, $2B\simeq \varepsilon^2\ll 1$, is consistent with the fact that the orbit is nearly circular. Since $\dot{\sigma}_{max}=\dot{\sigma}_{min}=0$ and $\rho_E$ is constant, we can equate $B$ with $\rho_E$ and $\rho_Q$ using \eqs{deltaeq}{condbound}:
\be{Avalue}
B=\sqrt{\frac{2(\rho_E-V_+(\sigma_{cr}))}{W^2\sigma^2_{cr}}}=\sigma_{cr}\sqrt{\frac{2(\rho_E-V_+(\sigma_{cr}))}{\left.(\sigma^4V^{\p\p})\right|_{\sigma_{cr}}+3\rho^2_Q}}\simeq \frac{\varepsilon^2}{2} \ll 1,
\ee
where a prime denotes the differentiation with respect to $\sigma$. Finally, we obtain
\be{sigsol}
\sigma^2(t) = \sigma^2_{cr}\bset{1+\varepsilon^2\cos(Wt) + \order{\varepsilon^4}},
\ee
where $W$ is given by \eq{condbound} (compare with \eq{tisig}). Now, we can define the period $\tau$
\be{periodgen}
\tau=\frac{2\pi}{W},
\ee
which reproduces the case for $W=2M$ in \eq{period}.
Using \eqs{phshomo}{sigcr}, we can also find
\be{dottheta}
\dot{\theta}\simeq\frac{\sqrt{\left.V^\p/\sigma\right|_{\sigma_{cr}}}}{1+\varepsilon^2\cos{(Wt)}}.
\ee
Using \eq{rhop}, let us compute the pressure of this AD condensate whose orbit is described by \eq{sigsol}. By expanding $V_-(\sigma)$ around $\sigma=\sigma_{cr}$ and using \eq{sigsol}, we obtain $V_-(\sigma)\simeq V_-(\sigma_{cr}) + \frac{\varepsilon^2 \rho^2_Q}{\sigma^2_{cr}} \cos{(Wt)} +\frac{\varepsilon^4\sigma^2_{cr}}{8}\bset{W^2-\frac{6\rho^2_Q}{\sigma^4_{cr}}}\cos^2{(Wt)} + \dots,$ where we assumed that the higher order terms in $V_-$ are negligible. Therefore,
\bea{}
\nb p &\simeq& \frac{W^2\sigma^2_{cr}\varepsilon^4}{8}\bset{1-2\cos^2{(Wt)}}-V(\sigma_{cr})+\frac{\rho^2_Q}{2\sigma^2_{cr}}\bset{1-2\varepsilon^2\cos{(Wt)}+ \frac{3}{2} \varepsilon^4 \cos^2{(Wt)}  },\\
\label{pressure} \lr\hspace*{5pt} \state{p} &\simeq& -V(\sigma_{cr})+\frac{\rho^2_Q}{2\sigma^2_{cr}}.
\eea
Here we defined an averaged value over an one rotation in the orbit, \eq{sigsol}, namely $\state{X}\equiv \frac{1}{\tau} \int^{\tau}_0 dt X(t)$ where $X$ is some quasi-periodic quantity and $\tau$ is determined by \eqs{periodgen}{condbound}. The result, \eq{pressure}, can be easily understood by the fact that the averaged pressure corresponds to the value at $\sigma=\sigma_{cr}$ since the orbit oscillate around $\sigma_{cr}$ and $\dot{\sigma}_{cr}=0$, c.f. a real scalar field case \cite{Turner:1983he}. Similarly, we can obtain the averaged energy density 
\be{avgengy}
\state{\rho_E} \simeq V(\sigma_{cr})+\frac{\rho^2_Q}{2\sigma^2_{cr}} + \frac{W^2\sigma^2_{cr}\varepsilon^4}{16},
\ee
where we can see the contribution from the term involving $\varepsilon^4$.
Hence, the averaged equation of state is given by
\be{eosg}
\state{w}\equiv \state{\frac{p}{\rho_E}}=\frac{\frac{\rho^2_Q}{2\sigma^2_{cr}}-V(\sigma_{cr})}{\frac{\rho^2_Q}{2\sigma^2_{cr}}+V(\sigma_{cr})+W^2\sigma^2_{cr}\varepsilon^4/16}.
\ee
\subsubsection{The orbit for $u(\theta)=\sigma^{-1}(\theta)$}\label{uorbitMIN}

In order to obtain a $\theta$-dependent expression of the orbit as \eq{uel1}, let us switch the variable $\sigma$ to $u(\theta)\equiv1/\sigma(\theta)$. In Minkowski spacetime, where we can again drop the tilde variables there, the orbit equation \eq{orbiteq} is
\be{orbitMink}
\frac{d^2u}{d\theta^2}+u=-\frac{1}{\rho^2_Q}\frac{dV}{du}\equiv J(u).
\ee
Let $u_0$, which is independent of $\theta$, be the value of a circular orbit (i.e. $u_0\equiv1/\sigma_{cr}$). We then consider an orbit $u(\theta)$ that deviates slightly from $u_0$ with a fluctuation $\eta(\theta)$, i.e. $u=u_0+\eta$, where $u_0\gg |\eta|$. Since $\frac{du_0}{d\theta}=0=\frac{d^2u_0}{d\theta^2}$, \eq{orbitMink} implies that $u_0=J(u_0)$. By expanding $J(u)$ around $u=u_0$, we obtain $J(u)\simeq u_0 + \eta \left.\frac{dJ}{du}\right|_{u_0}+\dots$, where we are evaluating the differentiations at $u_0$. Hence, we can obtain the perturbed orbit equation for $\eta(\theta)$
\be{pertorbit}
\frac{d^2\eta}{d\theta^2}+\beta^2\eta=0,
\ee
where $\beta^2\equiv 1-\left.\frac{dJ}{du}\right|_{u_0}$ which should be positive for bounded orbits. Note that this condition, $\beta^2>0$, is equivalent to the previous condition, \eq{condbound}, since 
\be{betasq}
\beta^2=\frac{\sigma^4_{cr}}{\rho^2_Q}W^2=\left.\frac{3V^\p+\sigma V^{\p\p}}{V^\p}\right|_{\sigma_{cr}},
\ee
where we used the fact $V^\p=\frac{\rho^2_Q}{\sigma^3}$ at $\sigma=\sigma_{cr}$ from \eq{sigcr}.
The solution of \eq{pertorbit} is 
\be{etasol}
\eta=u_0C\cos(\beta\theta+\theta_0),
\ee
where $C$ and $\theta_0$ are constants, and $0<C\ll 1$ due to the fact that $u_0\gg |\eta|$. We can then show $C= B$ by equating the value of $C$ with $\rho_Q$ and $\rho_E$. Substituting $u$ into $\rho_E$ and expanding $V(u)$ around $u=u_0$ up to second order, we can find
\be{Cvalue}
C=\frac{1}{u_0}\sqrt{\frac{2(\rho_E-V_+(1/u_0))}{\left.\frac{d^2V(1/u)}{du^2}\right|_{u_0}+\rho^2_Q}}=B\simeq \frac{\varepsilon^2}{2},
\ee
where we used $\left.\frac{dV_+(u)}{du}\right|_{u_0}=\left.\frac{dV(u)}{du}\right|_{u_0}+\rho^2_Qu_0=0$ from \eq{sigcr}. The relation, $C=A$, is obtained by changing the variable $u$ back to $\sigma$ (compare \eq{Cvalue} with \eq{Avalue}).

Let us choose $\theta_0=\pi$ in \eq{etasol}, then we obtain
\bea{uel2}
u&=&u_0\bset{1-C\cos(\beta\theta)},\\ 
\label{uel3} u^2 &\simeq& u^2_0\sbset{1-2C\cos(\beta\theta) +\order{C^2}}.
\eea
Notice that $0<C \ll 1$ which is consistent with the condition for nearly circular orbits, $\varepsilon^2\ll 1$ as we have seen in appendix \ref{sigorbitMIN} and \eq{Cvalue}. We can also find that $\sigma_{max}=\frac{\sigma_{cr}}{1-C}$ for $\beta\theta=0$ and $\sigma_{min}=\frac{\sigma_{cr}}{1+C}$ for $\beta\theta=\pi$. 

To show that the orbit $u(\theta)$ in \eq{uel3} has a similar form as \eq{uel}, let us compute the following relations: $\sigma^2_{max}+\sigma^2_{min}\simeq 2\sigma^2_{cr}\bset{1+\order{C^2}},\; \sigma^2_{max}-\sigma^2_{min}\simeq 4\sigma^2_{cr}C\bset{1+\order{C}}$ and $\sigma^2_{max}\sigma^2_{min}\simeq \sigma^4_{cr}\bset{1+\order{C^2}}$. Hence, $u^2_0\simeq \frac{\sigma^2_{max}+\sigma^2_{min}}{2\sigma^2_{max}\sigma^2_{min}}$ and $2C \simeq \varepsilon^2$, which imply that \eq{uel3} is of similar orbit form as \eq{uel}. As we computed going from \eq{uel1} to \eq{uel}, where for this case we deduce \eq{uel} from \eq{uel1}, we finally obtain
\be{ext7}
u^2 \simeq \frac{\cos^2\frac{\beta}{2}\theta}{\sigma^2_{max}}+\frac{\sin^2\frac{\beta}{2}\theta}{\sigma^2_{min}}.
\ee
In the next subsection, we obtain the conditions for closed orbits using \eq{ext7} \cite{Whittaker37}.

\subsubsection{Conditions for closed orbits and equations of state}\label{Sectclose}

Let us define an angle $\Phi$, which is the phase difference as the orbit goes from $\eta=u_0C$ to $\eta=-u_0C$,
\be{Phi}
\Phi \equiv \frac{\pi}{\beta}=\pi \sqrt{\left.\frac{V^\p}{3V^\p+\sigma V^{\p\p}}\right|_{\sigma_{cr}}},
\ee
where we used \eq{betasq}. For closed orbits, the angle must have the value that is $\pi$ multiplied by a rational number, i.e. $\Phi=\pi\frac{r}{q}$ where $q,\; r \in  \mathbb{Z}$; therefore, $\beta$ should be the rational number. In order to obtain the $\sigma$-independent value for $\Phi$, potentials can be of the forms, $\frac{M^2\sigma^l}{2}(+ const.)$, $m^4_{\phi}\ln\bset{\sigma/m_{\phi}}^2(+ const.)$, and etc. Here, $M$ and $m_{\phi}$ are constant real values, and we should have $l<-2,\; 0<l$ for bound orbits, whereas we may have $-2<l<0$ for bound orbits when $M^2<0$, recalling \eq{condbound}. The constant terms in the potentials add an extra energy for the orbits, and it does not play a significant role, so that we consider the potentials without the constant terms. The former power-law potential case, $V=\frac{M^2\sigma^l}{2}$, gives 
\be{powerPhi}
\Phi=\frac{\pi}{\sqrt{l+2}},
\ee
which implies that the closed orbits exist for $l=(-1),\; 2,\; 7,\; \dots$. Using \eqs{pressure}{avgengy} and \eq{condbound}, we obtain 
\be{powerpress}
\state{p}\simeq \frac{(l-2)M^2\sigma^l_{cr}}{4}, \spc \state{\rho_E} \simeq \frac{(l+2)M^2\sigma^l_{cr}}{4}, \spc W^2=\frac{l(l+2)M^2\sigma^{l-2}_{cr}}{2}, 
\ee
which implies that the bound orbits of the AD condensate has a negative pressure for $l<2$. In the computation of $\rho_E$, \eq{avgengy}, we safely ignored the $\varepsilon^4$ term. The bound orbits for $l=(-1,)\; 2$ are closed. For the quadratic potential case $l=2$, the averaged pressure is zero, in which the AD condensate corresponds to an example of nonrelativistic cold dark matter \cite{Turner:1983he}. In addition, using \eqs{powerpress}{eosg} we can find
\be{compress}
\state{w}\simeq \frac{l-2}{l+2}.
\ee

On the other hand, the latter logarithmic potential case, $m^4_{\phi}\ln\bset{\sigma/m_{\phi}}^2$, leads to 
\be{logPhi}
\Phi=\frac{\pi}{\sqrt{2}}\sim \frac{2\pi}{3},
\ee
which corresponds to the former power-law case with $l=0$. Similarly, using \eqs{pressure}{avgengy} and \eq{condbound}, we obtain 
\be{presslog}
\state{p}\simeq m^4_{\phi}\bset{1-2\ln{\frac{\sigma_{cr}}{m_{\phi}}}}, \spc \state{\rho_E} \simeq m^4_{\phi}\bset{1+2\ln{\frac{\sigma_{cr}}{m_{\phi}}}}, \spc W^2=\frac{4m^4_{\phi}}{\sigma^2_{cr}},
\ee
which implies that the AD condensate has a negative pressure for $\sigma_{cr}>m_{\phi}\exp{\bset{\half}}$. In the computation of $\rho_E$, \eq{avgengy}, we safely ignored the $\varepsilon^4$ term again. Using \eqs{eosg}{presslog}, we also find
\be{compresslog}
\state{w}\simeq \frac{1-2\ln\bset{\frac{\sigma_{cr}}{m_\phi}}}{1+2\ln\bset{\frac{\sigma_{cr}}{m_\phi}} }.
\ee
In \eq{presslog}, we cannot clearly see the correspondence with the case for $l=0$, but we can find $W^2\simeq 0$ for $m_{\phi}\ll \sigma_{cr}$ and $\state{w} \simeq -1$ for $m_\phi \ll \sigma$ as the case with $l=0$.

\vspace*{10pt}

Let us comment on the pressure when the AD orbit is exact radial, which corresponds to the zero-charge density case as for real fields \cite{Turner:1983he}. In this case, the field $\sigma(t)$ coherently oscillates around the vacuum if the potential follows a power-law, i.e. $V\propto \sigma^l$ for $l>1$, and $\state{w}$ is the same equation as \eq{compress}, but it gives a negative pressure for $1<l<2$. Note that the lower bound of $l$ ensures to be a coherent oscillation for the radially oscillating AD fields and real scalar fields.

\vspace*{10pt}

In summary, we have obtained analytically the explicit expressions, \eqs{tisig}{uel}, for the orbit of AD fields in a quadratic potential under an expanding universe, and approximately obtained the elliptic orbit expressions, \eqs{sigsol}{ext7}, for nearly circular orbits in Minkowski spacetime in potentials which satisfy the condition \eq{condbound} for bound orbits.


\end{document}